\documentclass[12pt,english]{article}
\usepackage[T1]{fontenc}
\usepackage[latin1]{inputenc}
\usepackage{a4wide}
\usepackage{amsmath}
\usepackage{color}
\usepackage{graphicx}
\usepackage{amssymb}

\usepackage{babel}

\newcommand {\beq}{\begin{equation}}
\newcommand {\eeq}{\end{equation}}
\newcommand {\bea}{\begin{eqnarray}}
\newcommand {\eea}{\end{eqnarray}}
\newcommand {\nn}{\nonumber \\}

\newcommand {\e}{{\rm e}}

\newcommand {\K}{{\bf K}}
\newcommand {\I}{{\bf I}}
\newcommand {\J}{{\bf J}}
\newcommand {\N}{{\bf N}}
\newcommand {\KBtil}{\tilde {\bf K}}
\newcommand {\IBtil}{\tilde {\bf I}}
\newcommand {\JBtil}{\tilde {\bf J}}
\newcommand {\NBtil}{\tilde {\bf N}}
\newcommand {\KBhat}{\hat {\bf K}}
\newcommand {\IBhat}{\hat {\bf I}}
\newcommand {\JBhat}{\hat {\bf J}}
\newcommand {\NBhat}{\hat {\bf N}}

\newcommand {\n}{\nu}
\newcommand {\pl}{\partial}
\newcommand {\p} {\phi}
\newcommand {\vp}{\varphi}

\newcommand {\al}{\alpha}
\newcommand {\be}{\beta}

\newcommand {\la}{\lambda}
\newcommand {\La}{\Lambda}
\newcommand {\si}{\sigma}

\newcommand {\om}{\omega}
\newcommand {\Om}{\Omega}
\newcommand {\ep}{\epsilon}
\newcommand {\vep}{\varepsilon}
\newcommand {\na}{\nabla}
\newcommand {\del}  {\delta}
\newcommand {\Del}  {\Delta}

\newcommand {\half}{ {\frac{1}{2}} }

\newcommand {\Lcal}{{\cal L}}

\newcommand {\Ktil}  {{\tilde K}}

\newcommand {\ptil} {{\tilde p}}

\newcommand {\Xtil}{{\tilde X}}
\newcommand {\Ytil}{{\tilde Y}}
\newcommand {\ztil}{{\tilde z}}

\newcommand {\Lhat}{{\hat L}}

\newcommand {\phat}{{\hat p}}

\newcommand {\delh} {{\hat \delta}}

\newcommand {\Kbar}  {{\bar K}}

\newcommand {\Rbar}  {{\bar R}}

\newcommand {\bfZ} {{\bf Z}}

\newcommand  {\xf}{{x^5}}
\newcommand  {\yf}{{y^5}}
\newcommand  {\Zt}{{Z$_2$}}

\newcommand {\intk} {{\int \frac{d^4k}{(2\pi)^4}}}
\newcommand {\intp} {{\int \frac{d^4p}{(2\pi)^4}}}

\newcommand {\change} {\leftrightarrow}
\newcommand {\ra} {\rightarrow}

\newcommand {\pr}   {{\quad .}}
\newcommand {\com}  {{\quad ,}}
\newcommand {\q}    {\quad}

\newcommand {\nl}    {\newline}

\newcommand {\NP}   {Nucl.Phys.}
\newcommand {\PL}   {Phys.Lett.}
\newcommand {\PR}   {Phys.Rev.}
\newcommand {\PRL}   {Phys.Rev.Lett.}

\newcommand {\Pla} {\frac{{\tilde p}}{\omega}}
\newcommand {\tPla} {\frac{{\hat p}}{\omega}}
\newcommand {\Tev} {\frac{{\tilde p}}{T}}
\newcommand {\tTev} {\frac{{\hat p}}{T}}
\newcommand {\PlaH} {\frac{{\hat p}}{\omega}}
\newcommand {\TevH} {\frac{{\hat p}}{T}}
\newcommand {\PlaMn} {\frac{{M_n}}{\omega}}
\newcommand {\TevMn} {\frac{{M_n}}{T}}
\newcommand {\vmp} {v_\mp}
\newcommand {\vm} {v_-}
\newcommand {\vpl} {v_+}

\newcommand {\npl}  {{\frac{n\pi}{l}}}
\newcommand {\mpl}  {{\frac{m\pi}{l}}}

\begin{document}

\title{Field Quantization in 5D Space-Time with Z$_2$-parity
and Position/Momentum Propagator}

\author{S. Ichinose$^{1}$ and A. Murayama$^{2}$}

\maketitle
\begin{center}\emph{$^{1}$
Laboratory of Physics, School of Food and Nutritional Sciences, 
University of Shizuoka\\
Yada 52-1, Shizuoka 422-8526, Japan;\ E-mail: ichinose@smail.u-shizuoka-ken.ac.jp
}\end{center}

\begin{center}\emph{$^{2}$
Department of Physics, Faculty of Education, Shizuoka University,
Shizuoka 422-8529, Japan;\ E-mail: a-murayama@mountain.ocn.ne.jp
}\end{center}

\begin{abstract}
Field quantization in 
5D flat and warped space-times with Z$_2$-parity 
is comparatively examined. 
We carefully and closely derive 
5D position/momentum(P/M) propagators.
Their characteristic behaviours depend on the
4D (real world) momentum in relation to the boundary parameter ($l$)
and the bulk curvature ($\om$). They also depend on
whether the 4D momentum is space-like or time-like.   
Their behaviours are graphically presented and  
the Z$_2$ symmetry, the "brane" formation and the singularities are
examined. It is
shown that the use of absolute functions is important
for properly treating the singular behaviour. 
The extra coordinate appears 
as a {\it directed} one like the temperature. The $\delta(0)$ problem, 
which is an important consistency check of the  
bulk-boundary system, is solved {\it without} the use of KK-expansion. 
The relation between P/M propagator (a closed expression
which takes into account {\it all} KK-modes)
and the KK-expansion-series propagator is clarified. 
In this process of comparison, two views on the extra space
naturally come up: orbifold picture and interval (boundary) picture. 
Sturm-Liouville expansion ( a generalized Fourier expansion ) 
is essential there. 
Both 5D flat and warped quantum systems are 
formulated by the Dirac's bra and ket vector formalism, 
which shows the warped model  can be regarded as a {\it deformation}
of the flat one with the {\it deformation parameter} $\om$.
We examine the meaning of the position-dependent cut-off
proposed by Randall-Schwartz.
\end{abstract}

PACS: 
PACS NO:
04.50.+h,\ 
11.10.Kk,\ 
11.25.Mj,\ 
12.10.-g 
11.30.Er,\ 

\section{Introduction\label{S.Intro}}
Since the discovery of the wall picture model of our space-time in 1999
by Randall and Sundrum\cite{RS9905,RS9906}, 
eight years have passed. The model has surely brought a new tool to
extend the standard model in the particle physics and in the cosmology.
The mass hierarchy between the Weak scale and the Planck scale can be
introduced more naturally than any other models before. 
The exponentially damping factor along the extra coordinate
cause the mass scale so rapidly damp that the widely-ranging
mass scales in nature could be naturally explained. 
Although it is
still regarded as one of promising candidates as a beyond-standard model, 
some fundamental points below are not clear.
\begin{enumerate}
\item Stableness of the Randall-Sundrum(RS) model I
\item Renormalizability
\item Physical observables, BRS structure
\end{enumerate}
The second and third ones are common problems in the higher dimensional
(field theory) models. 
At the beginning stage of the model, it may be allowed to disregard them 
for the reason that it is an effective theory which should be 
derived from a more fundamental
one such as the string theory, D-brane theory and M-theory. Recently, however, 
the above problems have gradually become serious because, inspired by the
soon-coming LHC experiment, we are compelled to estimate the extra-dimensional
effect in some physical quantities such as 
the B-physics experiments data,  
the electric dipole moment\cite{soni04} and the hadron spectroscopy\cite{brodsky05}.
The problems cited above make these calculations have some ambiguity. 

A source of (technical) difficulty of higher-dimensional models
is the summation of all KK modes. 
In Ref.\cite{GT9911,GKR0002,Pomarol0005,GP0012}, a new type propagator 
which takes into account {\it all} KK-modes was used. 
In the paper by Randall-Schwartz\cite{RS01}, it was closely examined and was called 
position/momentum(P/M) propagator. 
It was applied to the $\beta$-function calculation and the analysis of 
the unification of coupling in GUTs. 
We focus on the P/M propagator behavior
and the relation to the familiar KK-expansion approach. In the analysis two kinds of
standpoints about the extra axis naturally appear. They were pointed out by Ho\v{r}ava and Witten
\cite{Witten9602,HW9603} in the context of Calabi-Yau compactification of eleven-dimensional
supergravity and $E_8\times E_8$ heterotic string affairs. 
One is called {\it orbifold} approach. We regard the extra space as $S_1/Z_2$ by requiring
, in the real space ${\bf R}$, the periodicity and $Z_2$-parity symmetry. The other view
is called the {\it interval} approach. We simply regard the extra space a finite interval $[0,l]$. We will
examine the two approaches comparatively.

Another difficulty is the lack of the clear systematic treatment of singular functions
which have singularities at fixed points. We must handle functions involving
$\del(x)$ and $\ep(x)$ in relation to the symmetry requirement. 
We have been worried by the consistency with the field equation. 
We will develop a new method using the absolute function. 

P/M propagator approach also gives us some new treatment about the regularization
of divergences in quantization. It has a coordinate (position parameter), instead of
a momentum, for the extra space description. In the original paper\cite{RS01}, the real world 
4-momentum integral is regularized by the extra-space {\it position-dependent} cutoff. 
In fact the regularization successfully works in the Randall-Schwartz's paper and the finite $\beta$-function
of the gauge coupling is obtained. 
This kind of 
regularization had never been taken before Ref.\cite{RS01}. 

The content is organized as follows. We start by, in Sec.\ref{S.5Dprop}, the 5D massless scalar field 
propagator on ${\cal M}\times S^1/Z^2$. It serves as the firm reference 
that is compared later with the warped case. For the quantization 
of 5D space-time, we introduce Dirac's bra and ket vector formalism in Sec.\ref{S.braket}. 
This is again the preparation for the quantization of the 5D warped space-time. 
In Sec.\ref{P/Mapproach}, P/M propagator is explained carefully taking the simple model of Sec.\ref{S.5Dprop}. 
A systematic treatment of the extra coordinate, in relation to the symmetrization
of the P/M propagator, is explained. It is shown that, in Sec.\ref{S.KKvsPM}, the KK-expansion approach of Sec.\ref{S.5Dprop} 
and the P/M propagator approach of Sec.\ref{P/Mapproach} are related by the Fourier expansion. 
In Sec.\ref{AdS5}, the warped space-time is treated using the eigen-function expansion. 
Two alternative coordinates, $y$ and $z$, are used. Through the Dirac's formalism analysis
of the warped system, we see it is a deformation of the flat system 
with the deformation parameter (bulk curvature)$\om$. P/M propagator 
is obtained with much care for the Z$_2$ symmetry and the singularity 
in Sec.\ref{P/MproAdS5}. The z-coordinate is used there. In Sec.\ref{SLvsPM}, the relation between 
the eigen-function expansion approach (Sec.\ref{AdS5}) and the P/M propagator approach (Sec.\ref{P/MproAdS5}) 
is clarified using the Sturm-Liouville expansion. As the visual output 
of the present analysis, we present the graphical display of the P/M propagators
in Sec.\ref{S.P/Mpropagator}. We can clearly see how the characteristic scales and Z$_2$-symmetry 
appear, in particular, 
the distinct propagator behaviours between the flat and warped cases. 
The $\del(0)$-problem, which generally appear in the bulk-boundary system, 
is solved, in Sec.\ref{deltazero}, using the results obtained before. We conclude in Sec.\ref{Conc}. 
Three appendices are ready to supplement the text. App.A explains the Sturm-Liouville
expansion in relation to the familiar Fourier expansion. In App.B, a general treatment
of the propagator is given.  In App.C, we display the propagator graphs for various 
interesting cases:\ 
1)\ flat massless scalar with Z$_2$-parity even, Neumann-Neumann boundary condition (b.c.);\ 
2)\ flat massless scalar with Z$_2$-parity odd, Dirichlet-Neumann b.c.;\ 
3)\ z-representation, warped scalar with Z$_2$-parity odd, Dirichlet-Dirichlet b.c., space-like;\ 
4)\ warped massless vector with Z$_2$-parity even, Neumann-Neumann b.c., space-like;\ 
5)\ z-representation of 4);\ 
6)\ warped massless vector with Z$_2$-parity even, Neumann-Neumann b.c., time-like.
\section{5D propagator on flat geometry ${\cal M}_4\times S^1/Z^2$\label{S.5Dprop}}
\begin{figure}
\caption{The extra space ${\bf R}=(-\infty,\infty)$ in which the propagator is
periodically (periodicity 2l) defined.}
\includegraphics[height=2cm]{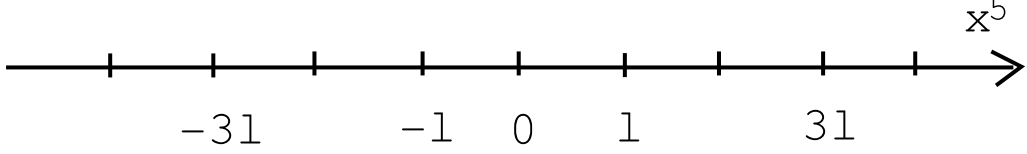}
\label{Rperiodic}
\end{figure}

The simplest and most popular higher-dimensional model is 5D model
with the circle as the one extra space-manifold. It began with 
the original ones by Kaluza\cite{Kal21} and Klein\cite{Klein26}. 
The bulk curvature vanishes, hence we call this model 'flat model'
in comparison with 'warped model' later explained. 
We first consider the 5D massless scalar, on S$^1$/Z$^2$, 
interacting with an external source $J(X)$.
%
\bea
S=\int d^5X(-\half\pl_M\Phi~\pl^M\Phi+J\Phi)\com\q
(\eta^{MN})=\mbox{diag}(-1,1,1,1,1)\ ,\nn
(X^M)=(x^m,\xf)\equiv (x,\xf)\ ,\ 
M,N=0,1,2,3,5;\ m,n=0,1,2,3.
\label{prop1}
\eea
The field $\Phi$ and the source $J$ 
have the properties:
\bea
\mbox{(1) Periodicity}\q\q
\Phi(x,x^5)=\Phi(x,x^5+2l)\ ,\ 
J(x,x^5)=J(x,x^5+2l)\com \nn
\mbox{(2) Z$_2$-property (5D parity)}\q\q
\Phi(x,x^5)=\mbox{P}\Phi(x,-x^5)\ , \
J(x,x^5)=\mbox{P} J(x,-x^5)\com
\label{prop2}
\eea
where P=$-$1 (odd) or +1 (even).  
The above choice comes from the requirement
that the 5D theory (\ref{prop1}) is Z$_2$-invariant.
The extra space manifold is shown in Fig.\ref{Rperiodic}. 
From the periodicity, we can express as
\bea
\Phi(x,x^5)=\sum_{n\in\bfZ}\phi_n(x)\e^{i\frac{n\pi}{l}\xf}\com
\label{prop3}
\eea
where $\bfZ$ is the set of all integers. 
This is the Kaluza-Klein (KK) expansion.  
The \Zt-property (\ref{prop2}) requires the above coefficients
to be
\bea
\phi_n(x)=\mp\phi_{-n}(x)\q\mbox{for}\q \mbox{P}=\mp
\pr
\label{prop4}
\eea
The plural signs $P=\mp$ used in the paper mean that
the upper sign case (lower sign case)
corresponds to that of another quantities.

In particular, $\phi_0=0$ for the odd case ($P=-1$).
\footnote{
There is {\it no} zero mode for the odd parity. This fact will be
utilized in some places later.
}
 Then we obtain
\bea
\Phi(x,\xf)=\left\{
\begin{array}{cc}
2i\sum_{n=1}^{\infty}\phi_n(x)\sin(\npl\xf),& P=-1\\
\phi_0(x)+2\sum_{n=1}^\infty\phi_n(x)\cos(\npl\xf),& P=+1
\end{array}
            \right.
\pr
\label{prop5}
\eea
The odd and even functions w.r.t. $\xf$ appear
for $P=-1$ case and for $P=1$ case respectively.
Similarly for $J(x)$.
\bea
J(x,\xf)=\left\{
\begin{array}{cc}
2i\sum_{n=1}^{\infty}j_n(x)\sin(\npl\xf),& P=-1\\
j_0(x)+2\sum_{n=1}^\infty j_n(x)\cos(\npl\xf),& P=+1
\end{array}
            \right.
\pr
\label{prop6}
\eea
Using the orthogonality relations:\ 
\bea
\int_{-l}^l d\xf\sin(\mpl\xf)\sin(\npl\xf)=
\left\{
\begin{array}{cc}
0 & m\neq n \\
l & m=n(\neq 0) \\
0 & m=n=0
\end{array}
\right.
\com\nn
\int_{-l}^l d\xf\cos(\mpl\xf)\cos(\npl\xf)=
\left\{
\begin{array}{cc}
0 & m\neq n \\
l & m=n(\neq 0) \\
2l & m=n=0
\end{array}
\right.
\com
\label{prop7}
\eea
the equation (\ref{prop6}) is "inverted" w.r.t. $j_n$.
\bea
P=-1\com\q
j_n(x)=\frac{1}{2il}\int_{-l}^l d\xf~J(x,\xf)\sin(\npl\xf) \ ,\ 
n=1,2,3,\cdots\\
P=+1\com\q
j_n(x)=\frac{1}{2l}\int_{-l}^l d\xf~J(x,\xf)\cos(\npl\xf) \ ,\ 
n=0,1,2,3,\cdots
\label{prop8}
\eea
With above properties purely from the boundary conditions,
let us solve the 5D field equation of (\ref{prop1}). 
\bea
\pl_M\pl^M\Phi(X)=-J(X)\pr
\label{prop9}
\eea
Inserting (\ref{prop5}) and (\ref{prop6}), we obtain
the 4D Klein-Gordon equation for the n-th Kaluza-Klein mode.
\bea
\{\pl_m\pl^m-(\npl)^2\}\phi_n(x)=-j_n(x)\com\q
\left\{
\begin{array}{cc}
n=1,2,\cdots & P=-1 \\
n=0,1,2,\cdots & P=1
\end{array}
\right.
\com
\label{prop10}
\eea
where $\pl_M\pl^M=\pl_m\pl^m+\pl_5\pl_5$\ 
($(\eta^{mn})=\mbox{diag}(-1,1,1,1)$) is used.
\footnote{
The set of eigenvalues $\{n\pi/l |n=1,2,3,\cdots \}$ for $P=-1$ and 
$\{n\pi/l |n=0,1,2,\cdots \}$ for $P=1$, are the same except the zero mode. 
They are {\it equally} spaced. 
This is contrasting with the warped case appeared later (\ref{adsfM4}). }
We note the massless mode appears for the even case (P=1)
 and does not for the odd case (P=-1).
$\phi_n(x)$ can be solved by the Feynman propagator.
\bea
\phi_n(x)=\int d^4y~\Del_F^n(x-y)j_n(y)\com\nn
\{\pl_m\pl^m-(\npl)^2\}\Del_F^n(x-y)=-\del^4(x-y)\ ,\ 
\Del_F^n(x)=\int\frac{d^4k}{(2\pi)^4}
\frac{\e^{-ikx}}{k^2+(\npl)^2-i\ep}
\com
\label{prop11}
\eea
where $k^2=k_mk^m=-(k^0)^2+(k^1)^2+(k^2)^2+(k^3)^2$. 
Using (\ref{prop5}), (\ref{prop11}) and (\ref{prop8}), we finally
obtain
\bea
\Phi(X)=\int d^4y\int_{-l}^{l}dy^5\Del_F(X,Y)J(Y)\ ,\ 
(X^M)=(x^m,x^5),(Y^M)=(y^m,y^5),\nn
\Del_F(X,Y)=\frac{1}{2l}\sum_{n\in\bfZ}\Del_F^n(x-y)
\half (\e^{-i\npl(\xf-\yf)}+P\e^{-i\npl(\xf+\yf)} )\nn
=\int d^5K
\frac{\e^{-ik(x-y)}}{k^2+(k^5)^2-i\ep}
\half (\e^{-ik^5(\xf-\yf)}+P\e^{-ik^5(\xf+\yf)})
\com
\label{prop12}
\eea
where $\int d^5K\equiv \frac{1}{2l}\sum_{k^5}\int \frac{d^4k}{(2\pi)^4}, 
\{k^5\}\equiv\{\frac{n\pi}{l}| n\in \bfZ\}$. 
Later we will come back here to confirm a result obtained
by the new approach is correct.
\footnote{
Note that the arguments in $\Del_F$ should not be $X-Y$, 
because P-part is the function of $x^5+y^5$, not of $x^5-y^5$. 
         }
5D propagator $\Del_F(X,Y)$ satisfies 
\bea
\pl_M\pl^M\Del_F(X,Y)=-\half (\del^5(X-Y)+P\del^5(X-\Ytil))\com
\label{prop13}
\eea
where $(Y)=(y^m, y^5), (\Ytil)=(y^m, -y^5)$ and 
$\delh(x^5\mp y^5)\equiv \frac{1}{2l}\sum_{n\in {\bf Z}}
\exp \{-i\frac{n\pi}{l}(x^5\mp y^5) \}$ is the {\it periodic} delta function
with the periodicity $2l$.

\section{Dirac's bra and ket vector formalism\label{S.braket}}
P.A.M. Dirac\cite{Dirac39} introduced "bra and ket vector formalism" to formulate
the quantum theory in the abstract way. The formalism clearly
presents the authogonality and the completeness relation
between quantum states.
In the completeness relation, Dirac's delta function generally appears. 
It is a singular function which should be properly treated. 
We show the 5D quantum field theory on
Z$_2$-orbifold (flat case and warped case) can be naturally  
expressed in this formalism.

We introduce bra-vectors $<K|, <X|$, and ket-vectors $|K>, |X>$, in the
Hilbert space of (abstract) quantum states labeled by 
{\it 5D momentum} $K$ and by
{\it 5D coordinate} $X$. 
\bea
\mbox{e}^{iK\cdot X}&=\exp \{ ik_m\cdot x^m+ik_5x^5 \}\equiv <K|X>\com  \nn
\mbox{e}^{-iK\cdot X}&=\exp \{ -ik_m\cdot x^m-ik_5x^5 \}\equiv <X|K>\com \nn
<K|X>^*\ =\ <X|K>\com
\label{bracket1}
\eea
where $(K_M)=(k_m, k_5=\frac{n\pi}{l})$, $(X^M)=(x^m,x^5)$, 
$(\eta^{MN})=\mbox{diag}(-1,1,1,1,1)$ and the symbol 
"$\mbox{}^*$" means the complex conjugate operation. 
From the {\it completeness} property of $\e^{ikx}$, we know
\bea
\int\frac{d^4k}{(2\pi)^4}\frac{1}{2l}\sum_{k_5}<X|K><K|Y>\equiv 
\int d^5K <X|K><K|Y>  \nn
=\del^4(x-y)\delh(x^5-y^5)\equiv \del^5(X-Y)\com\\
\int\frac{d^4x}{(2\pi)^4}\frac{1}{2l}\int_{-l}^{l}dx^5<K|X><X|P>
\equiv \frac{1}{2l}\int d^5X<K|X><X|P> \nn
=\del^4(k-p)\del_{k_5p_5}\equiv\frac{1}{2l}\del^5(K-P)\com
\label{bracket2}
\eea
where $\delh(x^5-y^5)\equiv \frac{1}{2l}\sum_{n\in {\bf Z}}
\exp \{-i\frac{n\pi}{l}(x^5-y^5) \}$ and 
$\delh(k_5-p_5)\equiv 2l\del_{k_5p_5}$ are used.
We {\it require} the orthogonality between the coordinate states $|X>$, 
and the momentum states $|K>$. 
\bea
<X|Y>=\del^5(X-Y)\com\q
<K|P>=\del^5(K-P)\com
\label{bracket3}
\eea
then the {\it completeness},
\bea
\int d^5K |K><K|=1\com\q
\int d^5X |X><X|=1\com
\label{bracket4}
\eea
is deduced. 

The 5D flat propagator $\Del_F(X,Y)$ of the previous section
can be expressed as
\bea
\Del_F(X,Y)=\frac{1}{2l}\sum_{n\in {\bf Z}}\intk
\frac{\e^{-ik(x-y)}}{k^2+(\npl)^2-i\ep}
\times\half(\e^{-i\npl(x^5-y^5)}+P\e^{-i\npl(x^5+y^5)}) \nn
=\int d^5K\frac{1}{K^2-i\ep}\half (\e^{-iK(X-Y)}+P\e^{-iK(X-\Ytil)})\com 
\label{bracket5}
\eea
where $\Ytil^M\equiv (y^m,-y^5)$. This can be further expressed as
\bea
\Del_F(X,Y)=\half\int d^5K\frac{<X|K><K|Y>+P<X|K><K|\Ytil>}
{K^2-i\ep}\notag\\
=\half<X|\int d^5K\frac{|K><K|}{K^2-i\ep}|Y>
+\half P<X|\int d^5K\frac{|K><K|}{K^2-i\ep}|\Ytil>\pr
\label{bracket6}
\eea
Note that Z$_2$ parity property is naturally presented 
by the Z$_2$-parity changed states $|\Ytil>$.  
From the relation $\pl_M\pl^M\Del_F(X,y)=-\half \{\del^5(X-Y)+P\del^5(X-\Ytil)\}$,
we can see the consistency with (\ref{bracket1}). 
\bea
\pl_M\pl^M <X|K>=-K^2 <X|K>\pr
\label{bracket7}
\eea
We will see, in Sec.\ref{AdS5}, this formalism holds true also for the {\it warped} case.

\section{Position/Momentum Propagator Approach\label{P/Mapproach}}
Let us do the previous analysis in the way free from the eigen-mode expansion.

\subsection{P/M propagator Approach}
  Although the extra coordinates are not observed at present, 
the coordinates could be different from others as its role in the quantum field theory (QFT). 
The treatment of Sec.\ref{S.5Dprop} and Sec.\ref{S.braket}
is just the 5 dimensional generalization of the ordinary (4 dimensional)
QFT. We have treated there the extra coordinate on the equal footing
with others. We introduce, in this section, the new approach to the propagator
where the extra coordinate is differently treated with others.

\begin{figure}
\caption{The interval $[-l, l]$ in which the propagator is defined}
\includegraphics[height=2cm]{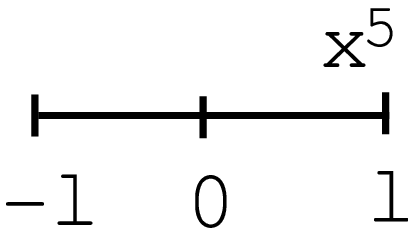}
\label{FiniteRegion}
\end{figure}

We start from eq.(\ref{prop9}).
\bea
\pl_M\pl^M\Phi(X)=-J(X)\pr
\label{PMapp1}
\eea
As for the region of the extra-coordinate and 
the boundary on $\Phi$, we take different ones. 
The extra space is the {\it interval}
\footnote{finite real region}
 [-l,l], and
Z$_{2}$ symmetry only is imposed.
\footnote{
The approach of Sec.\ref{S.5Dprop} is called "orbifold picture" or "up-stairs picture",
while that of this section "interval (boundary) picture" or "down-stairs picture".
\cite{Witten9602,HW9603}
For the recent discussion on the supersymmetric case, see Ref.\cite{Belyaev0509a,Belyaev0509b}
         } 
(Note that we do {\it not} consider the periodicity in this section.)
We take odd one, P=$-$1, for the explicit presentation.  
\bea
\mbox{Z$_2$-property (5D parity)}\q\q
\Phi(x,x^5)=-\Phi(x,-x^5)\ , \
J(x,x^5)=-J(x,-x^5)\com
\label{PMapp2}
\eea
The extra space manifold is shown in Fig.\ref{FiniteRegion}.
We will take into account the extension to {\bf R}=\{ $-\infty <x^5<\infty$\}
and the periodicity later(Sec.\ref{S.KKvsPM}). In order to solve (\ref{PMapp1}) we define
the 5D propagator $\Del(X,X')$ as follows. 
\bea
\Phi(X)=\int d^5X'\Del(X,X')J(X')\com\nn
\pl_M\pl^M\Del(X,X')={\pl_M}'{\pl^M}'\Del(X,X')=
-\half (\del^5(X-X')+P\del^5(X-\Xtil'))\com\q P=-1   \com
\label{PMapp3}
\eea 
where $(\Xtil)\equiv (x^m, -x^5)$ and 
$\Del(X,X')$ is defined in the symmetric way w.r.t. $X\leftrightarrow X'$.
Now we introduce the position/momentum propagator 
$G_p(y\equiv x^5,y'\equiv {x^5}')$ as follows.
\bea
\Del(X,X')\equiv \intp \e^{ip(x-x')}G_p(y,y')\com\nn
(-p^2+\frac{\pl^2}{\pl y^2})G_p(y,y')=
(-p^2+\frac{\pl^2}{\pl {y'}^2})G_p(y,y')=
-\half (\del(y-y')+P\del(y+y'))\com\q P=-1\com\nn
p^2\equiv p_mp^m\pr
\label{PMapp4}
\eea 
The symmetries of the defining equation of $G_p(y,y')$ above are
\begin{description}
\item[Sym(A)] $y\change y'$
\item[Sym(B)] $y\ra -y \ \mbox{and}\  y'\ra -y'$\com\q $P=(-1)\times (-1)=1$
\item[Sym(C)] $y\ra -y$\com\q $P=-1$
\item[Sym(C')] $y'\ra -y'$\com\q $P=-1$
\end{description}

Corresponding to the specific choice
of Z$_{2}$-parity, P=-1 in (\ref{PMapp2}), we must take the 
Dirichlet boundary condition (b.c.) at x$^5$=0.
We also take the same one at x$^5$=l.
\footnote{
We may take the Neumann b.c. at $x^5=l$. 
This choice is excluded in the case of Sec.\ref{S.5Dprop} (up-stairs picture), 
because
the periodicity and the continuity requires the vanishing
of the function at $x^5=l$. See App.C.2 for this case.
}
\bea
G_p(y=0,y')=0\com\q G_p(y,y'=l)=0\q\mbox{for}\q 0\leq y<y'\leq l\com\nn
G_p(y,y'=0)=0\com\q G_p(y=l,y')=0\q\mbox{for}\q 0\leq y'<y\leq l\pr
\label{PMapp5}
\eea
The corresponding conditions for others ($y,y'=-l$) are 
assumed in the same way. 
(When we take the even case of $Z_{2}$-parity in (\ref{PMapp2}), 
the Neumann b.c. is imposed at x$^5$=0. See App.C.1.). 

We consider first the case $p^2>0$, that is, $p^m$ is the {\it space-like} 
4-momentum. 

\begin{figure}
\caption{8 regions}
\includegraphics[height=8cm]{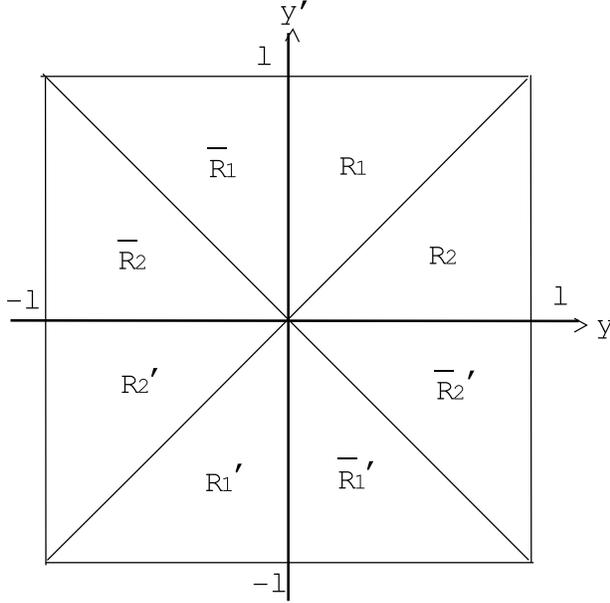}
\label{8regions}
\end{figure}

We divide the whole region into 8 ones $R_1, R_2, R_1', R_2', \Rbar_1,
\Rbar_2, \Rbar_1'$ and $\Rbar_2'$ as in Fig.\ref{8regions}. 

{\bf Step 1.}Region R$_1$ and R$_2$

We start by solving (\ref{PMapp4}) for the region
$0\leq y , y'\leq l$. \nl
(i)$y\neq y'$ .\nl
In this case the equation (\ref{PMapp4}) reduces to the
{\it homogeneous} one.
\bea
(-\ptil^2+\frac{\pl^2}{\pl y^2})G_p(y,y')=
(-\ptil^2+\frac{\pl^2}{\pl {y'}^2})G_p(y,y')=0\com\q
\ptil\equiv \sqrt{p^2}\pr
\label{PMapp6}
\eea 
The general solution is given by
\bea
K_p(y,y')=A(y')\cosh \ptil y +B(y')\sinh \ptil y
         =C(y)\cosh \ptil (y'-l)+D(y)\sinh \ptil (y'-l)
\com
\label{PMapp7}
\eea 
where  $A(y'), B(y'), C(y)$ and $D(y)$ are to be fixed by
the boundary conditions. We take the solution as
\bea
G_p(y,y')=K_p(y,y')=A(y')\cosh \ptil y +B(y')\sinh \ptil y \q 
\mbox{for}\q R_1\ :\ 0\leq y<y'\leq l\nn
G_p(y,y')=K_p(y',y)=C(y')\cosh \ptil (y-l)+D(y')\sinh \ptil (y-l) \q 
\mbox{for}\q R_2\ :\ 0\leq y'<y\leq l
\ ,
\label{PMapp8}
\eea 
where we have used the sym(A) of (\ref{PMapp4}). 
Note that the lower equation is the $y\change y'$ exchanged one of 
the upper equation. This will be utilized in the inhomogeneous case below. 
Applying the b.c. (\ref{PMapp5}) to the above result, we see
\bea
A(y')=0\com\q C(y)=0 \com
\label{PMapp9}
\eea 
in (\ref{PMapp7}).

(ii) $|y-y'|\ <\ +0$\nl
We must take into accout the {\it inhomogeous} term, 
the singularity $\half\del(y-y')$ in (\ref{PMapp4}). 
To do it, we note the following fact.
The absolute-value function $\half |y-y'|$ satisfies the following relation.
\bea
v(y,y')\equiv \half |y-y'|=
          \left\{ \begin{array}{ll}
                  \half (y'-y) & \mbox{for}\ y < y' \\
                  \half (y-y') =y\leftrightarrow y'\mbox{ in the above}& \mbox{for}\ y > y' 
                  \end{array}
          \right.\com
\nn
-\frac{\pl v(y,y')}{\pl y}=+\frac{\pl v(y,y')}{\pl y'}=\half \ep(y'-y)\equiv
\left\{\begin{array}{ll}
       +\half & \mbox{for}\q y<y' \\
         0    & \mbox{for}\q y=y' \\
       -\half & \mbox{for}\q y>y'
       \end{array}
\right.\com
\nn
\frac{\pl^2}{\pl y^2}v(y,y')=\frac{\pl^2}{\pl {y'}^2}v(y,y')=
\half\frac{\pl}{\pl y'}\ep (y'-y)=-\half\frac{\pl}{\pl y}\ep (y'-y)=\del(y-y')\com
\nn
\frac{\pl^2}{\pl y \pl y'}v(y,y')=-\del(y-y')\com
\label{PMapp10}
\eea
where $\ep(y)$ is the {\it sign function}. With the above relations and Sym(A)
, we take the following b.c..
\bea\mbox{Jump Condition:}\q
\left.
\left[ -\frac{\pl G_p(y,y')}{\pl y}+\frac{\pl G_p(y,y')}{\pl y'}\right]
\right|_{y'\ra y+0}
=\half\q \mbox{for}\q y<y'\pr
\label{PMapp11}
\eea 
(This b.c., {\it in combined with the $y\change y'$ exchanged definition
for $y'<y$ in (\ref{PMapp8})}, simply demands
$G_p(y,y')\sim +\frac{1}{4} |y-y'|=\half v(y,y')$ for $|y-y'|\ll 1$.
) 
This condition and the {\it continuity} 
of the 2nd and 3rd equation of (\ref{PMapp7})
at $y=y'$ 
fix the remainig two functions $B$ and $D$.
\footnote{
Putting $y=y'$ in (\ref{PMapp7}), the continuity condition leads to
$B(y)\sinh \ptil y=D(y)\sinh \ptil (y-l)$. As for the Jump condition (\ref{PMapp11}),
two $G_p(y,y')$'s appear in the left hand side. 
We take $B(y')\sinh \ptil y$ as the first $G_p(y,y')$ and 
$D(y)\sinh \ptil (y'-l)$ as the second $G_p(y,y')$.  
Then it leads to
$-B(y)\ptil \cosh \ptil y +D(y)\ptil \cosh \ptil (y-l)=1$.
} 
Finally we get
\bea
B(y')=\frac{\sinh \ptil (y'-l)}{2\ptil \sinh \ptil l}\com\q
D(y)=\frac{\sinh \ptil y}{2\ptil \sinh \ptil l}\com\nn
G_p(y,y')=\left\{
\begin{array}{ll}
\half\sinh \ptil y\sinh \ptil (y'-l)/\ptil\sinh \ptil l \equiv \Kbar_p(y,y')& \mbox{for}\q 0\leq y<y'\leq l\\
\half\sinh \ptil (y-l)\sinh \ptil y'/\ptil\sinh \ptil l \equiv \Kbar_p(y',y)& \mbox{for}\q 0\leq y'<y\leq l
\end{array}
\right.
\label{PMapp12}
\eea 

We can view the result of Step 1 as follows. The solution for
Region R$_1$ ($0\leq y<y'\leq l$) can be expressed as
\bea
\Kbar_p(y,y')=
\sinh \ptil y\sinh \ptil (y'-l)/2\ptil\sinh \ptil l \nn
=\frac{1}{4\ptil\sinh \ptil l}\left[
\{\cosh \ptil (y'+y)-\cosh \ptil (y'-y)\} \cosh \ptil l\right.\nn
\left. +\{-\sinh \ptil (y'+y)+\sinh \ptil (y'-y)\}\sinh \ptil l
                      \right]\pr
\label{PMapp13}
\eea 
As for the Region R$_2$, the solution is given by $\Kbar_p(y',y)$, 
which is given by changing $\sinh \ptil (y'-y)$ in (\ref{PMapp13})
by $\sinh \ptil (-y'+y)$. In the combined region R$_1$ and R$_2$, this change
is equivelent to change $\sinh \ptil (y'-y)$ by $\sinh \ptil |y'-y|$. 
This procedure of {\it taking} the {\it absolute value} of $(y'-y)$,
at the same time,  
makes the solution have the singularity $\del(y-y')$ and 
satisfy the {\it inhomogeneous} equation. (See eq.(\ref{PMapp10}))

Here we stress the {\it requirement} of 
the exchange symmetry (A) and the $Z_2$ symmetry, 
with the Jump Condition (\ref{PMapp11}), {\it demand} 
the delta-function
source at the fixed point(s). We should compare this with the situation
of the KK-expansion approach in Sec.\ref{S.5Dprop} and \ref{S.braket}, where the singularity
$\del(y-y')$ comes from the {\it completeness} of the eigen functions
$\delh(y-y')\equiv \frac{1}{2l}\sum_{n\in {\bf Z}}
\exp \{-i\frac{n\pi}{l}(y-y') \}$. ( This situation is represented in the
mathematical relation (\ref{PMapp14b}) derived below.)

{\bf Step 2.}Extension to Region R$_1'$ and R$_2'$\nl
\q Here we extend the solution to Regions R$_1'$ and R$_2'$. 
We make use of the symmetry Sym(B). In order to make
the solution (\ref{PMapp13})have the symmetry Sym(B), we must {\it take}
the {\it absolute value} of $(y'+y)$ in $-\sinh \ptil (y'+y)$ besides
$(y'-y)$ in $\sinh \ptil (y'-y)$. 
\bea
G_p(y,y')
=\frac{1}{4\ptil\sinh \ptil l}\left[
\{\cosh \ptil |y'+y|-\cosh \ptil |y'-y|\} \cosh \ptil l\right.\nn
\left. +\{-\sinh \ptil |y'+y|+\sinh \ptil |y'-y|\}\sinh \ptil l
                      \right]\nn
=\frac{1}{4\ptil\sinh \ptil l}
\{\cosh \ptil (|y'+y|-l)-\cosh \ptil (|y'-y|-l)\} \pr
\label{PMapp14}
\eea 
This expression is valid for $R_1'$ and $R_2'$ besides for $R_1$ and $R_2$. 

Taking the limit $\ptil\ra +0$ in the above and (\ref{prop12}), we obtain 
an interesting formula.
\bea
\sum_{n\in \bfZ, n\neq 0}\frac{1}{(\npl)^2}\sin\npl y\sin\npl y'
=\frac{1}{8l}
\{(|y'+y|-l)^2-(|y'-y|-l)^2\} \pr
\label{PMapp14b}
\eea 
In this $P=-1$ case, we can take the limit $\ptil\ra +0$ which means 
there is no massless mode.

{\bf Step 3.}Extension to Region ${\bar R}_1$, ${\bar R}_2$ 
${\bar R}_1'$ and ${\bar R}_2'$  \nl
\q The solution (\ref{PMapp14}) have the symmetries: 
Sym(C) ($y\ra -y$) and Sym(C') ($y'\ra -y'$) with P=-1 property. 
These allow us to use (\ref{PMapp14}) in {\it all regions}. 

The appearance $|y'+y|$ in (\ref{PMapp14}) 
makes the solution have the singularity $\del(y+y')$ 
and satisfy the inhomogenous equation.

We notice, in this 5D propagator treatment, the extra coordinate
behaves as a {\it directed} axis like the {\it temperature}. 
This is because the whole regions of ($y$,$y'$)-plane reduces
to the fundamental region $R_1 (0\leq y<y'\leq l)$.
The property comes from 
the requirement of Z$_2$ symmetry (and the singularity at $y=\pm y'$). 
Wave propagation in the continuum medium with the delta-function sources
 have a {\it fixed direction} in order to satisfy Z$_2$ symmetry.
\footnote{
Similar situation is stressed in the analysis of the fermion
chiral determinant.\cite{SI00PR}
}

\q For the time-like 4-momentum case, $p^2<0$, the solution is 
obtained in the same way and is given by
\bea
G_p(y,y')
=-\frac{1}{4\phat\sin \phat l}\left[
\{\cos \phat |y'+y|-\cos \phat |y'-y|\} \cos \phat l\right.\nn
\left. +\{\sin \phat |y'+y|-\sin \phat |y'-y|\}\sin \phat l
                      \right]\com\nn
=-\frac{1}{4\phat\sin \phat l}\left\{
\cos \phat (|y'+y|-l)-\cos \phat (|y'-y|-l) 
                               \right\}\com
\label{PMapp15}
\eea 
where $\phat=\sqrt{-p^2}$.

Later, in Sec.\ref{S.P/Mpropagator}.1,  
we will show the graphs of (\ref{PMapp14}) and of (\ref{PMapp15})
and examine their behaviour.

\subsection{Systematic treatment of symmetries of $G_p(y,y')$}
\q The defining equation of $G_p(y,y')$, (\ref{PMapp4}), has
the symmetries Sym(A),(B),(C) and (C'). The {\it fundamental region}
is R$_1$, and the solution in other regions can be expressed
by some change of arguments $y, y'$ in the P/M propagator
in the fundamental region, $\Kbar_p(y,y')$. 

In the previous subsection, we have done the procedure
of  taking the absolute value of $y'-y$ for the singularity
$\del(y'-y)$ and of $y'+y$ for $\del(y'+y)$. This procedure
says that the P/M propagator valid in all regions can be obtained
by changing arguments $y, y'$ in $\Kbar_p(y,y')$, which is
the P/M propagator defined in the fundamental region $R_1$ and has
no absolute-value quantities , in the following way.
\footnote{
This substitution is useful particularly in the warped case.
See Sec.\ref{P/MproAdS5}, (\ref{adspro18}). 
} 
\bea
y=\half (y+y')-\half (y'-y)\ra Y(y,y')=\half |y+y'|-\half |y'-y|=
\left\{
\begin{array}{ll}
y  & \mbox{for}\q R_1,\ \Rbar_1 \\
-y' & \mbox{for}\q \Rbar_2,\ R_2' \\
-y  & \mbox{for}\q R_1',\ \Rbar_1' \\
y' & \mbox{for}\q \Rbar_2',\ R_2 
\end{array}
\right.\nn
y'=\half (y+y')+\half (y'-y)\ra Y'(y,y')=\half |y+y'|+\half |y'-y|=
\left\{
\begin{array}{ll}
y'  & \mbox{for}\q R_1,\ \Rbar_1 \\
-y & \mbox{for}\q \Rbar_2,\ R_2' \\
-y'  & \mbox{for}\q R_1',\ \Rbar_1' \\
y & \mbox{for}\q \Rbar_2',\ R_2 
\end{array}
\right.
\label{PMapp16}
\eea 
Hence P/M propagator has the following form, for the space-like
4D momentum case, in each region.
\bea
\begin{array}{ll}
\Kbar_p(y,y')=\frac{\sinh \ptil y\sinh \ptil (y'-l)}{2\ptil\sinh \ptil l} 
                                         & \mbox{for}\q R_1\q\mbox{and}\q \Rbar_1 \com\\ 
\Kbar_p(-y',-y)=\frac{\sinh \ptil y'\sinh \ptil (y+l)}{2\ptil\sinh \ptil l} 
                                         & \mbox{for}\q \Rbar_2\q\mbox{and}\q R_2'\com\\ 
\Kbar_p(-y,-y')=\frac{\sinh \ptil y\sinh \ptil (y'+l)}{2\ptil\sinh \ptil l} 
                                         & \mbox{for}\q R_1'\q\mbox{and}\q \Rbar_1'\com\\ 
\Kbar_p(y',y)=\frac{\sinh \ptil y'\sinh \ptil (y-l)}{2\ptil\sinh \ptil l} 
                                         & \mbox{for}\q \Rbar_2'\q\mbox{and}\q R_2 \pr\\ 
\end{array}
\label{PMapp17}
\eea 
In Fig.4, the arrangement of above propagators is shown.
\begin{figure}
\caption{Arguments Arrangement for Propagators in 8 regions.
The explicit expressions are given in (\ref{PMapp17}).}
\includegraphics[height=8cm]{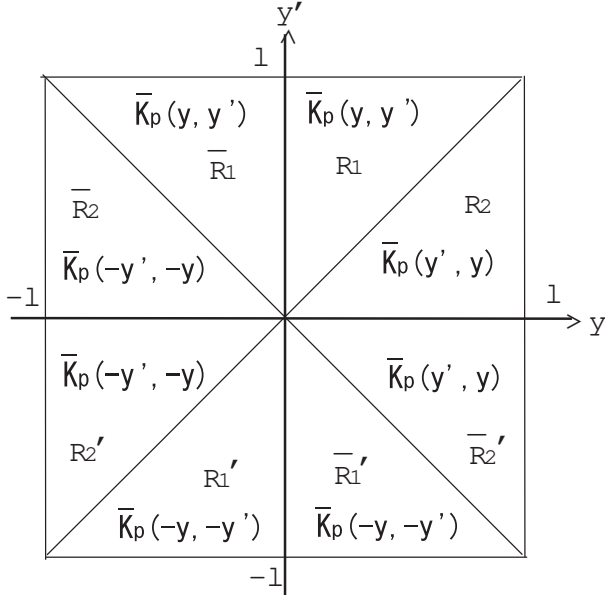}
\label{8regionsProp}
\end{figure}
We surely see the symmetries (A),(B),(C) and (C') are satisfied. 
Using the new variables $Y(y,y'), Y'(y,y')$ , the solution (\ref{PMapp14})
(space-like momentum case) is equivalently expressed as
\bea
G_p(y,y')=\Kbar_p(Y(y,y'),Y'(y,y'))
 \com
\label{PMapp18}
\eea 
where $\Kbar_p$ is given in (\ref{PMapp13}). For the time-like case,
$\Kbar_p$ is given by
$\Kbar_p(Y,Y')=\frac{\sin \phat Y\sin \phat (Y'-l)}{2\phat\sin \phat l}, 
\phat=\sqrt{-p^2}$.

As for the periodic property and the extension to $-\infty <y<\infty$, 
we examine in the next section.

We have given the P/M propagator in the flat 5D scalar theory (with $P=-1$)
in the two ways:\ (1) eq.(\ref{PMapp14}) or eq.(\ref{PMapp15}) 
(compactly by (\ref{PMapp18})) 
where
absolute functions appear, and this expression is valid for all regions;\ 
(2) eq.(\ref{PMapp17}) where {\it no} absolute-value function appears and
the arguments within $\Kbar_p$ change depending on each region. Both ways
are important. The first way is indispensable for 
analyzing the propagator at singular points. Practically it is
necessary to draw a graph which has singular behaviour, 
while the second one stresses the importance of $Z_2$-symmetry. 
(See the $\del(0)$-problem in Sec.\ref{deltazero}.)
\section{KK-expansion approach versus  P/M propagator 
approach\label{S.KKvsPM}}
We have examined the same propagator equation (\ref{prop13}) and (\ref{PMapp3})
in two approaches. One is utilizing the expansion with the eigen functions
(KK-modes) of $\exp(i\npl x^5)$. The other makes use of the P/M propagator
$G_p(y,y')$ in (\ref{PMapp4}). In the former, the extra space is 
{\bf R}=($-\infty < x^5 < \infty$) and the periodic b.c. and
$Z_2$ symmetry are imposed. The 4D space-time propagator is Feynman's. 
Ths is the orbifold approach. 
In the latter case, the extra space is the interval $[-l,l]$. $Z_2$
symmetry only is imposed. 
This is the interval approach. 
We now connect the two results.

The P/M propagator for $0\leq y<y'\leq l$ (Region $R_1$)is given by
\bea
\Kbar_p(y,y')=\nn
\left\{
\begin{array}{ll}
\frac{\sinh \ptil y\sinh \ptil (y'-l)}{2\ptil\sinh \ptil l} 
=\frac{1}{4\ptil \sinh \ptil l}\{\cosh\ptil(y'+y-l)-\cosh\ptil(y'-y-l) \}
\ ,\ \ptil=\sqrt{p^2}
& \mbox{for}\q p^2>0 \\  
\frac{\sin \phat y\sin \phat (y'-l)}{2\phat\sin \phat l} 
=\frac{1}{4\phat \sin \phat l}\{-\cos\phat(y'+y-l)+\cos\phat(y-y'+l) \}
\ ,\ \phat=\sqrt{-p^2}
& \mbox{for}\q p^2<0   
\end{array}
\right.
\label{KKvPM1}
\eea 
Using the {\it Fourier expansion} formulae,
\bea
\frac{\pi}{2a}\frac{\cosh a(\pi-x)}{\sinh a\pi}=\frac{1}{2a^2}
+\sum_{n=1}^\infty\frac{\cos nx}{n^2+a^2},\ 0\leq x\leq 2\pi,\ 
a:\ \mbox{arbitrary constant.}\nn
\frac{\pi}{2a}\frac{\cos ax}{\sin a\pi}=\frac{1}{2a^2}
+\sum_{n=1}^\infty(-1)^{n-1}\frac{\cos nx}{n^2-a^2},\ -\pi\leq x\leq \pi,\ 
a:\ \mbox{non-integer arbitrary constant.}
\label{KKvPM2}
\eea 
we obtain
\bea
\Kbar_p(y,y')=\left\{
\begin{array}{ll}
-\frac{1}{l}\sum_{n=1}^\infty\frac{1}{\ptil^2+(\npl)^2}\sin\npl y\sin\npl y' &\mbox{for}\q p^2>0\\
\frac{1}{l}\sum_{n=1}^\infty\frac{1}{\phat^2-(\npl)^2}\sin\npl y\sin\npl y' &\mbox{for}\q p^2<0
\end{array}
\right.
\label{KKvPM3}
\eea 
This propagator satisfies the all symmetries (A),(B),(C) and (C')
and is periodic ($y\ra y+2l$), hence we may take it as the propagator
valid for $y,y'\in${\bf R} where {\bf R}=$(-\infty, \infty)$. 
The last procedure is taking the {\it universal covering} of $S_1$. 
Both the space-like case and the time-like case 
are just the result (\ref{prop12}) with $P=-1$.


We reemphasize the following points. In the periodic approach of
Sec.\ref{S.5Dprop}, the delta function singularity comes from the {\it completeness} 
of the eigen-functions $\e^{in\pi y/l}$. In the interval approach of Sec.\ref{P/Mapproach}, 
the singularity comes from the exchange symmetry (A) and Z$_2$-symmetry 
(, that is, taking the absolute values of ($y\mp y'$)). The {\it position} space treatment
gives the intimate relation between the wave propagation with symmetries
and its source (singularity). 
 
\section{5D QFT on warped geometry AdS$_5$ \label{AdS5}}
\begin{figure}
\caption{Relation between coordinates $z$ and $y$}
\includegraphics[height=8cm]{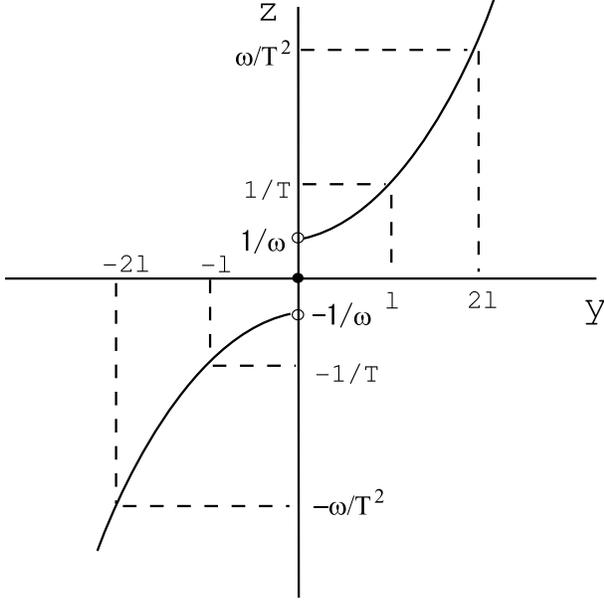}
\label{ZvsY}
\end{figure}

Let us consider the warped case. The space-time is AdS$_5$
manifold. 
\bea
ds^2=\e^{-2\si (y)}\eta_{ab} dx^a dx^b+dy^2\com\q -\infty <y<\infty\com\nn
(\eta_{ab})=\mbox{diag}(-1,1,1,1)\com\q \si(y)=\om |y|\com\q \om>0
\label{adsf1}
\eea 
The manifold has the negative cosmological constant and is
maximally symmetric with the curvature $\om$.  
In the limit $\om\ra 0$, the line element (\ref{adsf1}) goes to the flat one of Sec.\ref{S.5Dprop}. 
We {\it can} consider the symmetries.
\begin{description}
\item[(1) Periodicity]\q$y\ra y+2l$
\item[(2) Z$_2$-property]\q $y\change -y$,\q $P=\pm$ 1 
\end{description}
as in Sec.\ref{S.5Dprop}. When the periodicity condition (1) is assumed, 
$|y|$ in (\ref{adsf1}) is 
the periodic absolute-linear function\cite{IM04PLB}. Instead of $y$-coordinate,
another one $z$, defined below, is also important.
\bea
z=\left\{
\begin{array}{ll}
\frac{1}{\om}\e^{\om y} & \mbox{for}\q y>0\\
& \\
\ 0                      & \mbox{for}\q y=0\\
& \\
-\frac{1}{\om}\e^{-\om y} & \mbox{for}\q y<0
\end{array}
\right.
\q\q=\ \ep(y)\frac{\e^{\om |y|}}{\om}
\label{adsf2}
\eea 
See Fig.\ref{ZvsY}. 

In terms of $z$, the metric
can be expressed as
\bea
ds^2=\frac{1}{\om^2 z^2}(
\eta_{ab} dx^a dx^b+(dz)^2)\com\q 
z> \frac{1}{\om}\ \mbox{or}\ z< -\frac{1}{\om}\com\nn
\frac{dz}{\om z}=\left\{
\begin{array}{ll}
dy & \mbox{for}\q y>0\\
-dy & \mbox{for}\q y<0
\end{array}
\right.
\label{adsf3}
\eea 
$(-\frac{1}{\om},0)$U$(0,\frac{1}{\om})$ is the 'prohibited' region whose values
the z-coordinate {\it cannot} take. This metric is {\it conformal flat}.

The advantage of the choice (\ref{adsf2}) is as follows: 
(a) $z$ is the monotonously increasing function of $y$;\ 
(b) one-to-one ;\  
(c) the $Z_2$ symmetry
is expressed in the same way as $y$. 
\begin{description}
\item[(2') Z$_2$-property]\q $z\change -z$,\q $P=\pm$ 1 
\end{description}
The periodicity is expressed as
\begin{description}
\item[(1') Periodicity\ $y\ra y+2l$]\q
\begin{tabular}{ll}
$z\ra z\e^{2\om l}(>0)$ & for \q $z>\frac{1}{\om}$\\
$z=0\ra z=\frac{\e^{2\om l}}{\om}=\frac{\om}{T^2}$ & for \q$z=0$\\
$z(<0)\ra -z\e^{2\om l}(>0)$ & for \q$-\frac{\om}{T^2}=-\frac{\e^{2\om l}}{\om}<z< -\frac{1}{\om}$\\
$z=-\frac{\e^{2\om l}}{\om}=-\frac{\om}{T^2}\ra z=0$  &  for \q$z=-\frac{\e^{2\om l}}{\om}=-\frac{\om}{T^2}$\\
$z\ra z\e^{-2\om l}(<0)$ & for \q $z< -\frac{\e^{2\om l}}{\om}=-\frac{\om}{T^2}$
\end{tabular}
\end{description}
The {\it translation} in $y$-coordinate is the {\it scale} transformation
in $z$-coordinate. In the conformal coordinate (\ref{adsf3}), we can {\it not}
obtain the {\it flat limit} simply by $\om\ra 0$. 

We take 5D massive scalar theory as a simple example. The Lagrangian
and the field equation is written as
\bea
\Lcal=\sqrt{-G}(-\half \na^A\Phi\na_A\Phi-\half m^2\Phi^2+J\Phi)\com
\q G\equiv \det G_{AB}\com\nn
ds^2=G_{AB}dX^AdX^B\com\q\na^A\na_A \Phi-m^2\Phi+J=0\com
\label{adsf4}
\eea 
where $\Phi(X)=\Phi(x^a,z)$ is the 5D scalar field and $J(X)=J(x^a,z)$ is the
external source field. 
The background geometry is AdS$_5$ which takes the following form, 
in terms of $z$,  
\bea
(G_{AB})=
\left(
\begin{array}{ll}
\frac{1}{\om^2z^2}\eta_{ab} & 0 \\
0               & \frac{1}{\om^2z^2}
\end{array}
\right)
\com\q
\sqrt{-G}=\frac{1}{(\om |z|)^5}\nn
-\frac{1}{T}\leq z \leq -\frac{1}{\om}\q\mbox{or}\q
\frac{1}{\om}\leq z \leq \frac{1}{T}\q (-l\leq y \leq l)\com\q
\frac{1}{T}\equiv \frac{1}{\om}\e^{\om l}
\com
\label{adsf5}
\eea 
where, at present, we take into account {\it only} $Z_2$ symmetry in the interval $y\in [-l,\ l]$. 
Later we will discuss the periodicity condition. The field equation
(\ref{adsf4}) leads to
\bea
\om^2z^2\pl_a\pl^a\Phi+(\om z)^5\pl_z(\frac{1}{(\om z)^3}\pl_z\Phi)
-m^2\Phi+J=0
\com
\label{adsf6}
\eea 
which is $Z_2$-symmetric. We can consider two cases:
\bea
\Phi(x,z)=P\Phi(x,-z)\com\q J(x,z)=PJ(x,-z)\com\nn
P=+1\ (\mbox{even})\com\q P=-1\ (\mbox{odd})
\com
\label{adsf6b}
\eea 
Let us first solve the above equation by the KK-expansion method
as in Sec.\ref{S.5Dprop}. 
\bea
\Phi(x^a,z)=\sum_n\p_n(x)\psi_n(z)\com\q
J(x^a,z)=\sum_n j_n(x)\psi_n(z)\com\q
\psi_n(z)=\mbox{P}\psi_n(-z)\com\q \mbox{P}=\mp1\ ,
\label{adsf7}
\eea 
where $\psi_n(z)$ is the eigen-functions of the Bessel differential
equation. The variable region of $z$ is the intervals defined in (\ref{adsf5}). 
\bea
(\om z)^3\pl_z(\frac{1}{(\om z)^3}\pl_z\psi_n)
-\frac{m^2}{(\om z)^2}\psi_n=-M_n^2\psi_n\nn
(  \int^{-\frac{1}{\om}}_{-\frac{1}{T}} + \int_{\frac{1}{\om}}^{\frac{1}{T}})
\frac{dz}{(\om |z|)^3}\psi_n(z)\psi_k(z)=
2\int_{\frac{1}{\om}}^{\frac{1}{T}}\frac{dz}{(\om z)^3}\psi_n(z)\psi_k(z)
=\del_{nk}
\com
\label{adsf8}
\eea 
where $M_n$ is the eigenvalues to be determined by the b.c.. The above equation
is obtained by the requirement that the 4D field $\p_n(x),\ j_n(x)$ in (\ref{adsf7}) satisfy the ordinary
massive scalar field equation.
\footnote{
Note that we assume here the "4D part" satisfy ${M_n}^2\geq 0$. 
If ${M_n}^2 < 0$, the modified Bessel functions, instead of the Bessel
functions, appear in the following sentences.
} 
\bea
\pl_a\pl^a\p_n-M_n^2\p_n+j_n=0
\pr
\label{adsf9}
\eea 
From eq.(\ref{adsf7}), $\psi_n(z)$ are odd or even functions of $z$.
$z$ varies in  $-1/T \leq z\leq -1/\om$ and $1/\om \leq z\leq 1/T$.  
Hence instead of {\it ordinary} Bessel functions $(\J_\nu (z),\N_\nu)$ 
where $z> 0$, 
it is better to introduce {\it odd or even} Bessel functions where the argument $z$
is valid even for the negative region of $z$.
\footnote{
The Bessel equation
$\frac{d^2\psi}{dz^2}+\frac{1}{z}\frac{d\psi}{dz}+(1-\frac{\n^2}{z^2})\psi=0$ 
 has two independet solutions: Bessel function of the first
kind $J_\nu (z)$ and that of the second kind (Neumann function)$N_\nu(z)$. 
}
 It can be done
as follows because the {\it Bessel equation is invariant
for $Z_2$ symmetry}: $z\ra -z$.
\bea
\begin{array}{lll}
\JBtil_\nu(z)\equiv\ep(z)\J_\nu(|z|) & \NBtil_\nu(z)\equiv\ep(z)\N_\nu(|z|) & \mbox{for P=}-1\\ 
\JBhat_\nu(z)\equiv\J_\nu(|z|) & \NBhat_\nu(z)\equiv\N_\nu(|z|) & \mbox{for P=1}
                       \end{array}
\label{adsf9b}
\eea 
where $\ep(z)$ is the sign function introduced in(\ref{PMapp10}).
\footnote{
The odd or even Bessel functions (\ref{adsf9b}) satisfy (\ref{adsf8}) as far as 
$z$ is in the variable region $-1/T\leq z\leq -1/\om$ or $1/\om \leq z \leq 1/T$. 
}
Taking the case P=$-1$, 
eq.(\ref{adsf8}) has two "intermediate" solutions. 
\begin{eqnarray}
\vp^{(1/\om)}(z) &=& \frac{(\omega 
z)^2}{N^{(1/\om)}}\left[\JBtil_\nu(Mz) +
b_\nu(M)\NBtil_\nu(Mz)\right], \\
\vp^{(1/T)}(z) &=& \frac{(\omega 
z)^2}{N^{(1/T)}}\left[\JBtil_\nu(Mz) +
b_\nu(\omega M/T)\NBtil_\nu(Mz)\right],
\label{adsfM1}
\end{eqnarray}
where $N^{(1/\om)}$ and $N^{(1/T)}$ are some normalization constants to be determined. 
Similarly for the case P=1. $b_\nu$ is given by 
\begin{eqnarray}
b_\nu(t) &\equiv & \left\{\begin{array}{c}
-\mbox{\J}_\nu(t/\omega)/\mbox{\N}_\nu(t/\omega), 
\q\mbox{for} \ P=-1,\\
-\mbox{\J}'_\nu(t/\omega)/\mbox{\N}'_\nu(t/\omega),\q 
\mbox{for} \ P=1,
\end{array}\right.
\label{adsfM2}
\end{eqnarray}
and
\begin{eqnarray}
\nu = \sqrt{4+\frac{m^2}{\om^2}}.
\label{adsfM3}
\end{eqnarray}
These solutions $\vp^{(1/\om)}(z)$ and $\vp^{(1/T)}(z)$ satisfy 
the following boundary conditions at $z=\frac{1}{\om}$
and
$\frac{1}{T}$, respectively. 
\begin{eqnarray}
\begin{array}{cccc}
\mbox{Dirichlet b.c.} &
\vp^{(1/\om)}(z)|_{z=1/\om} = 0 & 
\vp^{(1/T)}(z)|_{z=1/T} = 0 & \mbox{for}\q P=-1 \\
\mbox{Neumann b.c.} &
\left.\frac{d}{dz}\vp^{(1/\om)}(z)\right|_{z=1/\om} = 0 &
\left.\frac{d}{dz}\vp^{(1/T)}(z)\right|_{z=1/T} = 0 & \mbox{for}\q P=1 
\end{array}
\label{adsfM3b}
\end{eqnarray}
The "final" solution which satisfies the b.c. both at $z=1/\om$ and $z=1/T$
is obtained by 
the condition that these two solutions are {\it not} independent.
\begin{eqnarray}
b_\nu(M_n) & = & b_\nu(\omega M_n/T) \equiv b_\nu^{(n)}, \nn
\begin{array}{c}
\frac{\mbox{\J}_\nu(\PlaMn)}{\mbox{\N}_\nu(\PlaMn)}=
\frac{\mbox{\J}_\nu(\TevMn)}{\mbox{\N}_\nu(\TevMn)}, 
\q\mbox{for} \ P=-1,\\
\frac{\mbox{\J}_\nu'(\PlaMn)}{\mbox{\N}_\nu'(\PlaMn)}=
\frac{\mbox{\J}_\nu'(\TevMn)}{\mbox{\N}_\nu'(\TevMn)}, 
\q 
\mbox{for} \ P=1,
\end{array}
\label{adsfM4}
\end{eqnarray}
which makes them identified,
\begin{eqnarray}
\left.\vp^{(1/\om)}(z)\right|_{M=M_n} = \left.\vp^{(1/T)}(z)\right|_{M=M_n} 
\equiv \psi_n(z),
\label{adsfM5}
\end{eqnarray}
and determines the eigenvalues $M_n$. The normalization constants are then 
expressed as
\begin{eqnarray}
N_n^{(1/\om)2} & = &N_n^{(1/T)2}  = \int^{1/T}_{1/\omega} \omega 
zdz \left[\mbox{\J}_\nu(M_n z) + b_\nu^{(n)}
\mbox{\N}_\nu(M_n z)\right]^2 \equiv N_n^2.
\label{adsfM6}
\end{eqnarray}
The solution of (\ref{adsf9}) is given by
\bea
\p_n(x)=\int \Del_F^n(x-x')j_n(x')d^4x'\com\q
\Del_F^n(x-x')=\intk\frac{\e^{-ik(x-x')}}{k^2+M_n^2-i\ep}\com\nn
(\pl_a\pl^a-M_n^2)\Del_F^n(x-x')=-\del^4(x-x')
\pr
\label{adsf10}
\eea 

5D propagator can be expected to be
\footnote{
General proof of this propagator form is given in App.B. 
}
, in the analogy of Sec.\ref{S.5Dprop}'s result (\ref{prop12}), 
\bea
\Phi(X)=\int d^5X'\Del_W(X,X')\sqrt{-G'}J(X')\com\nn
\Del_W(X,X')\equiv \sum_n\Del_F^n(x-x')\times
\half \{ \psi_n(z)\psi_n(z')+P\psi_n(z)\psi_n(\ztil')  \}\com\q
\ztil'\equiv -z'
\pr
\label{adsf11}
\eea 
In fact we can confirm the following propagator equation.
\bea
(\na^A\na_A-m^2)\Del_W(X,X')
=\{ \om^2z^2\pl_a\pl^a+(\om z)^5\frac{\pl}{\pl z}\frac{1}{(\om z)^3}\pl_z-m^2 \}\Del_W(X.X')\nn
=\sum_n\om^2z^2(\pl_a\pl^a-M_n^2)\Del_F^n(x-x')\half\{ \psi_n(z)\psi_n(z')+P\psi_n(z)\psi_n(\ztil') \}\nn
=-(\om |z|)^5\del^4(x-x')\half (\delh(z-z')+P\delh(z-\ztil'))
\com
\label{adsf12}
\eea 
where some relations in (\ref{adsf8}), (\ref{adsf10}) and the {\it completeness} relation: 
\bea
(\om |z|)^{-3}\sum_n\psi_n(z)\psi_n(z')\equiv 
                \left\{
\begin{array}{ll}
\ep(z)\ep(z')\delh (|z|-|z'|) & \mbox{for\ \ P=}-1 \\
\delh (|z|-|z'|) & \mbox{for\ \ P=}1 
\end{array}
                  \right.\com
\label{adsf12b}
\eea
are used. 

Now we have obtained 5D propagator $\Del_W(x^a,y; {x'}^a,y')$ which satisfies $Z_2$ symmetry and is valid
for $y,y'\in[-l, l]$. When we want to extend this to $y,y'\in \mbox{{\bf R}}$ and
impose the periodicity, we may take the {\it universal covering} of $\Del_W(x^a,y; {x'}^a,y')$ as
in the flat case. That is, first we Fourier-expand
\footnote{
Ordinary one using periodic functions. Not Bessel Fourier-expansion.
         }
 $\Del_W(x^a,y; {x'}^a,y')$ within
$y,y'\in[-l, l]$, and then, confirming the non-singular behaviour, extend the variable region
to {\bf R}. 

The above relations can be expressed in the Dirac's bra and ket vector formalism.
Let us introduce bra and ket vectors as follows.
\bea
\psi_n(z)\equiv (n|z)=(z|n)\com\q
\e^{-ikx}\equiv <x|k>\com\q \e^{ikx}\equiv <k|x>\com
\label{adsf13}
\eea 
Depending on the Z$_2$-property of $\psi_n(z)$, $|z)$ and $(z|$ have
the following properties.
\bea
|-z)=P |z)\com\q (-z|=P (z|\com\q P=\mp 1
\pr
\label{adsf13b}
\eea 
From the orthogonality relation (\ref{adsf8}), we know
\bea
\left(\int_{-\frac{1}{T}}^{-\frac{1}{\om}}+\int_{\frac{1}{\om}}^{\frac{1}{T}}\right)
\frac{dz}{(\om |z|)^3}(n|z)(z|k)=
2\int_{\frac{1}{\om}}^{\frac{1}{T}}\frac{dz}{(\om z)^3}(n|z)(z|k)=\del_{n,k}
\pr
\label{adsf14}
\eea 
We require the orthogonality between $(n|$ and $|k)$,
\bea
(n|k)=\del_{n,k}
\pr
\eea
\label{adsf15}
then the completeness relation between the coordinate states $|z)$
\bea
\left(\int_{-\frac{1}{T}}^{-\frac{1}{\om}}+\int_{\frac{1}{\om}}^{\frac{1}{T}}\right)
\frac{dz}{(\om |z|)^3}|z)(z|=
2\int_{\frac{1}{\om}}^{\frac{1}{T}}\frac{dz}{(\om z)^3}|z)(z|={\bf 1}
\pr
\label{adsf16}
\eea
is deduced.

The completeness relation (\ref{adsf12b})
is expressed as
\bea
\sum_n(z|n)(n|z')=\left\{
\begin{array}{ll}
(\om |z|)^3\ep(z)\ep(z')\delh (|z|-|z'|) & \mbox{for\ \ P=}-1 \\
(\om |z|)^3\delh (|z|-|z'|) & \mbox{for\ \ P=}1 
\end{array}
                  \right.
\pr
\label{adsf16b}
\eea
If we require the orthogonality between the coordinate states $(z|$ and $|z')$: 
\bea
(z|z')=\left\{
\begin{array}{ll}
(\om |z|)^3\ep(z)\ep(z')\delh (|z|-|z'|) & \mbox{for\ \ P=}-1 \\
(\om |z|)^3\delh (|z|-|z'|) & \mbox{for\ \ P=}1 
\end{array}
        \right.
\com
\label{adsf16c}
\eea
then the completeness:
\bea
\sum_n|n)(n|={\bf 1}
\com
\label{adsf16d}
\eea
is deduced. 
The 5D propagator (\ref{adsf11}) can be expressed as
\bea
\Del_W(X,X')=\sum_n\intk \frac{<x|k><k|x'>}{k^2+M_n^2-i\ep}\times
\half \{
(z|n)(n|z')+P(z|n)(n|\ztil')
\}\nn
=2l\int d^5K\half\{
\frac{(X|K)(K|X')+P(X|K)(K|\Xtil')}{K^2-i\ep}
\}\com\nn
(X)\equiv (x^a,z)\com\q (X')\equiv ({x^a}',z')\com\q (\Xtil')\equiv ({x^a}',-z')\com\nn
(K)\equiv (k_a,M_n)\com\q \int d^5K\equiv \frac{1}{2l}\sum_n\intk
\pr
\label{adsf17}
\eea
where 5D bra and ket vectors are introduced as
\bea
(X|K)\equiv <x|k>(z|n)\com\q (K|X')\equiv <k|x'>(n|z')
\pr
\label{adsf18}
\eea
This is the same form as in the flat case of Sec.\ref{S.braket}. The generalized
points are the appearance of 
the extra-space 'measure factor' $(\om |z|)^{-3}$ in 
(\ref{adsf14}), (\ref{adsf16}), (\ref{adsf16b}), (\ref{adsf16c}) and, in
the extra dimensional part, the periodic eigen functions are replaced
by the Bessel ones $(n|z)=(z|n)=\psi_n(z)$. 

In this section we have shown the basic quantum structure of the warped
system, in the Dirac's bra and ket vector formalism, is the same as the flat one. 
In this sense, the warped system
can be regarded as a {\it deformation} of the flat theory with
the {\it deformation parameter} $\om$. We will again point out the same interpretation
from the propagator behaviour in Sec.\ref{deltazero}.

\section{P/M Propagator approach to 5D QFT on AdS$_5$\label{P/MproAdS5}}
Let us solve the field equation (\ref{adsf6}) in the P/M propagator 
approach. 
\bea
\na^A\na_A \Phi-m^2\Phi+J=\nn
\om^2z^2\pl_a\pl^a\Phi+(\om z)^5\pl_z(\frac{1}{(\om z)^3}\pl_z\Phi)
-m^2\Phi+J=0\pr
\label{adspro1}
\eea 
We consider the Z$_2$-parity odd case:
\bea
\mbox{Z$_2$-property (5D parity)}\q\q
\Phi(x,z)=\mbox{P}\Phi(x,-z)\ , \
J(x,z)=\mbox{P} J(x,-z)\com\q P=-1\com
\label{adspro2}
\eea
The P/M propagator $G_p(z,z')$ is introduced as
\bea
\Phi(X)=\int d^5X'\Del_W(X,X')\sqrt{-G'}J(X')\com\q\nn
(\na^A\na_A-m^2)\Del_W(X,X')
=-\frac{1}{\sqrt{-G}}\del^4(x-x')\half (\del(z-z')+P\del(z-\ztil'))
\com\q P=-1\com\nn
\Del_W(X,X')=\intp \e^{ip(x-x')}G_p(z,z')\com
\label{adspro3}
\eea 
where $(X^A)=(x^a,z),({X^A}')=({x^a}',z')$ and $G=\mbox{det}G_{AB}$. 
In Sec.\ref{AdS5}, we have derived $\Del_W$ in the KK-expansion approach. Here we
rederive it in the P/M propagator approach. From the propagator equation
above, $G_p(z,z')$ must satisfy
\bea
\{ -\om^2z^2 p^2+(\om z)^5\frac{\pl}{\pl z}\frac{1}{(\om z)^3}\pl_z-m^2 \}G_p(z,z')\nn
=-\half (\om |z|)^5(\del(z-z')+P\del(z-\ztil'))
\com
\label{adspro4}
\eea 
(The absolute value $|z|$ comes from the space-time volume measure 
$\sqrt{-G}=\sqrt{-\mbox{det}G_{AB}}$.)
Hence $G_p$ is determined by the Bessel differential equation. 
\footnote{
For the high energy region $\sqrt{|p^2|}>> \om$, we expect (\ref{adspro5}) approaches
the 'flat' (z-)space equation:
\bea
\{ -p^2+{\pl_z}^2 \}G_p(z,z')
=-\half (\om |z|)^3(\del(z-z')+P\del(z-\ztil'))
\pr
\label{adspro5b}
\eea 
This corresponds to the 'flat space' limit (6.16) of Ref.\cite{RS01}. 
Note that the warp parameter $\om$ remains in this limit and 
it is different from the flat case of Sec.\ref{S.5Dprop}. 
See Fig.\ref{IM32sp5Z} and App.C.3(3S). See also Fig.\ref{RS61sp5z} and 
App.C.5(3S). 
}
\bea
\{ -p^2+{\pl_z}^2 -\frac{3}{z}\pl_z-\frac{m^2}{\om^2 z^2} \}G_p(z,z')
=-\half (\om |z|)^3(\del(z-z')+P\del(z-\ztil'))
\com\nn
-\frac{1}{T}\leq z,z'\leq -\frac{1}{\om}\q\mbox{or}\q
\frac{1}{\om}\leq z,z'\leq \frac{1}{T}\ 
\label{adspro5}
\eea 
The above equation has the symmetries.
\begin{description}
\item[Sym(A)] $z\change z'$
\item[Sym(B)] $z\ra -z \ \mbox{and}\  z'\ra -z'$\com\q $P=(-1)\times (-1)=1$
\item[Sym(C)] $z\ra -z$\com\q $P=-1$
\item[Sym(C')] $z'\ra -z'$\com\q $P=-1$
\end{description}
Corresponding to the choice of $P=-1$ in (\ref{adspro2}), we take
the Dirichlet b.c. at the fixed points:\ $z,z'=\pm \frac{1}{\om},\ \pm\frac{1}{T}$. 
Now let us solve (\ref{adspro5}) in the same way as in Sec.\ref{P/Mapproach}. 
We consider $p^2>0$ (space-like) case first.

We divide the whole region into 8 ones, $R_1, R_2, R_1', R_2', \Rbar_1,
\Rbar_2, \Rbar_1'$ and $\Rbar_2'$ as in Fig.6. 

\begin{figure}
\caption{8 regions in the z-coordinates plane.
This is the transformed version of Fig.\ref{8regions} through
the relation (\ref{adsf2})}
\includegraphics[height=8cm]{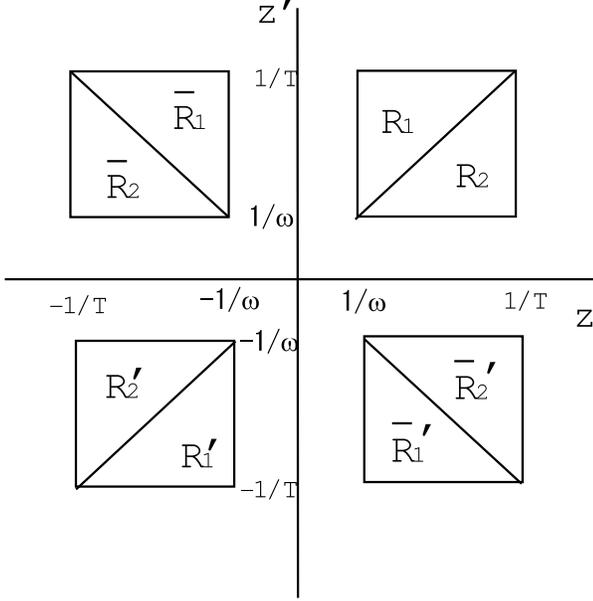}
\label{8regionsZ}
\end{figure}

{\bf Step 1.}Region R$_1$ and R$_2$

We start by solving (\ref{adspro5}) for the region
$\frac{1}{\om}\leq z , z'\leq \frac{1}{T}$. \nl
(i)$z\neq z'$ .\nl
In this case the equation (\ref{adspro5}) reduces to the
{\it homogeneous} one.
\bea
(-\ptil^2+{\pl_z}^2-\frac{3}{z}\pl_z-\frac{m^2}{\om^2z^2})G_p(z,z')\nn
=(-\ptil^2+{\pl_{z'}}^2-\frac{3}{z'}\pl_{z'}-\frac{m^2}{\om^2{z'}^2})G_p(z,z')
=0\com\q
\ptil\equiv \sqrt{p^2}\pr
\label{adspro6}
\eea 
The general solution is given by
\bea
K_p(z,z')=A(z')z^2\K_\n(\ptil z) +B(z')z^2\I_\n(\ptil z)
         =C(z){z'}^2\K_\n(\ptil z')+D(z){z'}^2\I_\n (\ptil z')
\com\nn
\n=\sqrt{4+\frac{m^2}{\om^2}}
\label{adspro7}
\eea 
where  $A(z'), B(z'), C(z)$ and $D(z)$ are to be fixed by
the boundary conditions. 
We consider the special case
\footnote{
We treat
the case of the general $\nu$ in App.B. 
}
 which is important in the
supersymmetry requirement\cite{GP0003}.
\bea
m^2=-4\om^2\com\q \n=0
\pr
\label{adspro8}
\eea 
We take the solution as
\bea
G_p(z,z')=
K_p(z,z')=A(z')z^2\K_0(\ptil z) +B(z')z^2\I_0(\ptil z)
         =C(z){z'}^2\K_0(\ptil z')+D(z){z'}^2\I_0 (\ptil z')
\com\nn
\mbox{for}\q \frac{1}{\om}\leq z<z'\leq\frac{1}{T}
\nn
G_p(z,z')=
K_p(z',z)=A(z){z'}^2\K_0(\ptil z') +B(z){z'}^2\I_0(\ptil z')
         =C(z'){z}^2\K_0(\ptil z)+D(z'){z}^2\I_0 (\ptil z)
\com\nn
\mbox{for}\q \frac{1}{\om}\leq z'<z\leq\frac{1}{T}
\pr
\label{adspro9}
\eea 
The latter equation is the $z\change z'$ exchanged one of 
the former equation. Here we use Sym.(A).  
The Dirichlet b.c. requires
\bea
A(z')\K_0(\Pla)+B(z')\I_0(\Pla)=0\com\nn
C(z')\K_0(\Tev)+D(z')\I_0(\Tev)=0
\pr
\label{adspro10}
\eea 

(ii) $|z-z'|\ <\ +0$\nl
In order to take into account the inhomogeneous term
$-\half (\om |z|)^3\del(z-z')$ in the RHS of (\ref{adspro5})
we put the following b.c..
\bea\mbox{Jump Condition:}\q
\left. -\frac{\pl G_p(z,z')}{\pl z}+\frac{\pl G_p(z,z')}{\pl z'}\right|_{z'\ra z+0}
=-\half (\om z)^3 \q\mbox{for}\q (\frac{1}{\om}\leq)z<z'\pr
\label{adspro11}
\eea 
(This b.c., {\it in combined with the $z\change z'$ exchanged definition
for $z'<z$ in (\ref{adspro9})}, simply demands
$G_p(z,z')\sim -\frac{1}{4}(\om z)^3|z-z'|$ for $|z-z'|\ll 1$.
)
Using the upper equation of (\ref{adspro9}), the above b.c. reduces to
\bea
\left.
-A(z')\frac{d}{dz}\{z^2\K_0(\ptil z)\}-B(z')\frac{d}{dz}\{z^2\I_0(\ptil z) \}
+C(z)\frac{d}{dz'}\{{z'}^2\K_0(\ptil z')\}+D(z)\frac{d}{dz'}\{{z'}^2\I_0(\ptil z')\}
\right|_{z'\ra z+0}\nn
=-\half (\om z)^3
\pr
\label{adspro12}
\eea 
The continuity at $z=z'$ requires, from eq.(\ref{adspro9}),
\bea
A(z)\K_0(\ptil z) +B(z)\I_0(\ptil z)
         =C(z)\K_0(\ptil z)+D(z)\I_0 (\ptil z)
\pr
\label{adspro13}
\eea 
The 4 relations in (\ref{adspro10}), (\ref{adspro12}) and (\ref{adspro13}), fix
$A, B, C$ and $D$ as
\bea
\left(
\begin{array}{l}
A(z)\\
\\
B(z)\\
\\
C(z)\\
\\
D(z)
\end{array}
\right)
\begin{array}{l}
\\
\\
\\
= \\
\\
\\
\end{array}
\begin{array}{l}
\\
\\
\\
\{\frac{\om^3z^2}{2}/ (\K_0(\Pla)\I(\Tev)-\K_0(\Tev)\I_0(\Pla) ) \} \\
\\
\\
\end{array}
\begin{array}{l}
\\
\\
\\
\times \\
\\
\\
\end{array}
\nn
\left(
\begin{array}{l}
-\I_0(\Pla)\{\I_0(\Tev)\K_0(\ptil z)-\K_0(\Tev)\I_0(\ptil z)\}\\
\\
+\K_0(\Pla)\{\I_0(\Tev)\K_0(\ptil z)-\K_0(\Tev)\I_0(\ptil z)\} \\
\\
-\I_0(\Tev)\{\I_0(\Pla)\K_0(\ptil z)-\K_0(\Pla)\I_0(\ptil z)\}\\
\\
+\K_0(\Tev)\{\I_0(\Pla)\K_0(\ptil z)-\K_0(\Pla)\I_0(\ptil z)\}
\end{array}
\right)
\label{adspro14}
\eea 
Hence $K_p(z,z')$ is completely fixed as
\bea
K_p(z,z')=-\frac{\om^3}{2}z^2{z'}^2
\frac{\{\I_0(\Pla)\K_0(\ptil z)-\K_0(\Pla)\I_0(\ptil z)\}  
      \{\I_0(\Tev)\K_0(\ptil z')-\K_0(\Tev)\I_0(\ptil z')\}
     }{\I_0(\Tev)\K_0(\Pla)-\K_0(\Tev)\I_0(\Pla)}
\pr
\label{adspro15}
\eea 

{\bf Step 2.}Extension to all other regions

We now know how to extend the above solution to all other regions
in the consistent way with Sym.(A)-(C'). Before that, we point out
the necessity of the extension of the modified Bessel functions to
the negative real axis. (same as in (\ref{adsf9b}).)
Following the procedure of Sec.\ref{P/Mapproach}.2, we see the solution in $R_1$ $\Rbar_1$
is given by $K_p(z,z')$ (\ref{adspro15}). This function must be odd for
$z\change -z$. (because of Sym(C)) It demands $\K_0(\ptil z)$ and $\I_0(\ptil z)$
is the odd function of $z$. 
\footnote{
This situation is the same as that in (\ref{adsf9b}) in the previous 
section. In the flat case (\ref{PMapp13}), the corresponding function
is $\sinh \ptil y$ which is odd for $y \change -y$. 
}
Hence the modified Bessel functions, in the $P=-1$ case,
is generalized to the corresponding odd functions.
\footnote{
For the case $P=1$, the generalized even functions are given by
\bea
\IBhat_\nu(z)\equiv \I_\nu(|z|)\com\q\KBhat_\nu(z)\equiv \K_\nu(|z|)
\pr
\label{adspro16b}
\eea 
These will be used for the vector propagator. 
} 
\bea
\IBtil_\nu(z)\equiv \ep(z)\I_\nu(|z|)\com\q\KBtil_\nu(z)\equiv \ep(z)\K_\nu(|z|)
\com
\label{adspro16}
\eea 
where $\I_\nu(z)$ and $\K_\nu(z)$ are the ordinary ones defined
in $z> 0$. 
Then the propagator (\ref{adspro15}) is improved to
\bea
\Ktil_p(z,z')=-\frac{\om^3}{2}z^2{z'}^2
\frac{\{\I_0(\Pla)\KBtil_0(\ptil z)-\K_0(\Pla)\IBtil_0(\ptil z)\}  
      \{\I_0(\Tev)\KBtil_0(\ptil z')-\K_0(\Tev)\IBtil_0(\ptil z')\}
     }{\I_0(\Tev)\K_0(\Pla)-\K_0(\Tev)\I_0(\Pla)}
\pr
\label{adspro15b}
\eea 
With these $Z_2$-odd quantities, the
final solution is given by
\bea
G_p(z,z')=\left\{
\begin{array}{ll}
\Ktil_p(z,z') & \mbox{for}\q R_1\q\mbox{and}\q \Rbar_1 \\ 
\Ktil_p(-z',-z)& \mbox{for}\q \Rbar_2\q\mbox{and}\q R_2'\\ 
\Ktil_p(-z,-z') & \mbox{for}\q R_1'\q\mbox{and}\q \Rbar_1'\\ 
\Ktil_p(z',z) & \mbox{for}\q \Rbar_2'\q\mbox{and}\q R_2 \\ 
\end{array}
          \right.
=-\frac{\om^3z^2{z'}^2/2}
{\I_0(\Tev)\K_0(\Pla)-\K_0(\Tev)\I_0(\Pla)}\times
\nn
\left\{
\begin{array}{ll}
\{\I_0(\Pla)\KBtil_0(\ptil z)-\K_0(\Pla)\IBtil_0(\ptil z)\}  
      \{\I_0(\Tev)\KBtil_0(\ptil z')-\K_0(\Tev)\IBtil_0(\ptil z')\}&
               \mbox{for}\ R_1\cup \Rbar_1\cup {R_1}'\cup {\Rbar_1}'\\
\{\I_0(\Pla)\KBtil_0(\ptil z')-\K_0(\Pla)\IBtil_0(\ptil z')\}  
      \{\I_0(\Tev)\KBtil_0(\ptil z)-\K_0(\Tev)\IBtil_0(\ptil z)\}&
               \mbox{for}\ R_2\cup \Rbar_2\cup {R_2}'\cup {\Rbar_2}'
\end{array}
\right.
\label{adspro17}
\eea 
As in Sec.\ref{P/Mapproach}, we can express $G_p(z,z')$ in the compact way using the absolute
functions.
\bea
z=\half (z+z')-\half (z'-z)\ra Z(z,z')\equiv\half |z+z'|-\half |z'-z|
\nn
z'=\half (z+z')+\half (z'-z)\ra Z'(z,z')\equiv\half |z+z'|+\half |z'-z|
\nn
G_p(z,z')=\Ktil_p(Z(z,z'),Z'(z,z'))
\label{adspro18}
\eea 
The last expression is valid for all regions and is indispensable for
the calculation in Sec.\ref{deltazero}  and is important for drawing graphs. This is because
the absolute functions can properly treat the singularities. 
 
For the time-like case $p^2<0$, the explanation goes in the same way
except the modified Bessel functions are replaced by the Bessel functions. 
The expression of $\Ktil_p(z,z')$ in (\ref{adspro18}) is given by
\bea
\Ktil_p(z,z')=-\frac{\om^3}{2}z^2{z'}^2
\frac{\{\J_0(\PlaH)\NBtil_0(\phat z)-\N_0(\PlaH)\JBtil_0(\phat z)\}  
      \{\J_0(\TevH)\NBtil_0(\phat z')-\N_0(\TevH)\JBtil_0(\phat z')\}
     }{\J_0(\TevH)\N_0(\PlaH)-\N_0(\TevH)\J_0(\PlaH)}
\com
\label{adspro19}
\eea 
where $\phat=\sqrt{-p^2}$. 

(\ref{adspro18}) and (\ref{adspro19}) will be used for z-coordinate graphical
representation in App.C.3. We will give the equivalent expression as the 
y-coordinate counter-part in Sec.\ref{S.P/Mpropagator}.2.

\section{Sturm-Liouville expansion approach versus P/M Propagator
approach\label{SLvsPM}}
We have solved the propagator of the 5D warped space-time (\ref{adsf6})
or (\ref{adspro1}) both in the expanded form (Sec.\ref{AdS5}) and 
in the closed form (Sec.\ref{P/MproAdS5}) . Here we relate them
as done for the flat case in Sec.\ref{S.KKvsPM}. 

Using the Strum-Liouville expansion formula\cite{Titchmarsh62}
($f(z)$: an arbitrary continuous (real) function defined in $a\leq z\leq b$)
\bea
f(z)=\sum_{n=1}^{\infty}\frac{k_n}{\Omega'(\la_n)}\psi_n(z)
\int_a^b r(\xi)f(\xi)\psi_n(\xi)d\xi,\nn
\Omega(\la)\equiv p(z)\left(\vp^a(z,\la){\vp^b}(z,\la)'-\vp^b(z,\la){\vp^a}(z,\la))'\right)\com\q
k_n\equiv \frac{\vp^b(z,\la_n)}{\vp^a(z,\la_n)}\com\q \Omega(\la_n)=0\com\nn
\psi_n(z)\equiv \vp^a(z,\la_n)\propto\vp^b(z,\la_n)\com
\label{SLvsPM1}
\eea 
where $\psi_n$ is the eigen functions 
of the general operator defined below ( such as 
appeared in (\ref{adsfM5})\ ), 
and $\vp^a(z,\la)$ and $\vp^b(z,\la)$ are "intermediate" solutions,
that is, they satisfy the differential equation below but the boundary
condition is imposed {\it only} at one of the two boundary points. 
\bea
\left( \Lhat+\la r(z)\right)\vp^a=\left( \Lhat+\la r(z)\right)\vp^b=0\com\q
\Lhat=\frac{d}{dz}p(z)\frac{d}{dz}-q(z)\com\nn
\vp^a(z,\la):\ \mbox{b.c. is satisfied only at\ }z=a\com\q
\vp^b(z,\la):\ \mbox{b.c. is satisfied only at\ }z=b\com
\label{SLvsPM2}
\eea 
where $\Lhat$ is the general kinetic operator (Sturm-Liouville differential
operator), and $a$ and $b$ are the boundary points on z-axis. 
In the present model, they are given by
\bea
p(z)=\frac{1}{(\om z)^3}\com\q q(z)=\frac{m^2}{(\om z)^5}\com\q
r(z)=\frac{1}{(\om z)^3}\com\q a=\frac{1}{\om}\com\q b=\frac{1}{T}\com\nn
\sqrt{\la}=\sqrt{-p^2}\q (p^2<0 \ \mbox{case})\pr
\label{SLvsPM3}
\eea 
(We also use the notation $M$ (=$\sqrt{\la}$) and $M_n$ (=$\sqrt{\la_n}$).) 
In this case, the equation (\ref{SLvsPM2}) is the sourceless ($J=0$) version
of (\ref{adsf6}) (the homogeneous differential equation). 
The formula (\ref{SLvsPM1}) reduces to the ordinary Fourier expansion
formula for the flat case. See App.A
. 

We can deduce, from the the P/M propagator,  
the expansion form by 
applying $G_p(z,z')$ of Sec.\ref{P/MproAdS5}, using
$K_p(z,z')$ of (\ref{adspro19}) (time-like case), to $f(z)$ above.
\bea
&& \frac{1}{\om}\ \leq z,\ z'\ \leq\ \frac{1}{T}\com\nn
G_p(z,z') 
&=& 
\sum^\infty_{n=1}\frac{k_n\psi_n(z)}{\Omega'(\la)|_{\la=M_n^2}}
\left[\int_{\frac{1}{\om}}^{z'}\xi^{-3}
K_p(\xi,z')\psi_n(\xi)d\xi 
+\int_{z'}^{\frac{1}{T}}\xi^{-3}K_p(z',\xi)\psi_n(\xi)d\xi\right]
\nn
& = & \frac{1}{2}\sum^\infty_{n=1}\frac{\psi_n(z)\psi_n(z')-\psi_n(z)\psi_n(-z')}
{p^2+M_n^2},
\label{SLvsPM4}
\eea
where we have used the following calculation results.
\begin{eqnarray}
k_n &\equiv& \frac{\vp^{(1/T)}(z, M_n)}{\vp^{(1/\om)}(z, M_n)} = 1, 
\nn
\Omega(\la)
 & \equiv & z^{-3}\left\{\vp^{(1/\om)}(z, 
\sqrt{\la})\frac{\pl\vp^{(1/T)}(z, \sqrt{\la})}{\pl z} -
\vp^{(1/T)}(z, \sqrt{\la})\frac{\pl\vp^{(1/\om)}(z, \sqrt{\la})}{\pl 
z}\right\} 
\nn
&=& \frac{2\om^4}{\pi 
N_n^2}\left\{b_0(\om\sqrt{\la}/T)-b_0(\sqrt{\la})\right\}, b_0\ \mbox{is defined in (\ref{adsfM2})}. 
\nn
\Omega'(\la)|_{\la=M_n^2} &=& -\frac{2\om^4\{\mbox{\boldmath
$J$}_0^2(\frac{M_n}{T})-\mbox{\boldmath
$J$}_0^2(\frac{M_n}{\om})\}}{\pi^2 N_n^2M_n^2\mbox{\boldmath
$N$}_0(\frac{M_n}{\om})\mbox{\boldmath
$N$}_0(\frac{M_n}{T})\mbox{\boldmath
$J$}_0(\frac{M_n}{\om})\mbox{\boldmath
$J$}_0(\frac{M_n}{T})},
\nn
N_n^2 &=& \frac{2\om\{\mbox{\boldmath
$N$}_0^2(\frac{M_n}{\om})-\mbox{\boldmath
$N$}_0^2(\frac{M_n}{T})\}}{\mbox{\boldmath
$N$}_0^2(\frac{M_n}{\om})\mbox{\boldmath
$N$}_0^2(\frac{M_n}{T})}.
\label{SLvsPM5}
\eea 
Particularly, we have used the Lommel's formula
$\J_\nu(z)\N_{\nu+1}(z)-\J_{\nu+1}(z)\N_\nu(z)=-2/\pi z$ and 
the following formula of indefinite integral
that makes the propagator $1/(p^2+{M_n}^2)$ appear in the
final expression of(\ref{SLvsPM4}).
\begin{eqnarray}
\int \xi \mbox{\boldmath$Z$}_0(\al\xi)\mbox{\boldmath$Z$}_0(\be\xi)d\xi = 
\frac{\xi}{\al^2-\be^2}
\{\al \mbox{\boldmath$Z$}_1(\al\xi)\mbox{\boldmath$Z$}_0(\be\xi)
-\be \mbox{\boldmath$Z$}_0(\al\xi)\mbox{\boldmath$Z$}_1(\be\xi)\},
\label{SLvsPM6}
\end{eqnarray}
where $\mbox{\boldmath$Z$}_\nu$ represents $\mbox{\boldmath$J$}_\nu$ and 
$\mbox{\boldmath$N$}_\nu$. 

The result (\ref{SLvsPM4}) is valid for other regions of $z$ and $z'$. 
The same result is obtained also for the space-like case ($p^2>0$). 
In this text the equivalence between SL-expansion and P/M propagator
is shown only for $\nu$=0. It is valid for general $\nu$. In App.B 
we give an alternative proof which is valid for general $\nu$.  

In this warped case, both SL-expansion and P/M-propagator approaches
are done in the interval $y\in [-l,l]$. It can be extended
to {\bf R}=$(-\infty, \infty)$ by the procedure of the {\it universal
covering}:\ we extend the solution, obtained for the interval $[-l,l]$, 
to {\bf R} by {\it requiring the periodicity} $y\ra y+2l$.
\bea
\Phi(x^a,z)=\sum_n\p_n(x)\psi_n(z)\com\q
\psi_n(z)=\sum_{m=1}^{\infty}c_{nm}\sin \mpl y\com\nn
c_{nm}\equiv \frac{1}{l}\int_{-l}^l\psi_n(z(y))\sin \mpl y dy
=\frac{2}{l}\int_0^l\psi_n(\frac{1}{\om}\e^{\om y})\sin \mpl y dy\com
\label{SLvsPM7}
\eea 
where the relation (\ref{adsf2}) is used. 
In this way, we can introduce the orbifold picture.  
This is important
to connect the warped solution and the flat solution by the
{deformation} parameter $\om$.

Whether we view the system in the orbifold picture or in the interval one
depends on our choice of the {\it infrared} regularization of the extra axis.  
For the former case, the extra axis is basically ${\bf R}=(-\infty, \infty)$ and
two identifications ($y\change -y,\ y\ra y+2l$) are imposed there. 
On the other hand, for the latter case, the extra axis is the interval [$-l,l$]
with $y\change -y$ identification. The importance of the infrared regularization
(of the extra axis) was stressed, in the context of the wall-anti-wall formation, 
in Ref.\cite{SI02PR}. In (\ref{SLvsPM7}), we see the importance of y-coordinate
as well as z.

\section{Behaviour of P/M Propagator\label{S.P/Mpropagator}}
The P/M propagator involves all KK modes contributions. Its behaviour 
characteristically changes depending on the 4-momentum $p_a$ in relation
to its absolute value $|p_ap^a|=|p^2|$ and 2 mass parameters, 
$l$ (boundary parameter, periodicity) and $\om$ (bulk curvature). 
The $p$-dependence for the various $(y,y')$ (or $(z,z')$) was examined in Ref.\cite{RS01}.
Here we show the $(y,y')$-dependence for various $p_a$. 
Characteristic 'brane' structures manifestly appear. 

Note:\ In all figures in the following, the vertical plot is cut at some appropriate value
when the graph height is too large. 

\subsection{
Flat 5D Massless Scalar Propagator ($Z_2$-parity Odd, Dirichlet-Dirichlet b.c.)}
The behaviour of 5D massless scalar in the flat geometry, 
space-like case (\ref{PMapp14}) and time-like case (\ref{PMapp15}), 
is shown in Fig.\ref{Flat2sp003}-\ref{Flat2tm1p7}. 
$Z_2$-parity is taken to be odd: P=$-1$. Diriclet b.c. is imposed for
all fixed points$(y,y'=0,\pm l)$. 

We take the boundary parameter value as\nl
$l$=$\pi$,\q 1/$l$$\sim$0.3\nl
We use the following notation. \nl
$\ptil\equiv \sqrt{p^2}$ for $p^2>0$(space-like);\q
$\phat\equiv \sqrt{-p^2}$ for $p^2<0$(time-like)

In this case, the scale parameter is the periodicity parameter $l$ only.
We can characterize the behaviours by the momentum $p$ in comparison
with 1/$l$. \nl

(A) Space-Like

(1S) $\ptil$ << 1/$l$\com\q Fig.\ref{Flat2sp003}\nl
Upheaval and downheaval surfaces front each other at sharp edges
which correspond the singularities at $y\pm y'=0$. The size of the
slope is $l$. Boundary constraint is strong. This is the 'Boundary phase'. 
The scale $\ptil$ does not appear in the graph.

(2S) $\ptil$ $\sim$ 1/$l$\com\q Fig.\ref{Flat2sp03}\nl
The gross shape is similar to (1S).

(3S) $\ptil$ >> 1/$l$\com\q Fig.\ref{Flat2sp3}\nl
Walls and valleys run along the diagonal axes. The configuration
is free from the boundary constraint. This is the 'Dynamical phase'.  
The size of the wall (valley) 
thickness is 1/$\ptil$.  Absolute value of the effective height decreases clearly. 
In the point of the wall (valley) formation, this situation is common to the warped case.
See the (3S) of Sec.\ref{S.P/Mpropagator}.2. \nl

(B) Time-Like

(1T) $\phat$ << 1/$l$\com\q Fig.\ref{Flat2tm003}\nl
Shape is similar to the space-like case. This is the 'Boundary phase'.

The situation that the propagator configuration, for the small $|p|$, 
is almost same both for the space-like case and for the time-like case, 
is generally valid in the following ( even for the warped case). 
\footnote{
A simple reason for this similarity is the relations:\ 
$\sin x\sim \sinh x\ ,\ \cos x\sim \cosh x $ for $|x|\ll 1$. 
}

(2T) $\phat$ $\sim$ 1/$l$\com\q Fig.\ref{Flat2tm03}\nl
The absolute value of the height increases and decreases
by changing $\phat$ within this region. 
The global shape does not change.

(3T) $\phat$ >> 1/$l$\com\q Fig.\ref{Flat2tm1p7}\nl
The wavy behaviour appears. This is the contrasting point compared with
the space-like case. 
The singularity-lines are buried in the waves. 
Boundary constraint is not effective. This is the 'Dynamical phase'. 
The size of the wave length is 1/$\phat$. 

This situation of wave formation, for the large $\phat$ in the time-like case, is
generally seen in the following.
Compared with the space-like case, the absolute value
does not so much change for the time-like case.

Other flat propagators are displayed in Appendixes C.1 and C.2. 

\begin{figure}
\caption{
Flat 5D Massless Scalar Propagator ($Z_2$-odd, Dirichlet-Dirichlet b.c.), 
$\ptil$=0.03 << 1/$l$ , space-like, S$^1$-boundary
is strongly influenced. Sec.\ref{S.P/Mpropagator}.1(1S) 
}
\includegraphics[height=8cm]{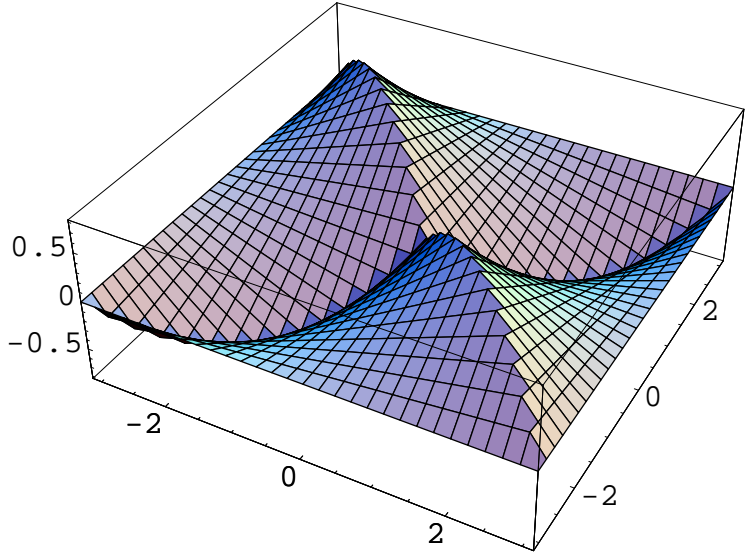}
\label{Flat2sp003}
\end{figure}


\begin{figure}
\caption{
Flat 5D Massless Scalar Propagator ($Z_2$-odd, Dirichlet-Dirichlet b.c.), 
$\ptil$=0.3 $\sim$ 1/$l$ , space-like, Sec.\ref{S.P/Mpropagator}.1(2S) 
}
\includegraphics[height=8cm]{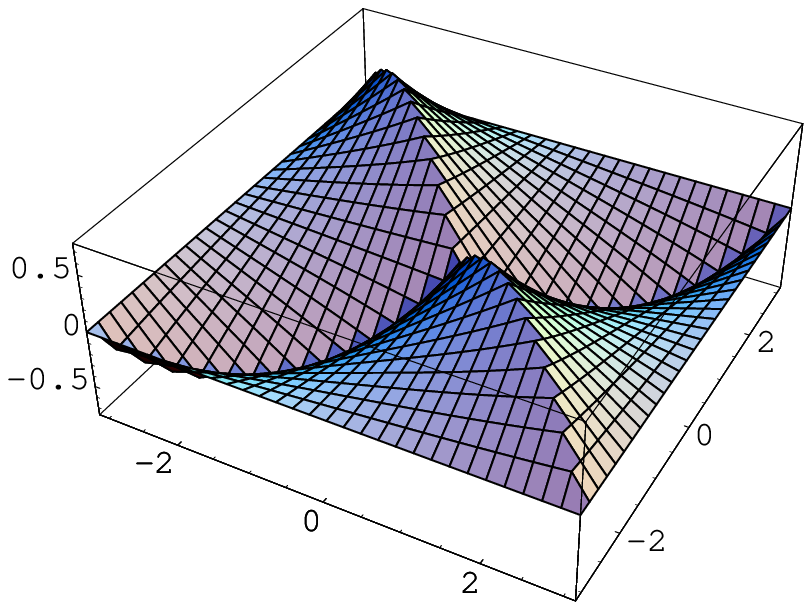}
\label{Flat2sp03}
\end{figure}

\begin{figure}
\caption{
Flat 5D Massless Scalar Propagator ($Z_2$-odd, Dirichlet-Dirichlet b.c.), 
$\ptil$=3 >> 1/$l$ , space-like, S$^1$-boundary
is not influenced, Sec.\ref{S.P/Mpropagator}.1(3S) 
}
\includegraphics[height=8cm]{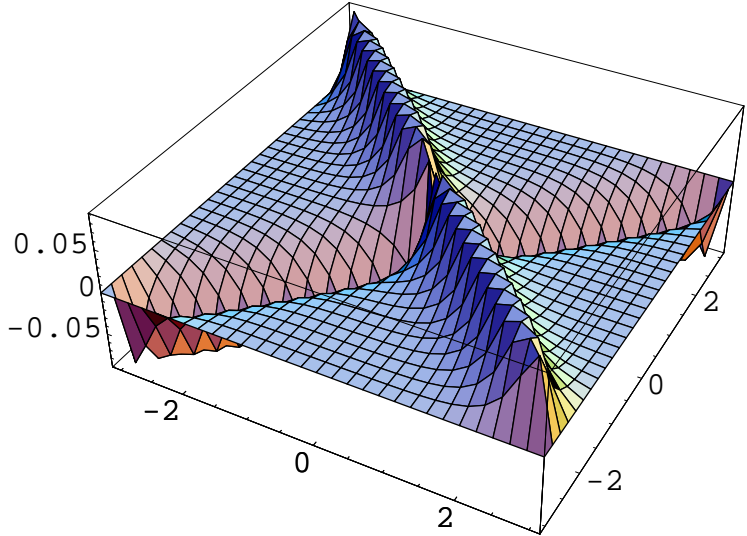}
\label{Flat2sp3}
\end{figure}

\begin{figure}
\caption{
Flat 5D Massless Scalar Propagator ($Z_2$-odd, Dirichlet-Dirichlet b.c.), 
$\phat$=0.03 << 1/$l$ , time-like, Sec.\ref{S.P/Mpropagator}.1(1T), S$^1$-boundary is strongly influenced 
}
\includegraphics[height=8cm]{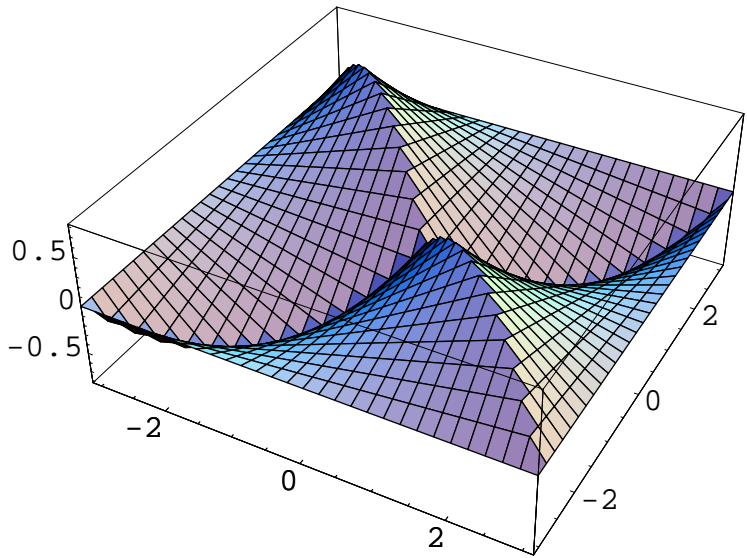}
\label{Flat2tm003}
\end{figure}

\begin{figure}
\caption{
Flat 5D Massless Scalar Propagator ($Z_2$-odd, Dirichlet-Dirichlet b.c.), 
$\phat$=0.3 $\sim$ 1/$l$ , time-like, Sec.\ref{S.P/Mpropagator}.1(2T) 
}
\includegraphics[height=8cm]{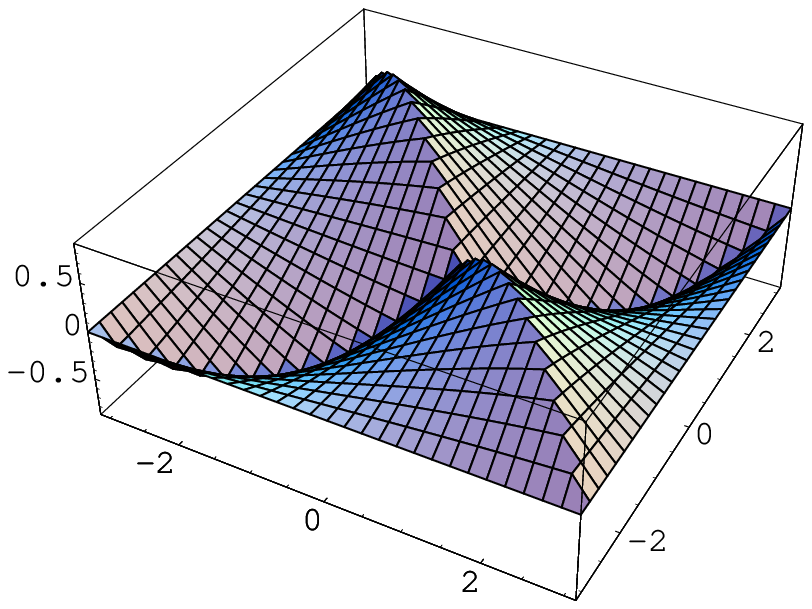}
\label{Flat2tm03}
\end{figure}


\begin{figure}
\caption{
Flat 5D Massless Scalar Propagator ($Z_2$-odd, Dirichlet-Dirichlet b.c.), 
$\phat$=1.7 > 1/$l$ , time-like, Sec.\ref{S.P/Mpropagator}.1(3T) 
}
\includegraphics[height=8cm]{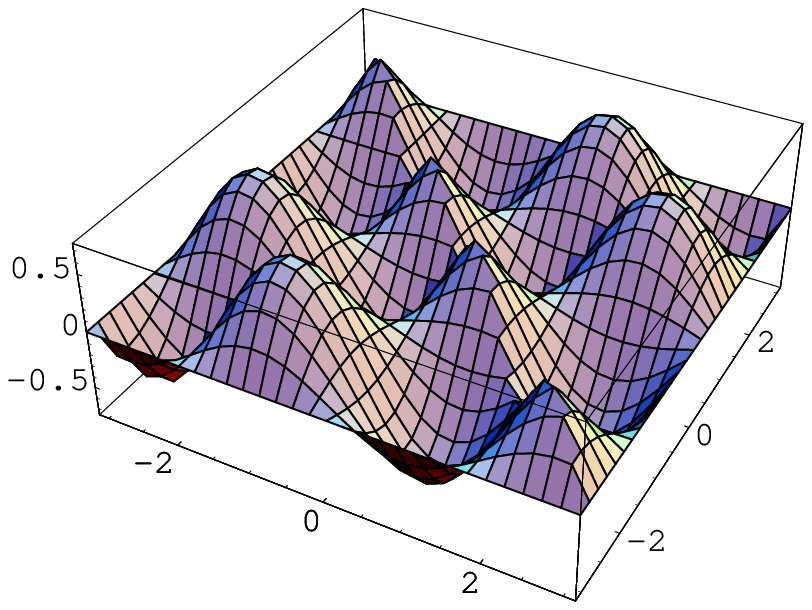}
\label{Flat2tm1p7}
\end{figure}


\subsection{
Warped 5D Scalar Propagator (Z$_2$-Odd, Dirichlet-Dirichlet b.c.)}
We graphically show the P/M propagator behaviour of 5D scalar
, with P=$-1$, in the warped geometry: (\ref{adspro18}) with (\ref{adspro15b}) for $p^2>0$, 
(\ref{adspro19}) for $p^2<0$.
\footnote{
The expressions, however, are written in z-coordinate. They are equivalently
rewritten in y-coordinate as in (\ref{ScalarOddSP}) for the space-like case, 
and in (\ref{ScalarOddTM}) for the time-like case. 
}  
Dirichlet b.c. is imposed on all fixed points $(y,y'=0,\pm l)$. 
As for the choice of the value of $\om$, we have the following possibilities:\ 
1) $\om\ll 1/l$,\ \ 2) $\om < 1/l$,\ \ 3) $\om\sim 1/l$,\ \ 4) $\om > 1/l$,\ \ 
5) $\om\gg 1/l$.\ \ Case 1) is important for approaching the flat limit from
the warped geometry. Case 5) is the situation in the real world because we know
Tev(Weak)-scale/Planck-scale =10$^{-16}$=$\exp(-\om l)$. For simplicity, 
we take the case 4) for the display of graphs. The values are\newline
$l$=$\pi$,\q 1/$l$$\sim$0.3,\q$\om$=1,\q T=$\om$ $\exp (-\om l)$ $\sim$ 0.04\nl
We use the following notation. \nl
$\ptil\equiv \sqrt{p^2}$ for $p^2>0$(space-like);\q
$\phat\equiv \sqrt{-p^2}$ for $p^2<0$(time-like)

(A) Space-Like Case

For the y-coordinate presentation, we use the following propagator
function. 
\bea
G_p(y,y')=K_p(Y(y,y'),Y'(y,y'))\com\nn
K_p(Y,Y')=\frac{1}{2\om}\exp(2\om|Y|+2\om|Y'|)\ep(Y)\ep(Y')\times\nn
\frac{\{\I_0(\Pla)\K_0(\Pla \mbox{e}^{\om|Y|})-\K_0(\Pla)\I_0(\Pla \mbox{e}^{\om|Y|})\}  
      \{\I_0(\Tev)\K_0(\Pla \mbox{e}^{\om|Y'|})-\K_0(\Tev)\I_0(\Pla \mbox{e}^{\om|Y'|})\}
     }{\I_0(\Tev)\K_0(\Pla)-\K_0(\Tev)\I_0(\Pla)}
\com
\label{ScalarOddSP}
\eea 
where $Y(y,y')$ and $Y'(y,y')$ are defined in (\ref{PMapp16}). 

(1S) $\ptil$ << 1/$l$ < $\om$\com\q Fig.\ref{IM32sp0005}\nl
2 sharp up-ward spikes and 2 sharp down-ward spikes appear
at corners. The effective thickness of the spikes is 1/$\om$. 
(Note again that the top surface is cut at an appropriate height.)
The size of the global upheaval and downheaval is $l$. The boundary
constraint is dominant. This is the "boundary phase". There is a flat region around
the center $(y=y'=0)$. The propagator vanishes there. 
This means that the bulk propagation, near the Planck brane, 
gives no contribution to the amplitude. On the other hand, near the Tev brane
$(y,y'=\pm l)$, it gives a sizable effect.

We will see the warped scale parameter $\om$ appears in all "phases".  

(2S) 1/$l$ < $\ptil$ $\sim$ $\om$\com\q Fig.\ref{IM32sp1}\nl
2 walls with sharp edges and 2 sharp valleys develop along
the diagonal axes 
from the corners to the center. Their thickness is 1/$\ptil$ $\sim$ 1/$\om$. 
The flat region near the center disappears. 
Absolute value of the effective height decreases.

(3S) 1/$l$ < $\om$ < $\ptil$\com\q Fig.\ref{IM32sp5}\nl
Walls and valleys develop from the coners almost to the center. 
The effective thickness
of them is 1/$\ptil$ near the corners and is 1/$\om$ near the center. 
There is no boundary effect. This is the "dynamical phase". 
There is
no flat region near the center, whereas in the off-diagonal region
there appears the zero-value flat region.  This means the bulk propagation
takes place only for the case $y'\pm y\sim 0$. 
Absolute value of the effective height decreases rapidly as $\ptil$ increases.

\begin{figure}
\caption{
Warped 5D Scalar Propagator ($Z_2$-odd, Dirichlet-Dirichlet b.c.), 
$\ptil$=0.005 << T < 1/$l$ <$\om$, space-like, Sec.\ref{S.P/Mpropagator}.2(1S), "boundary phase"
}
\includegraphics[height=8cm]{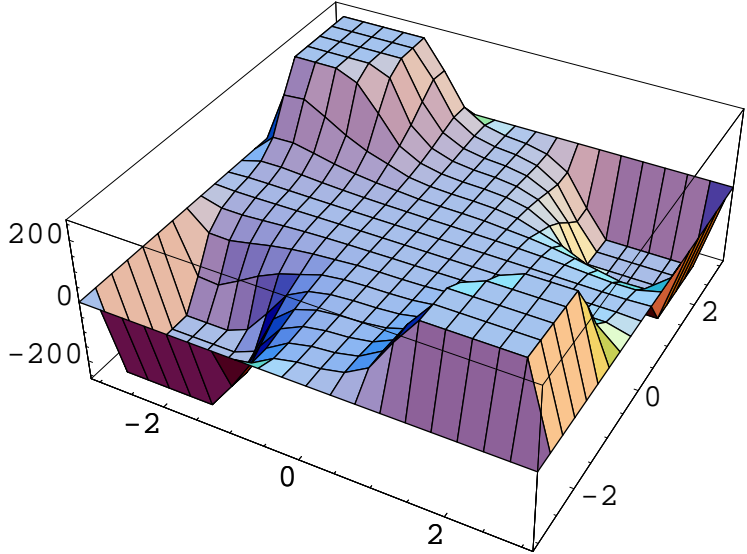}
\label{IM32sp0005}
\end{figure}


\begin{figure}
\caption{
Warped 5D Scalar Propagator ($Z_2$-odd, Dirichlet-Dirichlet b.c.), 
T << 1/$l$ < $\ptil$=1=$\om$, space-like, Sec.\ref{S.P/Mpropagator}.2(2S) 
}
\includegraphics[height=8cm]{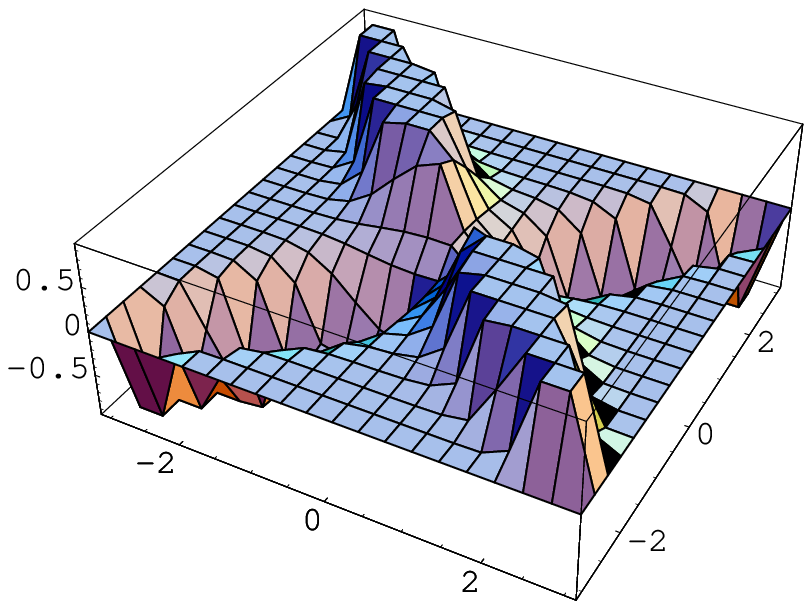}
\label{IM32sp1}
\end{figure}

\begin{figure}
\caption{
Warped 5D Scalar Propagator ($Z_2$-odd, Dirichlet-Dirichlet b.c.), 
T << 1/$l$ < $\om$ < $\ptil$=5, space-like, Sec.\ref{S.P/Mpropagator}.2(3S) 
}
\includegraphics[height=8cm]{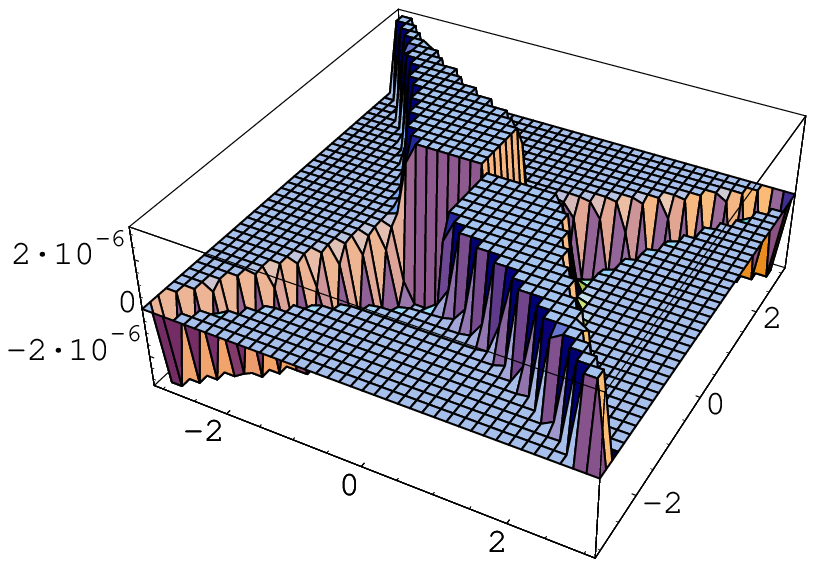}
\label{IM32sp5}
\end{figure}

(B) Time-Like Case

For the y-coordinate presentation, we use the following propagator
function. 
\bea
G_p(y,y')=K_p(Y(y,y'),Y'(y,y'))\com\nn
K_p(Y,Y')=\frac{1}{2\om}\exp(2\om|Y|+2\om|Y'|)\ep(Y)\ep(Y')\times\nn
\frac{\{\J_0(\tPla)\N_0(\tPla \mbox{e}^{\om|Y|})-\N_0(\tPla)\J_0(\tPla \mbox{e}^{\om|Y|})\}  
      \{\J_0(\tTev)\N_0(\tPla \mbox{e}^{\om|Y'|})-\N_0(\tTev)\J_0(\tPla \mbox{e}^{\om|Y'|})\}
     }{\J_0(\tTev)\N_0(\tPla)-\N_0(\tTev)\J_0(\tPla)}
\com
\label{ScalarOddTM}
\eea 
where $Y(y,y')$ and $Y'(y,y')$ are defined in (\ref{PMapp16}).

(1T) $\phat$ << 1/$l$ < $\om$\com\q Fig.\ref{IM32tm0005}\nl
The behaviour is quite similar to the space-like case (1S) above. 
The fact that, for low 4D momentum ($\phat$), the P/M propagator behaviours
of the space-like case and of the time-like case are similar, 
is widely valid. See other cases below. \nl

(2T) 1/$l$ < $\phat$ $\sim$ $\om$\com\q Fig.\ref{IM32tm1}\nl
Wavy behaviour appears. Sharp spikes gather near the 4 corners
and form groups within the range of order $1/\om=1/\phat$. 
Two types of waves are there. 
One type has
the small wave-length of order 1/$\phat$=1/$\om$, and the waves of this 
type appear along the 4 rims. The other type has the long wave-length
of order $l$, which comes from the boundary constraint. It appears
in the center and forms 
a very moderate hill. 
The propagator takes nearly 0 value there. 
This is contrasting with the space-like case.
As the whole configuration, $Z_2$ oddness disappears.\nl

(3T) 1/$l$ < $\om$ < $\phat$\com\q Fig.\ref{IM32tm5}\nl
Spikes and two types of waves are there. 
It is roughly similar to (2T).  
The plain in the center forms clearly a disk. 
The propagator takes nearly 0 value there. 
This is contrasting with the space-like case.
The overall height decreases. 
As in (2T), $Z_2$ oddness disappears. 
Although the scale $l$ looks to appear as the radius
of the plain around the center, the main configuration
is free from the boundary effect. 
It is nearly the "dynamical phase".

\begin{figure}
\caption{
Warped 5D Scalar Propagator ($Z_2$-odd, Dirichlet-Dirichlet b.c.), 
$\phat$=0.005 << T < 1/$l$ <$\om$, time-like, Sec.\ref{S.P/Mpropagator}.2(1T) 
}
\includegraphics[height=8cm]{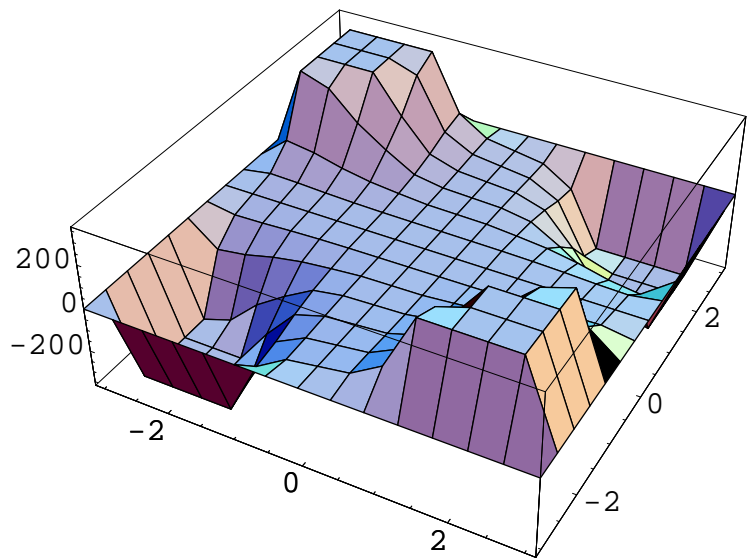}
\label{IM32tm0005}
\end{figure}

\begin{figure}
\caption{
Warped 5D Scalar Propagator ($Z_2$-odd, Dirichlet-Dirichlet b.c.), 
T << 1/$l$ < $\phat$=1=$\om$, time-like, Sec.\ref{S.P/Mpropagator}.2(2T) 
}
\includegraphics[height=8cm]{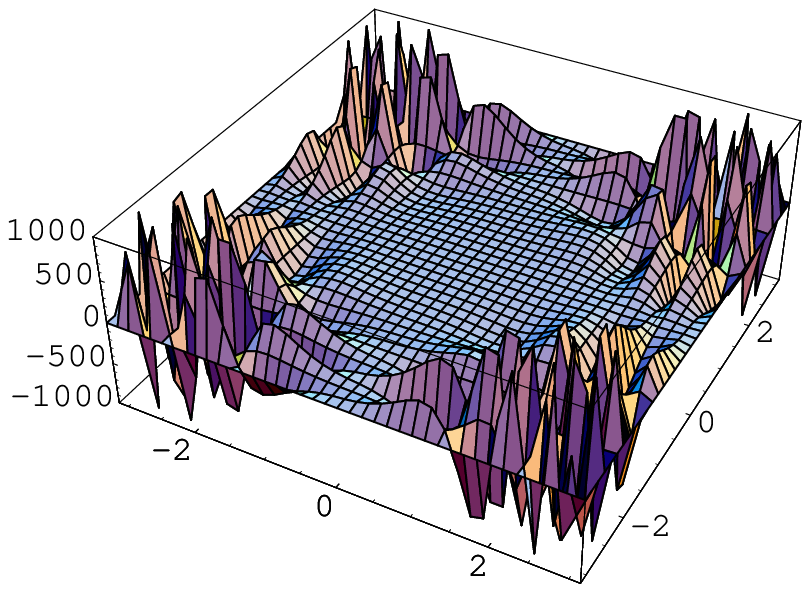}
\label{IM32tm1}
\end{figure}

\begin{figure}
\caption{
Warped 5D Scalar Propagator ($Z_2$-odd, Dirichlet-Dirichlet b.c.), 
T << 1/$l$ < $\om$ < $\phat$=5, time-like, Sec.\ref{S.P/Mpropagator}.2(3T)
}
\includegraphics[height=8cm]{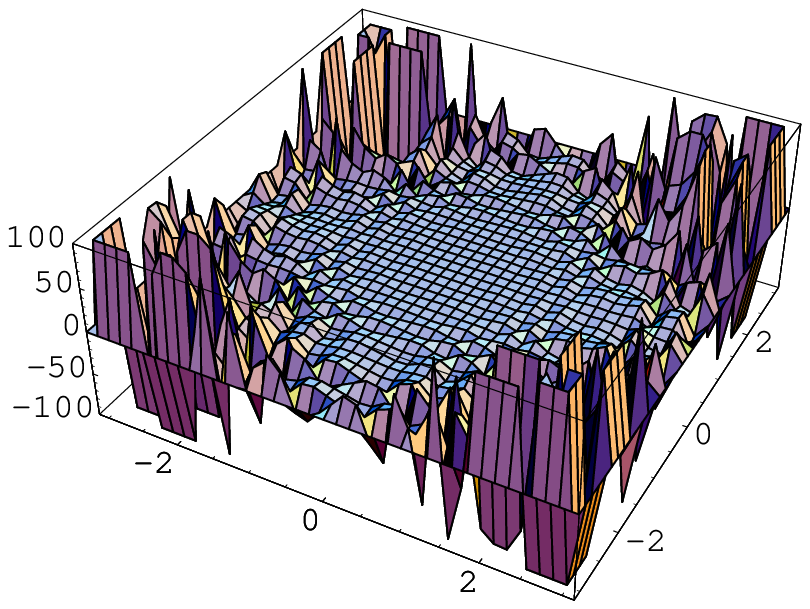}
\label{IM32tm5}
\end{figure}

\vspace{5mm}
From (1S) to (3S), from (1T) to (3T), the ratio $\om/p$ decreases. 
We cannot take the flat limit in this way. In fact the scale $\om$
remains in all "phases". The correct condition for the flat limit
is both $\om\ll 1/l$\ and $\om\ll p$\ are satisfied. 
We have confirmed that, taking the values $l=\pi,\ \om=0.001\ (\om\ll 1/l)$, 
the warped propagators (\ref{ScalarOddSP}) and (\ref{ScalarOddTM}) produce
graphs quite similar to the flat results of Sec.\ref{S.P/Mpropagator}.1 (Fig.\ref{Flat2sp003}-\ref{Flat2tm1p7}).

\section{Solving $\del(0)$-problem and the deformation of propagator\label{deltazero}}
P/M propagator for 5D scalar is, 
for the space-like 4 momentum case, 
given by (\ref{PMapp14}). It is expressed as
\bea
G_p(y,y')
=\frac{1}{4\ptil\sinh \ptil l}\left[
\{\cosh 2\ptil v_+(y,y')-\cosh 2\ptil v_-(y,y')\} \cosh \ptil l\right.\nn
\left. +\{-\sinh 2\ptil v_+(y,y')+\sinh 2\ptil v_-(y,y')\}\sinh \ptil l
                      \right]\pr
\label{del1}
\eea 
where $\vm$ and $\vpl$ are defined by
\bea
\vmp(y,y')\equiv \half|y'\mp y|\q :\q y\ \change \ y'\ \ \mbox{symmetric}
\pr
\label{del2}
\eea 
Using the following relations:
\bea
\frac{\pl\vm}{\pl y}=-\half\ep(y'-y)\com\q\frac{\pl\vm}{\pl y'}=\half\ep(y'-y)\nn 
\frac{\pl\vpl}{\pl y}=\half\ep(y'+y)\com\q\frac{\pl\vpl}{\pl y'}=\half\ep(y'+y)\nn
\frac{\pl^2\vm}{\pl y^2}=\del (y'-y)\com\q\frac{\pl^2\vm}{\pl y'\pl y}=-\del (y'-y)\com\q
\frac{\pl^2\vm}{\pl {y'}^2}=\del (y'-y)\nn
\frac{\pl^2\vpl}{\pl y^2}=\del (y'+y)\com\q\frac{\pl^2\vpl}{\pl y'\pl y}=\del (y'+y)\com\q
\frac{\pl^2\vpl}{\pl {y'}^2}=\del (y'+y)
\label{del3}
\eea 
we obtain, for the flat case (\ref{del1}), 
\bea
\left.\frac{\pl^2G_p}{\pl y\pl y'}\right|_{y=y'=0}=
-\del(0)+\half\ptil\coth \ptil l\com\nn
\left.\frac{\pl^2G_p}{\pl y\pl y'}\right|_{y=0,y'=\pm l}=
\half\frac{\ptil}{\sinh \ptil l}\com\nn
\left.\frac{\pl^2G_p}{\pl y\pl y'}\right|_{y=y'=\pm l}=
-\del(0)+\half\ptil\coth \ptil l\pr
\label{del4}
\eea 
The above result says the bulk scalar propagation starting and ending at the Planck brane
and that starting and ending at Tev brane  are the same. The propagation starting
at the Planck(Tev) brane and ending at the Tev(Planck) brane does not have
$\del(0)$ singularity.  

The $\del(0)$ problem first appeared in the analysis of 11D supergravity 
on a manifold with boundary in relation to the $E_8\times E_8$ heterotic string
and M-theory\cite{HW9603}. It was shown, using a simpler model, that the problem
generally occurs in the bulk-boundary theory\cite{MP9712}. When the bulk scalar
$\Phi$ has a derivative ($\partial_y$) coupling with other field (this is the case
of Mirabelli-Peskin model), the $\Phi$ propagator part in a quantum-loop amplitude
appears as the form  $\pl^2G_p/\pl y\pl y'$. In Ref.\cite{MP9712}, the cancellation
of $\del(0)$ was shown in a self energy calculation using KK-expansion. 
The cancellation was further
confirmed in an improved way in Ref.\cite{IM05NPB,IM04PLB2}. 
The first equation in (\ref{del4}) exactly coincides with 
the results in these papers. In the papers, however, the equation was obtained by 
summing all KK-modes contribution. 
In this paper the same equation is obtained
{\it without} doing the KK-summation. 

For the warped case,  the corresponding propagator (5D scalar, space-like 4 momentum, Dirichlet b.c.)
is given in (\ref{adspro18}). After calculation using (\ref{del3}), we obtain
\bea
\left.\frac{\pl^2G_p}{\pl y\pl y'}\right|_{y=y'=0}=
-\del(0)-2\om+
\ptil\frac{    
      \I_0(\Tev)\K_1(\frac{\ptil}{\om})+\K_0(\Tev)\I_1(\frac{\ptil}{\om})
           }{I_0(\Tev)\K_0(\Pla)-\K_0(\Tev)\I_0(\Pla)}
\pr
\label{del5}
\eea 
\footnote{
The appearance of $\del(0)$ was known in Ref.\cite{IM0606} where 
KK-summation was used. 
}
We confirm eq.(\ref{del5}) leads to the first equation of eq.(\ref{del4}) in the
limit: $\om/\ptil \ra +0,\ \ptil l=$fixed. This shows the warped case is 
, at the propagator level, continuously connected with the flat one. 
The result (\ref{del5}) says that the {\it warped version} of the Mirabelli-Peskin model
does {\it not} suffer from the $\del(0)$ problem. 

In Ref.\cite{IM05NPB}, it is shown that the finite part of above expressions 
, (\ref{del4}) and (\ref{del5}), can be 
regarded as a "deformation" factor from the ordinary 4D theory 
propagator. The linear divergence, for $\ptil\ra\infty$, of the finite
part just gives the UV-divergence due to 5D quantum fluctuation. 
Both flat and warped cases are non-renormalizable in this sense. 
\footnote{
See eq.(54) of Ref.\cite{IM05NPB} for detail. 
}
See Sec.\ref{Conc} for further discussion about the renormalizability.



\section{Discussions and Conclusion\label{Conc}}
We have treated QFT in the 5D flat and warped space-time. The Z$_2$-parity
is respected. The P/M propagator is closely analyzed. Its singular 
properties are systematically treated by the use of the absolute
functions. We have obtained the visual output of various P/M propagators, 
which enables us to know the various "phases" depending upon 
the choice of fields (scalar, vector, $\cdots$), boundary conditions
(Dirichlet, Neumann), space-time geometry (flat, warped), the 4D momentum 
property(space-like, time-like) and its magnitude ($\sqrt{|p^2|}$ in relation
to $1/l$ and $\om$). It is shown that the eigen-mode expansion approach is
equivalent to the P/M propagator approach. They are related by the Fourier-expansion
for the flat case and by the Strum-Liouville expansion for the warped case. 
The Dirac's bra and ket vector formalism is naturally introduced for quantizing 
the 5D (flat and warped) space-time with Z$_2$-parity.  

We add some comments and discussions as follows.

1)\ The Feynman rule for the present approach is straightforward and is 
given in Ref.\cite{RS01}. The characteristic points are the appearance of
the metric factors at vertices and the extra-axis integral form 
restricted by the directedness of the extra coordinate. 

2)\ BRS structure is important for defining physical quantities
in gauge theories. It is very successful in the 4D renormalizable
theories. For the present model of higher dimensions, the structure 
is missing. The formal higher dimensional extension is possible, but 
the treatment of the extra dimension part is quite obscure. 
In Ref.\cite{RS01}, some Ward identities seem to work.  

3)\ Generally the string theory is regarded advantageous over the QFT because 
the fundamental unit of the string tension $\al'$ are there and the
extendedness "softens" the singularities. 
The present approach is based on the higher-dimensional QFT. The extendedness
parameter appears as the thickness $\om$. The situation of the boundary conditions
and the "brane" formation looks similar to that in the string theory. 
In particular, the role of the extra-axis looks to correspond to that of the open string
which is used to define the D-brane. Of course, these similarities come from 
the fact that the original models are invented and examined, triggered by the string
theory development. 
We point out the present
approach could reveal some important {\it regularization} aspect of the string theory 
in a {\it simplified} way. 
In this respect, it is worth discussing the regularization in the present approach. 
In Fig.\ref{(z,p)INTregion}, the integration space is shown. 
The horizontal axis is z-coordinate, and the range is $1/\om \leq z \leq 1/T$. 
The vertical one is (the absolute value of) the 4D momentum. 
It runs in the range $\mu$($\equiv (T/\om)\cdot\La$,\ infrared cutoff)$\leq p \leq \La$(ultraviolet cutoff). 
What region of (z,p)-space, shown in Fig.\ref{(z,p)INTregion}, should be integrated is the 
present discussion point. Ref.\cite{RS01} proposed the region:
\bea
p\ z \leq \frac{\La}{\om}\com\q \frac{1}{\om}\leq z\leq \frac{1}{T}
\com
\label{conc1}
\eea 
based on the concrete behaviour of the P/M propagator. (See Fig.\ref{RS61sp1z} 
and Fig.\ref{RS61sp5z} and their explanation in App.C.5.)
Adapting the region (\ref{conc1}), the $\be$-function of the gauge coupling was {\it finitely} obtained\cite{RS01}. 
Clearly this procedure is still primitive and some persuasive explanation is required. 
We propose here a new definition of the integral region. For the explanation we look the integral
region in the space of (z-coordinate, 4D coordinate $x^a$), that is, the 5D coordinate space. 
See Fig.\ref{(z,x)INTregion}. 
This is an equivalent description. For simplicity we take 5D Euclidean space. 
On the Planck-brane ($z=1/\om$), the 4D space integral region is taken to be
$1/\La \leq x\equiv\sqrt{x^ax^a}\leq 1/\mu$. This region is 4 dimensional and forms 
the {\it thick} sphere-shell bounded by two $S^3$'s:\ one ($S^3_{\mbox{UV,Pla}}$) has the radius $1/\La$ and 
the other ($S^3_{\mbox{IR,Pla}}$) has the radius $1/\mu$. 
On the Tev-brane ($z=1/T$), the 4D space integral region is 
$(1-\vep)/\mu \leq x \leq 1/\mu$, where 
$\vep$ is a new regularization parameter which tends to the positive 0, $\vep\ra +0$. 
As on the Planck-brane, the integral region is 4 dimensional and forms 
the {\it thin} sphere-shell bounded by two $S^3$'s:\ one ($S^3_{\mbox{UV,Tev}}$) has the radius $(1-\vep)/\mu$ and 
the other ($S^3_{\mbox{IR,Tev}}$) has the radius $1/\mu$. 
Between the two branes ($1/\om < z < 1/T$), the integral region
is the 5D volume bounded by 
two 4D regularization surfaces, $B_{UV}$ and $B_{IR}$, 
which can be determined by the {\it minimal area principle} and
the boundary condition 
($S^3_{UV,Pla}$ at $z=1/\om$,$S^3_{UV,Tev}$ at $z=1/T$ for $B_{UV}$; 
$S^3_{IR,Pla}$ at $z=1/\om$,$S^3_{IR,Tev}$ at $z=1/T$ for $B_{IR}$).  
Two regularization $S^3$ spheres, the UV sphere and the IR sphere, "flow" along the z-axis
changing their radii:\ 
$B_{UV}$ describes the change of the UV sphere and 
$B_{IR}$ describes the change of the IR sphere.  
The advantage of the new definition is that, only at the fixed points, the artificial cutoffs 
are introduced. This is the same situation as taken in the ordinary 4D renormalizable theories.
Between the fixed points, the regularization surfaces ($B_{UV}, B_{IR}$) are not introduced by hand but 
determined by the {\it bulk geometrical dynamics}. If we view this new integral region in 
the (z,p)-space, the similar region to (\ref{conc1}) is expected to be obtained. 
It is quite interesting that the regularization surfaces, $B_{UV}$ and $B_{IR}$, have 
similarity to the {\it tree propagation} of the {\it closed string}. The necessity of restriction 
on the integral region (\ref{conc1}) 
strongly suggests the requirement of a new type "quantization". 
The integral region condition (\ref{conc1}) looks a sort of the {\it uncertainty relation}. 
If this view is right, it is quite notable that 
the conjugate variable (in the quantum phase space) of the extra coordinate z
is played by the absolute value of the 4D momentum, $\sqrt{|p^2|}$.  
The present standpoint described above is that the new relation comes from the {\it minimal area 
principle}. Hence the behaviour of the boundary surfaces, $B_{UV}$ and $B_{IR}$, plays an 
important role.

4)\ The flat system is characterized by the cyclic functions, while 
the warped one is characterized by the Bessel functions. 
Although the periodicity is {\it lost} in the latter system, 
both sets of
functions constitute the complete orthonormal system and 
sufficiently deserve describing the quantum Hilbert space. 
The present analysis strongly suggests the Bessel function system
can be regarded as a one-parameter {\it deformed system} of the cyclic functions. 

5)\ As for the phenomenology application, besides the ones mentioned 
in the introduction, Higgs sector analysis based on 5D model 
is active. The Higgs field is identified with the extra-component
of the bulk gauge field and the effective action is calculated
under the name of "holographic pseudo-Goldstone boson"\cite{OW0410,ACP0412,Falko0610}. The form
factors there correspond to P/M propagators $G_p(y,y')$ of the present
work. 

\begin{figure}
\caption{
Space of (z,p) for the integration. $\La/\om u$ is the
position-dependent cutoff\cite{RS01}. 
}
\includegraphics[height=8cm]{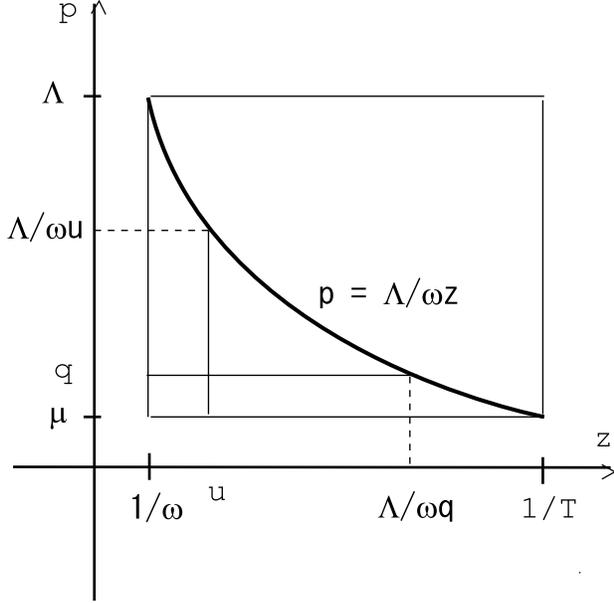}
\label{(z,p)INTregion}
\end{figure}

\begin{figure}
\caption{
Space of (z,$x^a$) for the integration. $B_{IR}$, $S^3_{IR,Pla}$ and $S^3_{IR,Tev}$ 
can be similarly shown. 
}
\includegraphics[height=15cm]{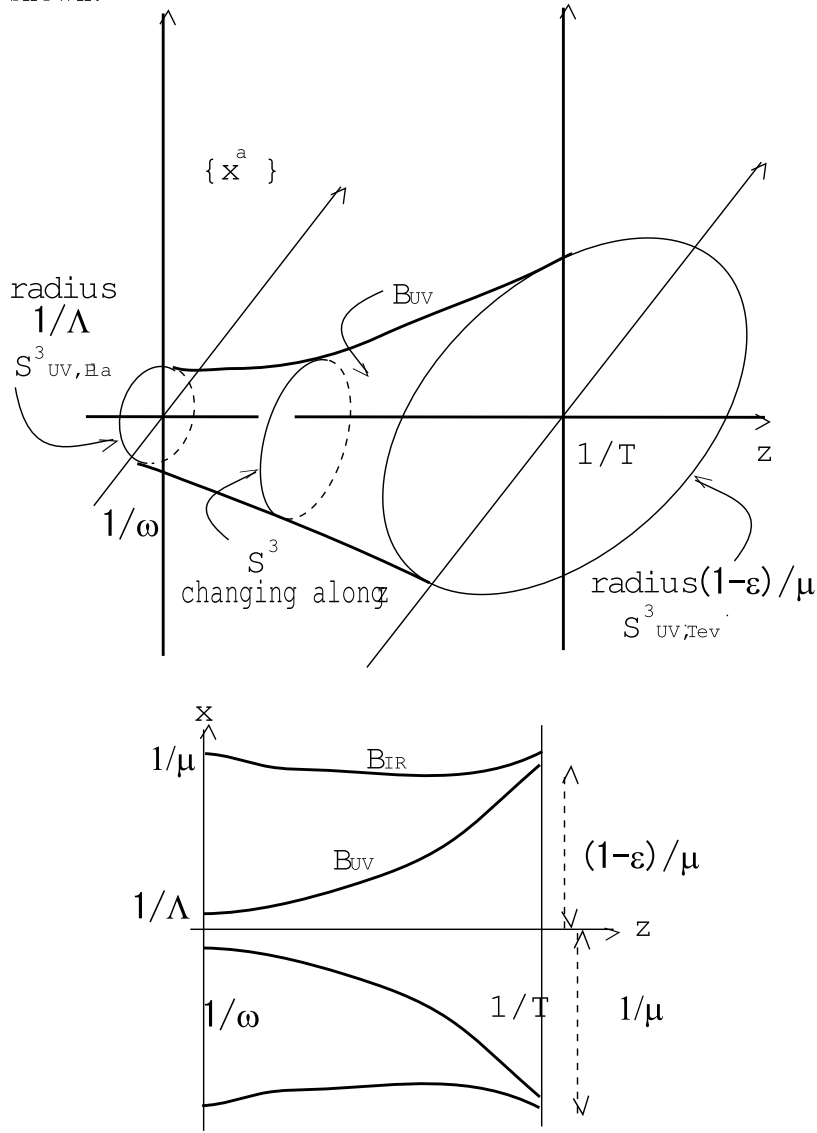}
\label{(z,x)INTregion}
\end{figure}

\section{App. A\ :\ Sturm-Liouville expansion formula\label{SLexpansion}}
We apply the Strum-Liouville expansion formula (\ref{SLvsPM1}) to
the flat case of Sec.\ref{S.5Dprop} and Sec.\ref{P/Mapproach}. 
\bea
\pl_M\pl^M\Phi(X)=-J(X)\com\nn
\Phi(X)=\intp \e^{ipx}\Phi_p(y)\com\q J(X)=\intp \e^{ipx}J_p(y)\com\nn
(-p^2+\frac{\pl^2}{\pl y^2})\Phi_p(y)=-J_p(y)
\pr
\label{SLexp1}
\eea 
Hence the general operator $\Lhat$ (\ref{SLvsPM2}) and the boundary points are
 given by
\bea
p(y)=1\com\q q(y)=0\com\q r(y)=1\com\q \la=-p^2\com\nn
a=0\com\q b=l
\pr
\label{SLexp2}
\eea 
We consider the odd Z$_2$-parity, P=$-1$, as in the text. 
Time-like case, $p^2<0$, is taken. The homogeneous equation
and its "intermediate" solutions are given as
\bea
(\la+\frac{\pl^2}{\pl y^2})\psi(y)=0\com\q -l\leq y \leq l\nn
\vp(y,\la)=(1/\sqrt{\la})\sin \sqrt{\la}y\q \mbox{Dirichlet b.c. is satisfied at}\ y=0\nn
\chi(y,\la)=(1/\sqrt{\la})\sin \sqrt{\la}(l-y)\q \mbox{Dirichlet b.c. is satisfied at}\ y=l\nn
\label{SLexp3}
\eea 
The quantity $\Om$, its zeros $\la_n$ and others are obtained as
\bea
\Om(\la)=p(\vp \chi'-\chi\vp')=-(1/\sqrt{\la})\sin(\sqrt{\la}l)\com\nn
\Om(\la_n)=0\com\q \la_n=(\frac{\pi}{l})^2n^2,\ n=1,2,\cdots\ ,\nn
\Om'(\la_n)=-(l^3/2\pi^2n^2)(-1)^n\com\q
k_n=\chi(y,\la_n)/\vp(y,\la_n)=(-1)^{n+1}.
\label{SLexp4}
\eea 
Using the above results and the eigen function 
$\psi_n(y)=\vp(y,\la_n)=(l/\pi n)\sin (\pi ny/l)$ , 
the Strum-Liouville expansion formula (\ref{SLvsPM1}) reduces to
\bea
f(y)=\frac{2}{l}\sum_n (\sin \frac{n\pi y}{l})\int_0^l f(\xi)\sin(\frac{\pi n\xi}{l})d\xi
\pr
\label{SLexp5}
\eea 
This is the familiar Fourier expansion formula for the odd function $f(y)$. 
Its Z$_2$ parity
even version is used in (\ref{KKvPM2}) and the equivalence
of the expansion approach and P/M approach is shown for the flat case.

\section{App. B\ :\ General treatment of the propagator\label{general}}
In this appendix, we treat the propagator of AdS$_5$ space-time
(Sec.\ref{AdS5} and Sec.\ref{P/MproAdS5}) in the general way
valid for the wide-range dynamics, that is, the Strum-Liuoville
differential operator. The field equation (\ref{adsf6}) can be expressed, 
using 4D-Fourier transformed field $\Phi_p(z), J_p(z)$ defined in 
(\ref{SLexp1}), as 
\bea
\left\{ \frac{d}{dz}p(z)\frac{d}{dz}-q(z)\right\}\Phi_p -p^2r(z)\Phi_p
\equiv\Lhat\Phi_p-p^2r(z)\Phi_p=-\frac{J_p(z)}{(\om z)^5}\ ,\nn
p(z)=\frac{1}{(\om z)^3}\com\q q(z)=\frac{m^2}{(\om z)^5}\com\q
r(z)=\frac{1}{(\om z)^3}
\pr
\label{gen1}
\eea 
This is the Strum-Liouville differential equation with source term
(the inhomogeneous case). We consider $p^2<0$ case (time-like). 
($p^2>0$ case is similarly treated.) Z$_2$ parity is 
taken to be odd, P=$-1$, and 
the Dirichlet b.c. is taken both at $z=1/\om$ and at $z=1/T$. 
(P=+1 case is similarly treated.) 

(i) Homogeneous Solution\ ($J_p=0$)\nl
\bea
\Lhat\Phi_p-p^2r(z)\Phi_p=0
\pr
\label{gen2}
\eea 
This is the eigenvalue equation for the operator $r(z)^{-1}\Lhat$. 
Two independent solutions are given by
\footnote{
Bessel functions here are generalized as defined in (\ref{adsf9b}).
}
\bea
(\om z)^2\JBtil_\nu (Mz)\com\q (\om z)^2\NBtil_\nu (Mz)\com\q
M=\sqrt{-p^2}\com\q \nu=\sqrt{4+\frac{m^2}{\om^2}}
\pr
\label{gen3}
\eea 
We introduce two "intermediate" solutions:\ 
$\vp_M(z)$ which satisfies the b.c. only at $z=1/\om$, and 
$\chi_M(z)$ which satisfies the b.c. only at $z=1/T$. 
\bea
\vp_M(z)=\frac{1}{N^{(1/\om)}}(\om z)^2(\JBtil_\nu(Mz)\N_\nu(\frac{M}{\om})-
\NBtil_\nu(Mz)\J_\nu(\frac{M}{\om}))/\N_\nu(\frac{M}{\om})\com\nn
\chi_M(z)=\frac{1}{N^{(1/T)}}(\om z)^2(\JBtil_\nu(Mz)\N_\nu(\frac{M}{T})-
\NBtil_\nu(Mz)\J_\nu(\frac{M}{T}))/\N_\nu(\frac{M}{T})
\pr
\label{gen4}
\eea 
The final solution is obtained by the requirement that
$\vp_M(z)$ and $\chi_M(z)$ become linearly-dependent each other.
\bea
\mbox{Wronskian}\q W[\vp_M, \chi_M]\equiv\vp_M\frac{d}{dz}\chi_M-\chi_M\frac{d}{dz}{\vp_M}
\propto \left\{ \J_\nu(\frac{M}{\om})\N_\nu(\frac{M}{T})- 
\J_\nu(\frac{M}{T})\N_\nu(\frac{M}{\om})\right\}=0
\pr
\label{gen5}
\eea 
This is because the solution must satisfy the Dirichlet b.c. at both points. 
The condition (\ref{gen5}) fixes the set of eigenvalues $\{M_n \}$. 
The eigen function $\psi_n(z)$ is obtained as
\bea
\psi_n(z)=
\vp_M(z)|_{M=M_n}
=\frac{1}{N^{(1/\om)}_n}(\om z)^2(\JBtil_\nu(M_nz)\N_\nu(\frac{M_n}{\om})-
\NBtil_\nu(M_nz)\J_\nu(\frac{M_n}{\om}))/\N_\nu(\frac{M_n}{\om})\ ,\nn
=\chi_M(z)|_{M=M_n}\com\nn
2\int_{\frac{1}{\om}}^{\frac{1}{T}}r(z)\psi_n(z)\psi_m(z)dz=\del_{nm}
\pr
\label{gen6}
\eea 

(ii) Solution of (\ref{gen1}), Inhomogeneous solution\nl
\q\q (iia) Expansion form\nl
First we obtain the solution in the expansion form using
the homogeneous solutions $\{\psi_n\}$ obtained in (i). 
\bea
\Phi_p(z)=\sum_{n=1}^{\infty}c_n(p)\psi_n(z)\com\q
\Lhat \psi_n=-{M_n}^2r(z)\psi_n
\pr
\label{gen7}
\eea 
Putting (\ref{gen7}) into (\ref{gen1}), we obtain
\bea
\sum_{n=1}^{\infty}c_n(p)(-p^2-{M_n}^2)r(z)\psi_n(z)=
-\frac{J_p(z)}{(\om z)^5}
\pr
\label{gen8}
\eea 
From the orthogonality (\ref{gen6}), we can read the
coefficient. 
\bea
c_n(p)=\frac{f_n}{p^2+{M_n}^2}\com\q
f_n=\int_{1/\om}^{1/T}\psi_n(z)\frac{J_p(z)}{(\om z)^5}dz
\pr
\label{gen9}
\eea 
Hence we obtain the solution.
\bea
\Phi_p(z)=\sum_{n=1}^{\infty}\frac{f_n}{p^2+{M_n}^2}\psi_n(z)
\pr
\label{gen10}
\eea 
This result corresponds to (\ref{adsf11}) of Sec.\ref{AdS5}. 
\nl
\q\q (iib) Closed form\nl
We can also obtain the solution in the closed form 
using the "intermediate" solutions $\vp_M(z), \chi_M(z)$ (\ref{gen4}). 
The solution $\Phi_p(z)$ of (\ref{gen1}) can be obtained as
\bea
\Phi_p(z)=-\chi_M(z)\int_{1/\om}^{z}\frac{\vp_M(\xi)}{[\vp_M(\xi),\chi_M(\xi)]}
\frac{J_p(\xi)}{(\om \xi)^5}d\xi
-\vp_M(z)\int_{z}^{1/T}\frac{\chi_M(\xi)}{[\vp_M(\xi),\chi_M(\xi)]}
\frac{J_p(\xi)}{(\om \xi)^5}d\xi\ ,\nn 
\q
[
\vp_M(z),\chi_M(z)
]
\equiv p(z)(\vp_M(z){\chi_M}'(z)-{\vp_M}'(z)\chi_M(z))
\com
\label{gen11}
\eea 
where $p(z)$ appears in (\ref{gen1}). 
Because of $(d/dz)[\vp_M(z),\chi_M(z)]=0$, we can express (\ref{gen11}) as
\bea
[\vp_M(z),\chi_M(z)]\equiv \Om(M)\com\nn 
\Phi_p(z)=-\frac{\chi_M(z)}{\Om(M)}\int_{1/\om}^{z}\vp_M(\xi)
\frac{J_p(\xi)}{(\om \xi)^5}d\xi
-\frac{\vp_M(z)}{\Om(M)}\int_{z}^{1/T}\chi_M(\xi)
\frac{J_p(\xi)}{(\om \xi)^5}d\xi
\pr
\label{gen12}
\eea 
Let us here introduce the P/M propagator
\footnote{
From
the result (\ref{gen13}), we understand the appearance of the "direction" 
property of the extra coordinate $z$ (or $y$), originates from the characteristic
structure of the solution of the inhomogeneous (source $J_p$ attached) differential
equation.
}
%
\bea
G_p(z,\xi)\equiv
\left\{
\begin{array}{cc}
-\frac{\chi_M(z)\vp_M(\xi)}{[\vp_M(\xi),\chi_M(\xi)]}\ ,& 
                   \frac{1}{\om}\leq \xi\leq z\leq \frac{1}{T}\ (\mbox{R$_2$ of Fig.6})\\
-\frac{\vp_M(z)\chi_M(\xi)}{[\vp_M(\xi),\chi_M(\xi)]}\ ,& 
                   \frac{1}{\om}\leq z\leq \xi \leq \frac{1}{T}\ (\mbox{R$_1$ of Fig.6})
\end{array}
\right.
\label{gen13}
\eea 
where $M=\sqrt{-p^2}$. 
For other regions of Fig.\ref{8regionsZ}, $G_p$ is defined following the $Z_2$-parity property
as done in Sec.\ref{P/MproAdS5}. 
In terms of the above propagator, (\ref{gen12}) can be written as
\bea 
\Phi_p(z)=\int_{1/\om}^{1/T}G_p(z,\xi)
\frac{J_p(\xi)}{(\om \xi)^5}d\xi
\pr
\label{gen14}
\eea 
The propagator (\ref{gen13}) is the same one that is introduced in (\ref{adspro3}) of the text. 
Note that $K_p(z,\xi)$ of (\ref{adspro19})
corresponds to the lower part of (\ref{gen13}). 

Taking into account the expression (\ref{gen10}) and the relation:
\bea
\frac{J_p(\xi)}{(\om \xi)^5}
=\sum_{n=1}^{\infty}f_n\psi_n(\xi)\com\q f_n\ \mbox{is defined in (\ref{gen9})}
\com
\label{gen15}
\eea 
we can read off, from (\ref{gen14}), the following relation between
the P/M propagator and the eigen functions $\{\psi_n\}$. 
\bea
G_p(z,\xi)=\sum_{n=1}^{\infty}\frac{1}{p^2+{M_n}^2}\half 
\left( \psi_n(z)\psi_n(\xi)-\psi_n(z)\psi_n(-\xi) \right)
\pr
\label{gen16}
\eea 
This is the same as (\ref{adsf11}) of Sec.\ref{AdS5}.

\section{App. C\ :\ Behaviour of Various P/M propagators\label{propbehaviour}}
The values for $l$ (half period), $\om$ (thickness) are taken as\newline
$l$=$\pi$,\q 1/$l$$\sim$0.3,\q$\om$=1,\q T=$\om$ $\exp (-\om l)$ $\sim$ 0.04\nl
We note the following notation used in the text. \nl
$\ptil\equiv \sqrt{p^2}$ for $p^2>0$(space-like);\q
$\phat\equiv \sqrt{-p^2}$ for $p^2<0$(time-like)

\subsection{App. C.1\ :\  
Flat 5D Massless Scalar Propagator ( 
$Z_2$-parity Even, Neumann-Neumann b.c.)}
The behaviour of 5D massless scalar propagator with Z$_2$-parity even (P=1)	
is shown in Fig.\ref{yFlatE(NN)sp003}-\ref{yFlatE(NN)tm17}. Neumann b.c. is imposed for
all fixed points. The P/M propagator is given by
\bea
G_p(y,y')=\Kbar_p(Y(y,y'),Y'(y,y'))\com\nn
\Kbar_p(Y,Y')=\left\{
\begin{array}{ccc}
-\frac{\cosh \ptil Y\cosh \ptil (Y'-l)}{2\ptil\sinh \ptil l} & \ptil=\sqrt{p^2} & \mbox{for}\ p^2>0 \\
\frac{\cos \phat Y\cos \phat (Y'-l)}{2\phat\sin \phat l} & \phat=\sqrt{-p^2} & \mbox{for}\ p^2<0
\end{array}
              \right.
\label{ScaEnn1}
\eea 
where $Y(y,y')$ and $Y'(y,y')$ are defined in (\ref{PMapp16}). 

In this case, the scale parameter is the periodicity parameter $l$ only.
We can characterize the behaviours by the momentum $\ptil$ or $\phat$ in comparison
with 1/$l$. 

(A) Space-Like

(1S) $\ptil$ << 1/$l$\com\q Fig.\ref{yFlatE(NN)sp003}\nl
Upheaval and downheaval surfaces front each other at sharp edges
which correspond to the singularities at $y\pm y'=0$. The size of the
slope is $l$. The boundary constraint is strong. 
This is the 'boundary phase'. 
The scale p does not appear in the graph.

(2S) $\ptil$ $\sim$ 1/$l$\com\q Fig.\ref{yFlatE(NN)sp03}\nl
The gross shape is similar to (1S). The height decreases.

(3S) $\ptil$ >> 1/$l$\com\q Fig.\ref{yFlatE(NN)sp3}\nl
Valleys run along the diagonal axes. The configuration
is free from the boundary constraint. 
This is the 'dynamical phase'.  
The size of the valley-width is 1/$\ptil$. 
In the off-diagonal region ($y\pm y'\neq 0$), flat planes appear and
the propagator takes nearly 0 there. 
The height decreases furthermore.

(B) Time-Like

(1T) $\phat$ << 1/$l$\com\q Fig.\ref{yFlatE(NN)tm003}\nl
Shape and height are similar to the space-like case. 
This is the 'boundary phase'.

(2T) $\phat$ $\sim$ 1/$l$\com\q Fig.\ref{yFlatE(NN)tm03}\nl
The absolute value of the height increases and decreases. 
Shape is similar to the space-like case.

(3T) $\phat$ >> 1/$l$\com\q Fig.\ref{yFlatE(NN)tm17}\nl
The wavy behaviour appears.  
The singularity-lines are buried in the waves. 
Boundary constraint is not effective.
This is the 'dynamical phase'. 
The size of the wave length is 1/$\phat$. 

Compared with the space-like case, the height
does not so much change for the time-like case.

\begin{figure}
\caption{
Flat 5D Massless Scalar Propagator ( Z$_2$-parity even, Neumann-Neumann b.c.), 
$\ptil$=0.03 << 1/$l$  , space-like, App.C.1(1S). "Boundary phase".  
}
\includegraphics[height=8cm]{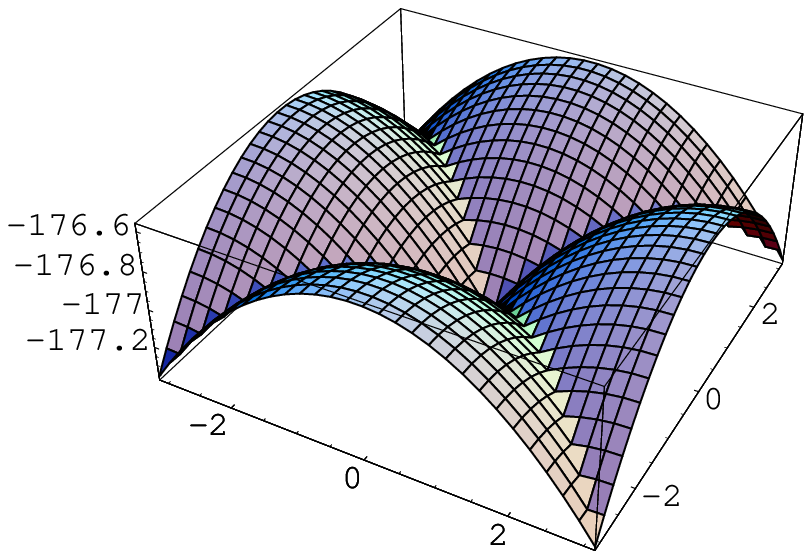}
\label{yFlatE(NN)sp003}
\end{figure}

\begin{figure}
\caption{
Flat 5D Massless Scalar Propagator ( Z$_2$-parity even, Neumann-Neumann b.c.), 
$\ptil$=0.3 $\sim$ 1/$l$, space-like, App.C.1(2S) 
}
\includegraphics[height=8cm]{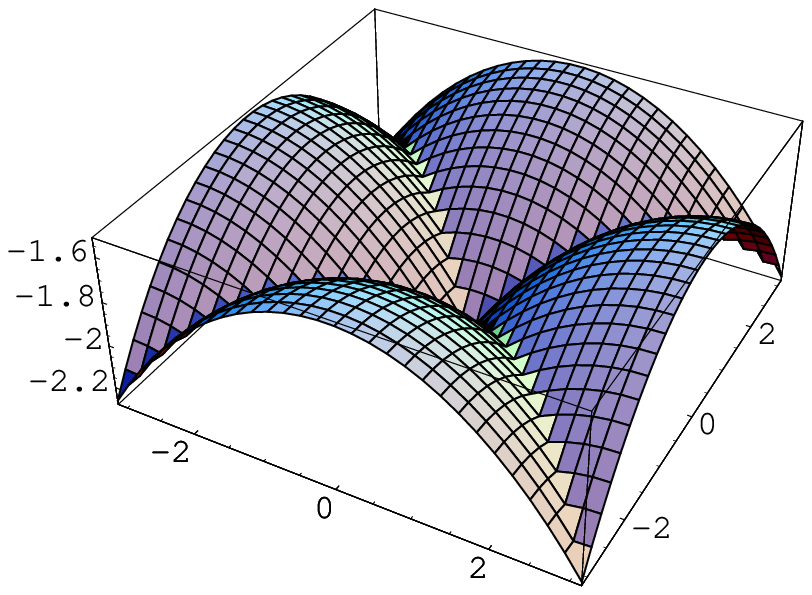}
\label{yFlatE(NN)sp03}
\end{figure}

\begin{figure}
\caption{
Flat 5D Massless Scalar Propagator ( Z$_2$-parity even, Neumann-Neumann b.c.), 
$\ptil$=3 >> 1/$l$, space-like, App.C.1(3S). 
The "dynamical phase".  
}
\includegraphics[height=8cm]{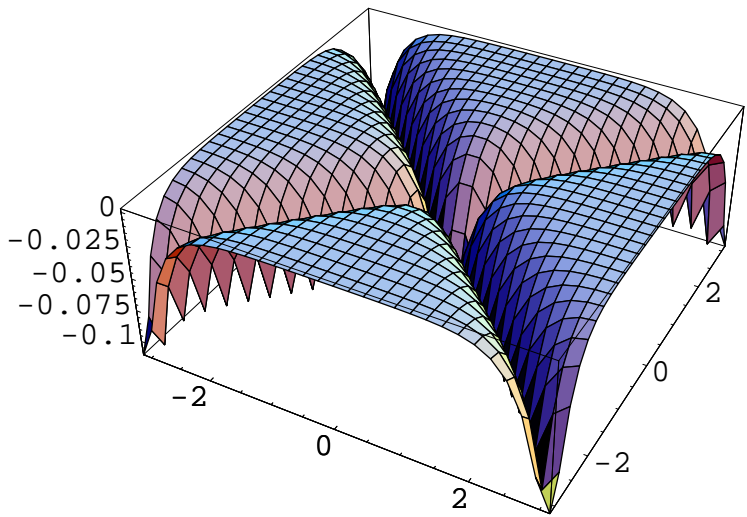}
\label{yFlatE(NN)sp3}
\end{figure}

\begin{figure}
\caption{
Flat 5D Massless Scalar Propagator ( Z$_2$-parity even, Neumann-Neumann b.c.), 
$\phat$=0.03 << 1/$l$, time-like, App.C.1(1T). The "boundary phase". 
}
\includegraphics[height=8cm]{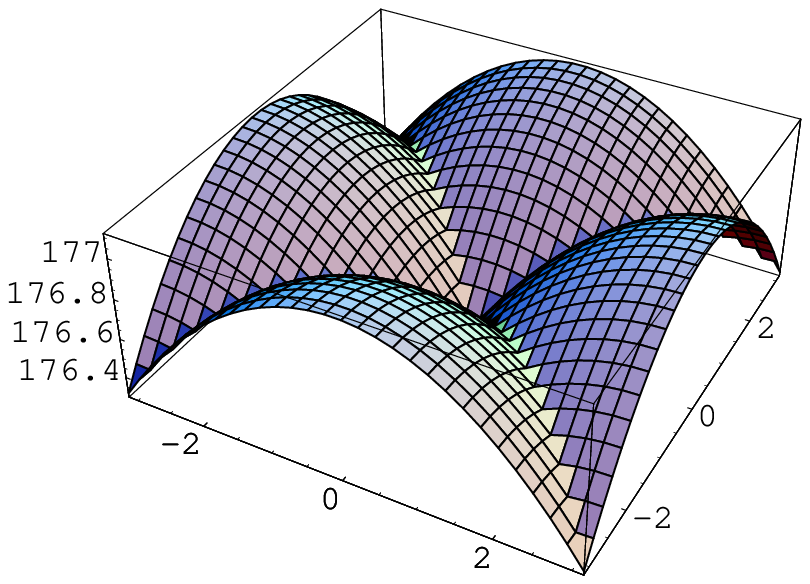}
\label{yFlatE(NN)tm003}
\end{figure}

\begin{figure}
\caption{
Flat 5D Massless Scalar Propagator ( Z$_2$-parity even, Neumann-Neumann b.c.), 
$\phat$=0.3 $\sim$ 1/$l$, time-like, App.C.1(2T). 
}
\includegraphics[height=8cm]{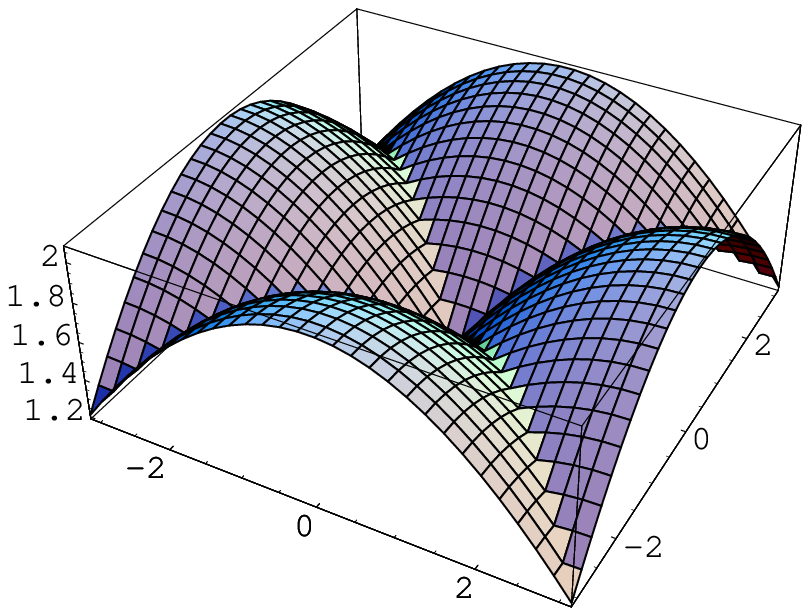}
\label{yFlatE(NN)tm03}
\end{figure}

\begin{figure}
\caption{
Flat 5D Massless Scalar Propagator ( Z$_2$-parity even, Neumann-Neumann b.c.), 
$\phat$=1.7 >> 1/$l$, time-like, App.C.1(3T). 
}
\includegraphics[height=8cm]{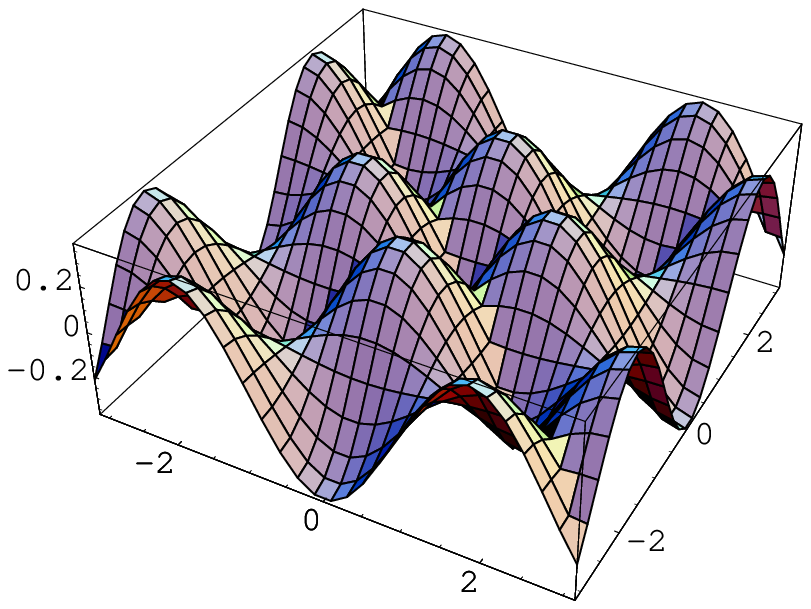}
\label{yFlatE(NN)tm17}
\end{figure}

\subsection{App. C.2\ :\  
Flat 5D Massless Scalar Propagator  
($Z_2$-parity Odd, Dirichlet-Neumann b.c.) }
The behaviour of 5D massless scalar flat propagators with the mixed b.c.
are shown in Fig.\ref{yFlatO(DN)sp003}-\ref{yFlatO(DN)tm17}.
$Z_2$-parity is taken to be odd: P=$-$1. Dirichlet b.c. is imposed for
y=0, while Neumann b.c. for y=$\pm l$. The P/M propagator is given by
\bea
G_p(y,y')=\Kbar_p(Y(y,y'),Y'(y,y'))\com\nn
\Kbar_p(Y,Y')=\left\{
\begin{array}{ccc}
-\frac{\sinh \ptil Y\cosh \ptil (Y'-l)}{2\ptil\cosh \ptil l} & \ptil=\sqrt{p^2} & \mbox{for}\ p^2>0 \\
-\frac{\sin \phat Y\cos \phat (Y'-l)}{2\phat\cos \phat l} & \phat=\sqrt{-p^2} & \mbox{for}\ p^2<0
\end{array}
              \right.
\label{ScaOdn1}
\eea 
where $Y(y,y')$ and $Y'(y,y')$ are defined in (\ref{PMapp16}). 

In this case, the scale parameter is the periodicity parameter $l$ only.
We characterize the behaviours by the momentum $\ptil$ or $\phat$ in comparison
with 1/$l$. 

(A) Space-Like

(1S) $\ptil$ << 1/$l$\com\q Fig.\ref{yFlatO(DN)sp003}\nl
Slanted flat surfaces front each other along the diagonal lines 
($y\pm y'=0$, the singularities). 
The size of the
surface is $l$. Boundary constraint is strong. 
This is the 'boundary phase'. 
The scale p does not appear in the graph.

(2S) $\ptil$ $\sim$ 1/$l$\com\q Fig.\ref{yFlatO(DN)sp03}\nl
The shape and the height are similar to (1S).

(3S) $\ptil$ >> 1/$l$\com\q Fig.\ref{yFlatO(DN)sp3}\nl
Walls and valleys run along the diagonal axes. The configuration
is free from the boundary constraint. 
This is the 'dynamical phase'.  
The size of the wall (valley) 
thickness is 1/p.  Absolute value of the effective height decreases.  

(B) Time-Like

(1T) $\phat$ << 1/$l$\com\q Fig.\ref{yFlatO(DN)tm003}\nl
Shape and height are similar to the space-like case. 
This is the 'boundary phase'.

(2T) $\phat$ $\sim$ 1/$l$ \com\q Fig.\ref{yFlatO(DN)tm03}\nl
The absolute value of the height increases and decreases
by changing p within this region. 
The global shape does not change.

(3T) $\phat$ >> 1/$l$\com\q Fig.\ref{yFlatO(DN)tm17}\nl
The wavy behaviour appears.  
The singularity-lines are buried in the waves. 
Boundary constraint is not effective. 
This is the 'dynamical phase'. 
The size of the wave length is 1/$\phat$.

\begin{figure}
\caption{
Flat 5D Massless Scalar Propagator ($Z_2$-parity Odd, Dirichlet-Neumann b.c.), 
$\ptil$=0.03 << 1/$l$ , space-like, App.C.2(1S). The "boundary phase".  
}
\includegraphics[height=8cm]{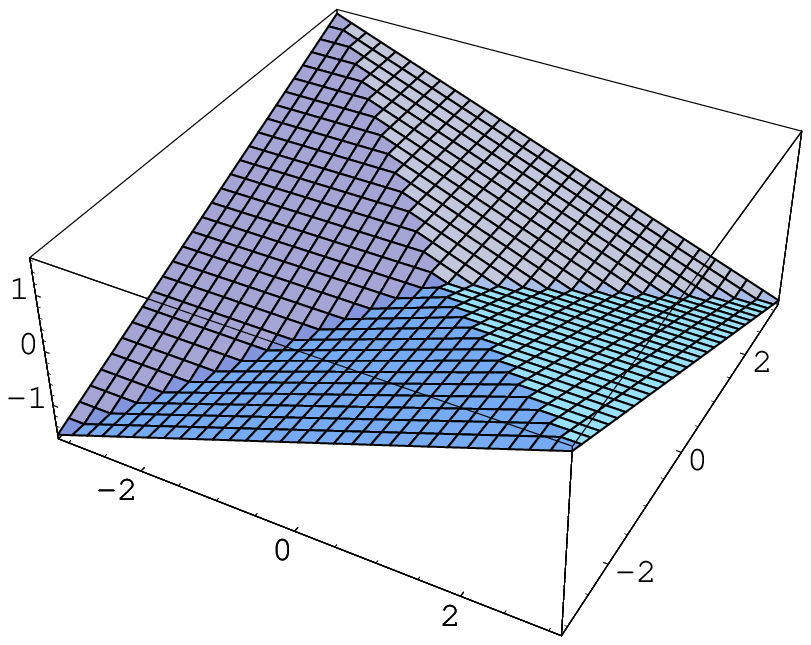}
\label{yFlatO(DN)sp003}
\end{figure}

\begin{figure}
\caption{
Flat 5D Massless Scalar Propagator ($Z_2$-parity Odd, Dirichlet-Neumann b.c.), 
$\ptil$=0.3 $\sim$ 1/$l$, space-like, App.C.2(2S). 
}
\includegraphics[height=8cm]{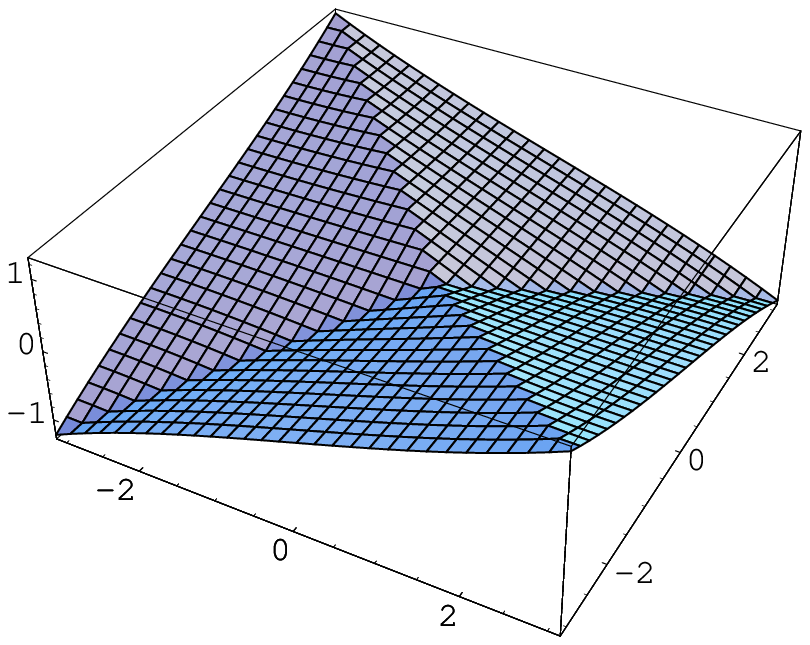}
\label{yFlatO(DN)sp03}
\end{figure}

\begin{figure}
\caption{
Flat 5D Massless Scalar Propagator ($Z_2$-parity Odd, Dirichlet-Neumann b.c.), 
$\ptil$=3 >> 1/$l$, space-like, App.C.2(3S). The "dynamical phase". 
}
\includegraphics[height=8cm]{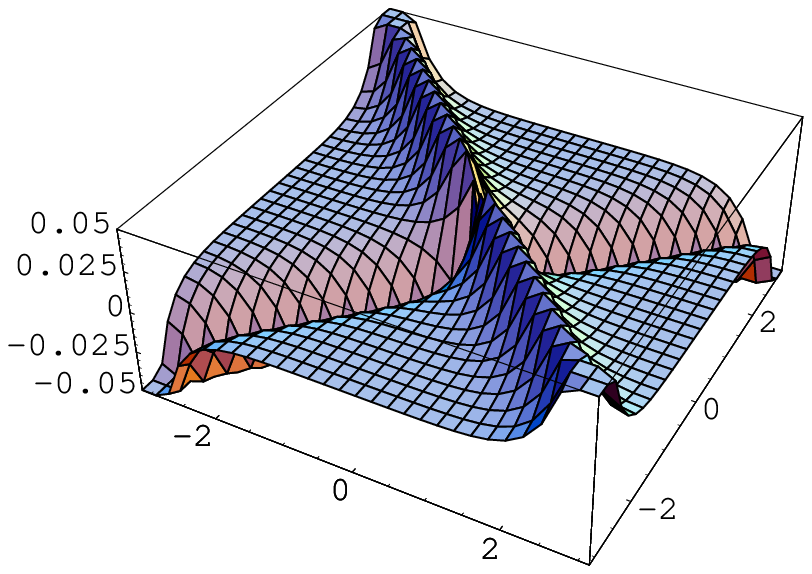}
\label{yFlatO(DN)sp3}
\end{figure}

\begin{figure}
\caption{
Flat 5D Massless Scalar Propagator ($Z_2$-parity Odd, Dirichlet-Neumann b.c.), 
$\phat$=0.03 << 1/$l$ , time-like, App.C.2(1T). The "boundary phase". 
}
\includegraphics[height=8cm]{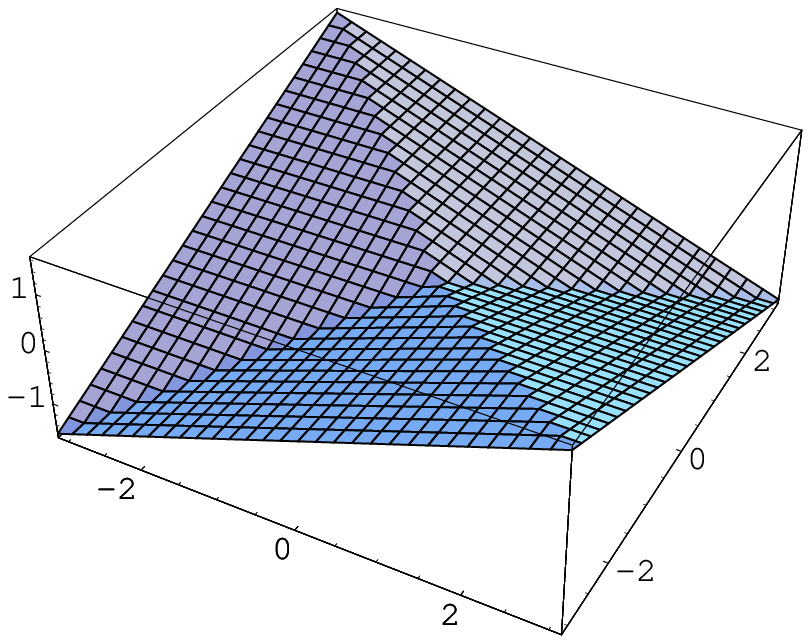}
\label{yFlatO(DN)tm003}
\end{figure}

\begin{figure}
\caption{
Flat 5D Massless Scalar Propagator ($Z_2$-parity Odd, Dirichlet-Neumann b.c.), 
$\phat$=0.3 $\sim$ 1/$l$, time-like, App.C.2(2T). 
}
\includegraphics[height=8cm]{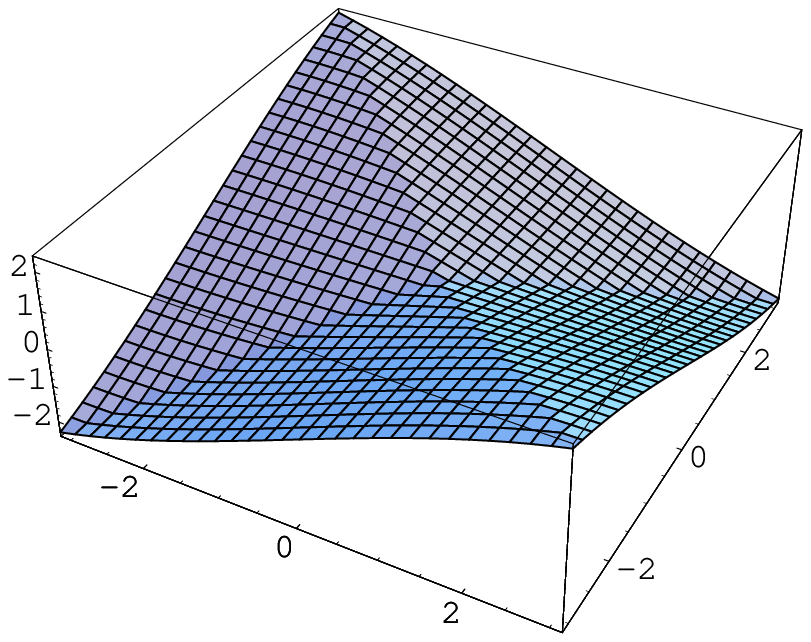}
\label{yFlatO(DN)tm03}
\end{figure}

\begin{figure}
\caption{
Flat 5D Massless Scalar Propagator ($Z_2$-parity Odd, Dirichlet-Neumann b.c.), 
$\phat$=1.7 >> 1/$l$, time-like, App.C.2(3T). The "dynamical phase". 
}
\includegraphics[height=8cm]{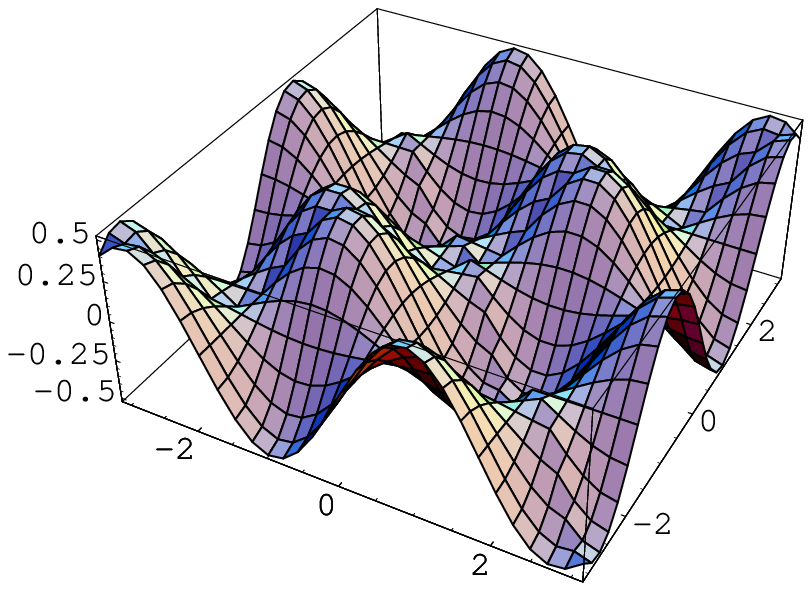}
\label{yFlatO(DN)tm17}
\end{figure}

\subsection{App. C.3\ :\ 
z-Coordinate Representation and Warped 5D Scalar Propagator (Z$_2$-parity odd, 
Dirichlet-Dirichlet b.c., space-like 4-momentum)}
We give here the P/M propagator behaviour in terms of
z-coordinate. Its relation to y is given in (\ref{adsf2}). 
We take 5D scalar propagator 
with P=$-1$.  Dirichlet b.c. is imposed on all fixed points.  
4D momentum is space-like.
The propagator function is given in (\ref{adspro18}) with (\ref{adspro15b}). 
It can be 
reexpressed as follows using the ordinary Bessel functions and
the sign function.
\bea
G_p(z,z')=K_p(Z(z,z'),Z'(z,z'))\com\nn
K_p(Z,Z')=\frac{\om^3}{2}Z^2{Z'}^2\ep(Z)\ep(Z')\times\nn
\frac{\{\I_0(\Pla)\K_0(\ptil |Z|)-\K_0(\Pla)\I_0(\ptil |Z|)\}  
      \{\I_0(\Tev)\K_0(\ptil |Z'|)-\K_0(\Tev)\I_0(\ptil |Z'|)\}
     }{\I_0(\Tev)\K_0(\Pla)-\K_0(\Tev)\I_0(\Pla)}
\com
\label{ScaOddZ1}
\eea 
where $Z(z,z')$ and $Z'(z,z')$ are defined in (\ref{adspro18}). 

As we see the following graphs, large part of the whole image is naturally
displayed. This means the z-coordinate is more suitable than y-coordinate
for the warped geometry description. (Compare the height-region 
in Sec.\ref{S.P/Mpropagator}.2 (y-coordinate is used)
and in this subsection for corresponding graphs.) 
From the z-coordinate property, however, it is {\it hard to detect
characteristic scales} in the graphs.  

The following graphs are re-drawing of those of Sec.\ref{S.P/Mpropagator}.2. 
They are displayed
for $\Rbar_1U\Rbar_2$ region of the $(z,z')$-plane (See Fig.\ref{8regionsZ}).  

(1S) $\ptil$ << 1/$l$ < $\om$\com\q Fig.\ref{IM32sp0005Z}\nl

(2S) 1/$l$ < $\ptil$ $\sim$ $\om$\com\q Fig.\ref{IM32sp1Z}\nl

(3S) 1/$l$ < $\om$ < $\ptil$\com\q Fig.\ref{IM32sp5Z}\nl
If we go further larger $\ptil$ ($1/l< \om \ll \ptil$), the situation is 
the "flat (z-)space limit" represented by eq.(6.16) of Ref.\cite{RS01}.
Compare with Fig.\ref{Flat2sp3}.
\footnote{
Note that the propagator in this flat limit is different from 
the flat propagator given in Sec.\ref{S.5Dprop}-\ref{S.KKvsPM} of this paper. The present flat propagator
satisfies the free propagator equation in y-coordinate (\ref{PMapp4}), 
whereas eq.(6.16) of Ref.\cite{RS01} satisfies that in z-coordinate 
(\ref{adspro5b}). 
}

\begin{figure}
\caption{
z-Coordinate Representation, Warped 5D Scalar Propagator (Z$_2$-parity odd, 
Dirichlet-Dirichlet b.c., 
$\ptil$=0.005 << T < 1/$l$ <$\om$, space-like, App.C.3(1S). 
}
\includegraphics[height=8cm]{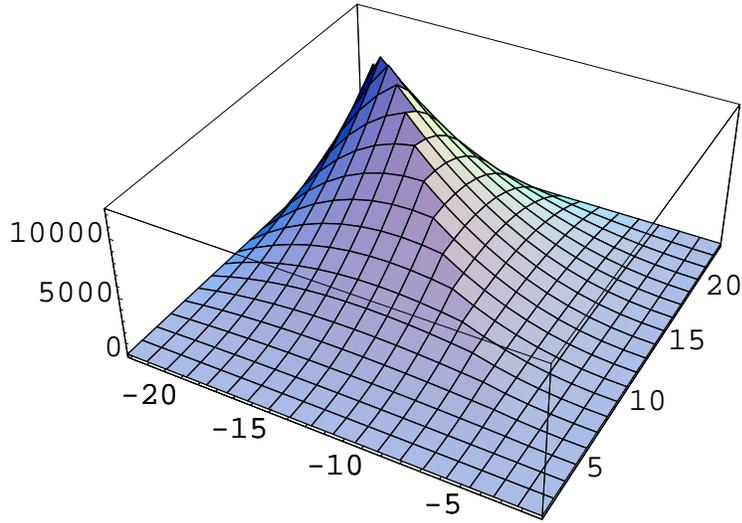}
\label{IM32sp0005Z}
\end{figure}

\begin{figure}
\caption{
z-Coordinate Representation, Warped 5D Scalar Propagator (Z$_2$-parity odd, 
Dirichlet-Dirichlet b.c., 
T << 1/$l$ < $\ptil$=1=$\om$, space-like, App.C.3(2S). 
}
\includegraphics[height=8cm]{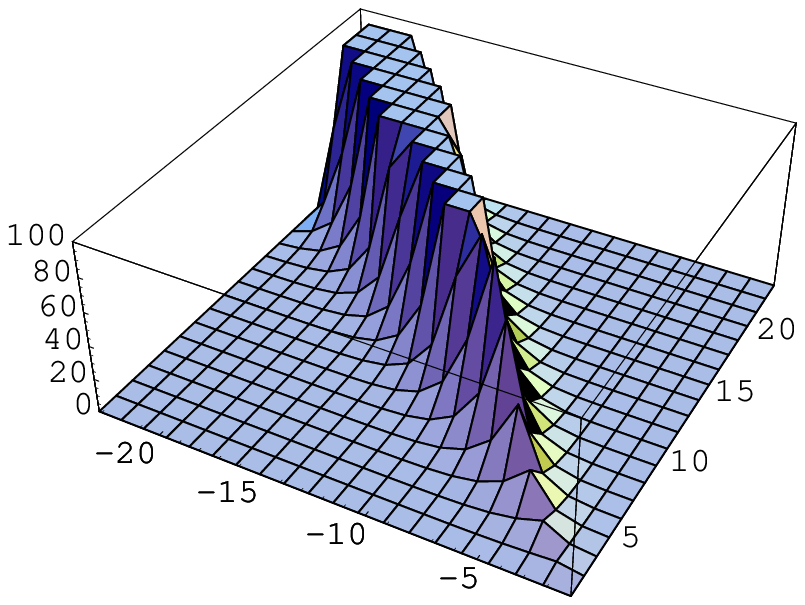}
\label{IM32sp1Z}
\end{figure}

\begin{figure}
\caption{
z-Coordinate Representation, Warped 5D Scalar Propagator (Z$_2$-parity odd, 
Dirichlet-Dirichlet b.c.), 
T << 1/$l$ < $\om$ < $\ptil$=5, space-like, App.C.3(3S).
}
\includegraphics[height=8cm]{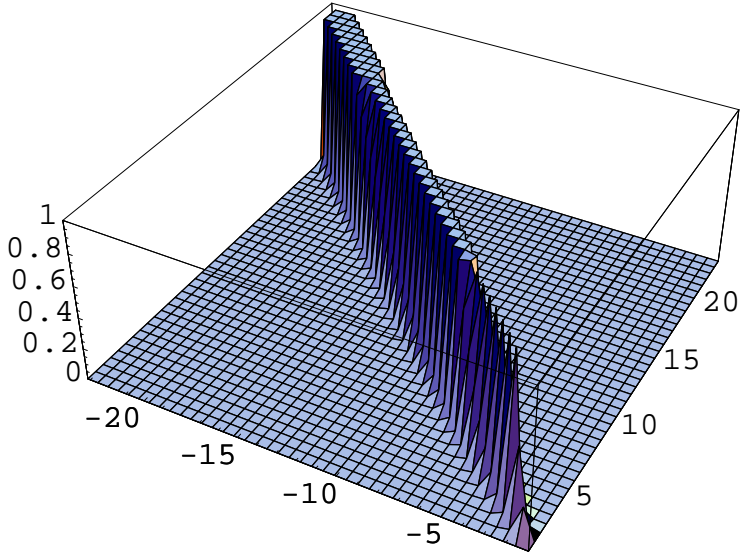}
\label{IM32sp5Z}
\end{figure}

\subsection{
App. C.4\ :\ 
Warped 5D Massless Vector ($Z_2$-parity Even, 
Neumann-Neumann b.c., space-like 4 momentum)}
In the warped case the theory has two scale parameters:
thickness $\om$ and the periodicity parameter $l$. The propagator behaviour
is characterized by the relation between $\sqrt{|p^2|}$, $\om$ and $1/l$.  

In this subsection, the 5D vector propagator with $Z_2$-parity even (P=1) 
is examined. The Neumann b.c. is imposed on
all fixed points. The P/M propagator is given by
\bea
G_p(y,y')=K_p(Y(y,y'),Y'(y,y'))\com\nn
K_p(Y,Y')=\frac{1}{\om}\exp(k|Y|+k|Y'|)\times\nn
\frac{\{\I_0(\Pla)\K_1(\Pla \mbox{e}^{\om|Y|})+\K_0(\Pla)\I_1(\Pla\mbox{e}^{\om|Y|})\}  
      \{\I_0(\Tev)\K_1(\Pla\mbox{e}^{\om|Y'|})+\K_0(\Tev)\I_1(\Pla\mbox{e}^{\om|Y'|})\}
     }{\I_0(\Pla)\K_0(\Tev)-\K_0(\Pla)\I_0(\Tev)}
\com
\label{VecEnnSP1}
\eea 
where $Y(y,y')$ and $Y'(y,y')$ are defined in (\ref{PMapp16}).

(1S) $\ptil$ << 1/$l$ < $\om$\com\q Fig.\ref{RS61sp0005}\nl
4 notches 
appear at 4 corners. 
The effective width of the notch is 1/$\om$. 
The size of the global upheaval and downheaval is $l$. The boundary
constraint is dominant. 
This is the "boundary phase". 
There is a flat region around
the center $(y=y'=0)$. The propagator takes a non-zero constant there. 
This means that the bulk propagation, near the Planck brane, 
simply gives a common constant ,
as the extra-space contribution, 
to the amplitude. 
On the other hand, near the Tev brane
$(y,y'=\pm l)$, it gives a "sizable" effect.  

(2S) 1/$l$ < $\ptil$ $\sim$ $\om$\com\q Fig.\ref{RS61sp1}\nl
4 valleys develop along the diagonal axes 
from the corners to the center. Their width is 1/$\ptil$ $\sim$ 1/$\om$. 
The flat region near the center disappears. 
In the off-diagonal region ($y\pm y'\neq 0$) , flat planes begin to appear 
and the propagator takes nearly 0 value there.  
The height decreases.

(3S) 1/$l$ < $\om$ < $\ptil$\com\q Fig.\ref{RS61sp5}\nl
Valleys develop furthermore. The width 
of them is 1/$\ptil$ near the corners and is 1/$\om$ near the center. 
There is no boundary effect. This is the "dynamical phase". 
There is no flat region near the center, whereas in the off-diagonal region
there appears the flat region. The propagator value is 0 in this flat region. 
 This means the bulk propagation
takes place only for the case $y'\pm y\sim 0$. 
Absolute value of the effective height decreases rapidly as $\ptil$ increases. 
If we take further larger $\ptil$ ($1/l< \om \ll \ptil$), the situation is 
the "flat limit". Compare with Fig.\ref{yFlatE(NN)sp3}. 

Time-like case is given in App.C.6.

\begin{figure}
\caption{
Warped 5D Massless Vector ($Z_2$-parity Even, 
Neumann-Neumann b.c.)
$\ptil$=0.005 << T$\sim$0.04 << 1/$l$$\sim$0.3 << $\om$=1, space-like., App.C.4(1S)
}
\includegraphics[height=8cm]{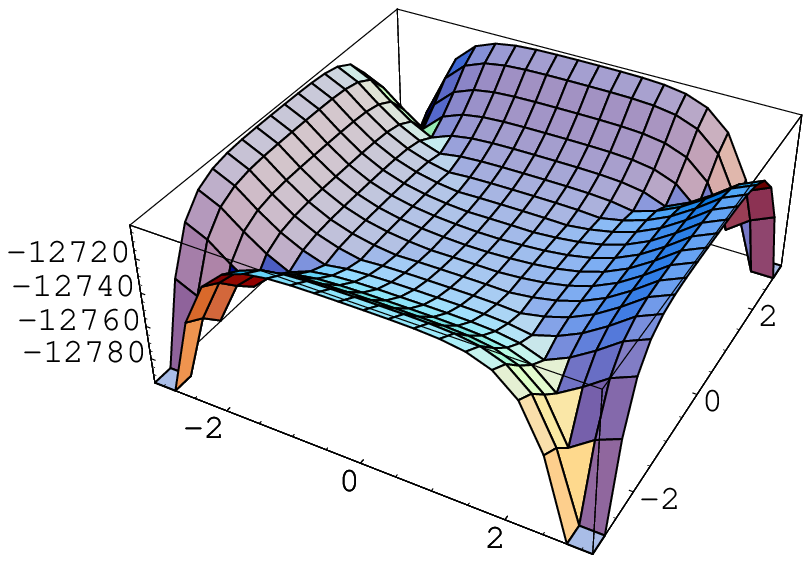}
\label{RS61sp0005}
\end{figure}


\begin{figure}
\caption{
Warped 5D Massless Vector ($Z_2$-parity Even, 
Neumann-Neumann b.c.) 
T << 1/$l$ < $\ptil$=1=$\om$, space-like, App.C.4(2S)
}
\includegraphics[height=8cm]{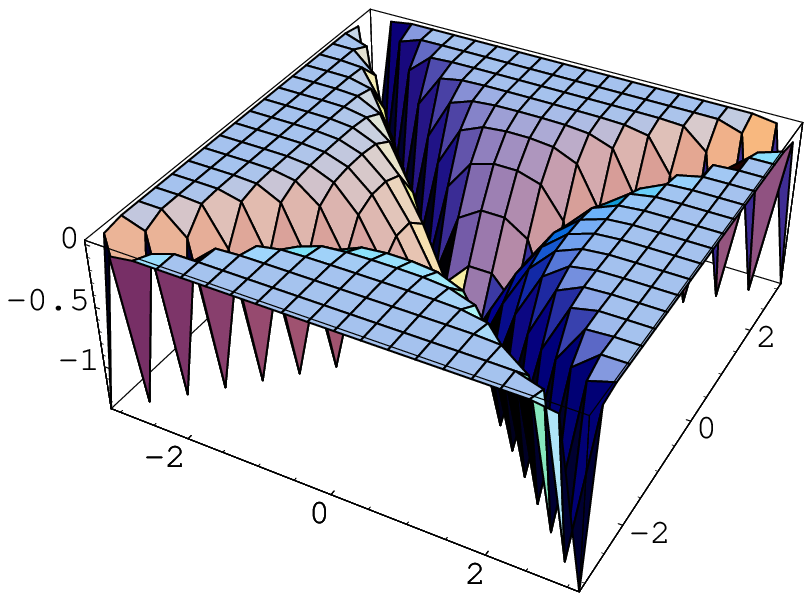}
\label{RS61sp1}
\end{figure}

\begin{figure}
\caption{
Warped 5D Massless Vector ($Z_2$-parity Even, 
Neumann-Neumann b.c.), 
T << 1/$l$ < $\om$ < $\ptil$=5, space-like,  App.C.4(3S).  
}
\includegraphics[height=8cm]{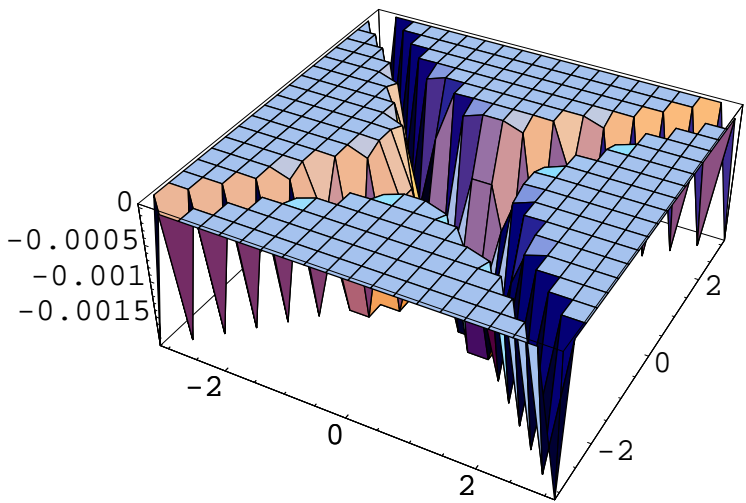}
\label{RS61sp5}
\end{figure}

\subsection{App. C.5\ :\ 
z-Coordinate Representation for 
Warped 5D Massless Vector ($Z_2$-parity Even, Neumann-Neumann b.c.,space-like)}
We give here the P/M vector ($Z_2$-parity even, Neumann-Neumann b.c., space-like) 
propagator behaviour in terms of z-coordinate. The propagator expression
is given in (\ref{VecEnnSP1}) using the y-coordinate. Here its z-coordinate
expression is given.
\bea
G_p(z,z')=K_p(Z(z,z'),Z'(z,z'))\com\nn
K_p(Z,Z')=\om |Z||Z'|\times\nn
\frac{\{\I_0(\Pla)\K_1(\ptil |Z|)+\K_0(\Pla)\I_1(\ptil |Z|)\}  
      \{\I_0(\Tev)\K_1(\ptil |Z'|)+\K_0(\Tev)\I_1(\ptil |Z'|)\}
     }{\I_0(\Pla)\K_0(\Tev)-\K_0(\Pla)\I_0(\Tev)}
\com
\label{VecEnnSPz1}
\eea 
where $Z(z,z')$ and $Z'(z,z')$ are defined in (\ref{adspro18}). 

The following graphs are re-drawing of those of App.C.4 using the z-coordinate. 
They are displayed
for $R_1$U$R_2$ region of the $(z,z')$-plane (See Fig.\ref{8regionsZ}).  

(1S) $\ptil$ << 1/$l$ < $\om$\com\q Fig.\ref{RS61sp0005z}\nl

(2S) 1/$l$ < $\ptil$ $\sim$ $\om$\com\q Fig.\ref{RS61sp1z}\nl

(3S) 1/$l$ < $\om$ < $\ptil$\com\q Fig.\ref{RS61sp5z}\nl
If we take further larger $\ptil$, the configuration becomes the 
"flat (z-)space limit". Compare with Fig.\ref{yFlatE(NN)sp3}. 
( The same situation has already appeared for the scalar propagator case,
Fig.\ref{IM32sp1Z} and Fig.\ref{IM32sp5Z} in App.C.3. )

In the graphs of (2S) and (3S), we see, in their off-diagonal regions,  
the top surfaces very gradually approach the 0-value surface as they
deviate from the diagonal axis: $G_p(z,z')\sim -\exp (-\ptil |z'-z|)$. 
This is the "untrusted" region pointed out in ref.\cite{RS01}. 

\begin{figure}
\caption{
z-Coordinate Representation for 
Warped 5D Massless Vector ($Z_2$-parity Even, Neumann-Neumann b.c.), 
$\ptil$=0.005 << T$\sim$0.04 << 1/$l$$\sim$0.3 << k=1, space-like, App.C.5(1S).
}
\includegraphics[height=8cm]{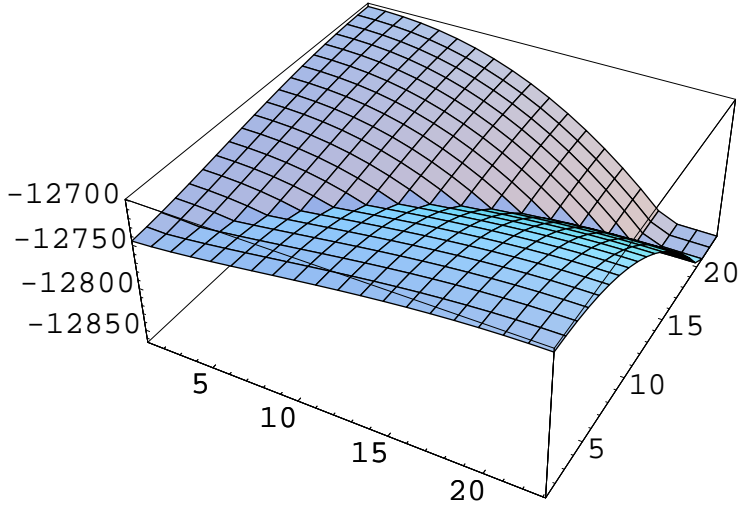}
\label{RS61sp0005z}
\end{figure}

\begin{figure}
\caption{
z-Coordinate Representation for 
Warped 5D Massless Vector ($Z_2$-parity Even, Neumann-Neumann b.c.), 
T << 1/$l$ < $\ptil$=1=$\om$, space-like, App.C.5(2S). 
}
\includegraphics[height=8cm]{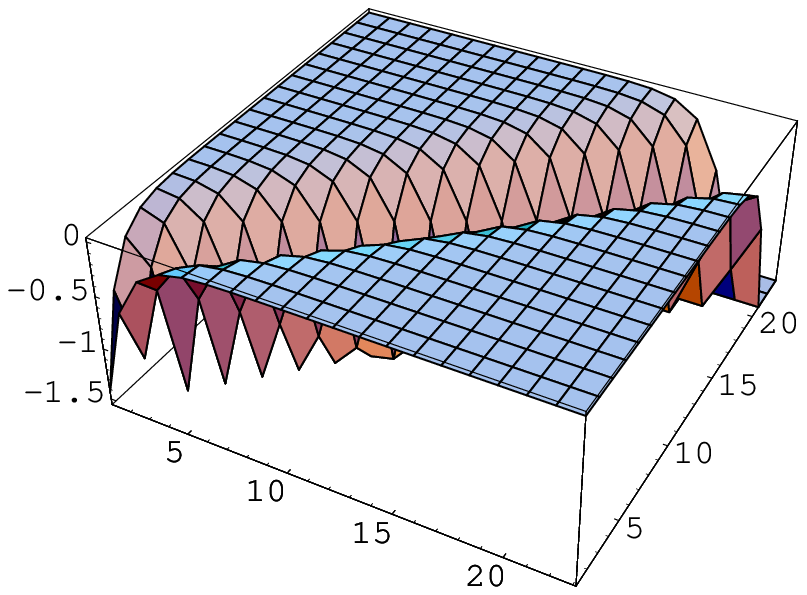}
\label{RS61sp1z}
\end{figure}

\begin{figure}
\caption{
z-Coordinate Representation for 
Warped 5D Massless Vector ($Z_2$-parity Even, Neumann-Neumann b.c.), 
T << 1/$l$ < $\om$ < $\ptil$=5, space-like, App.C.5(3S)
}
\includegraphics[height=8cm]{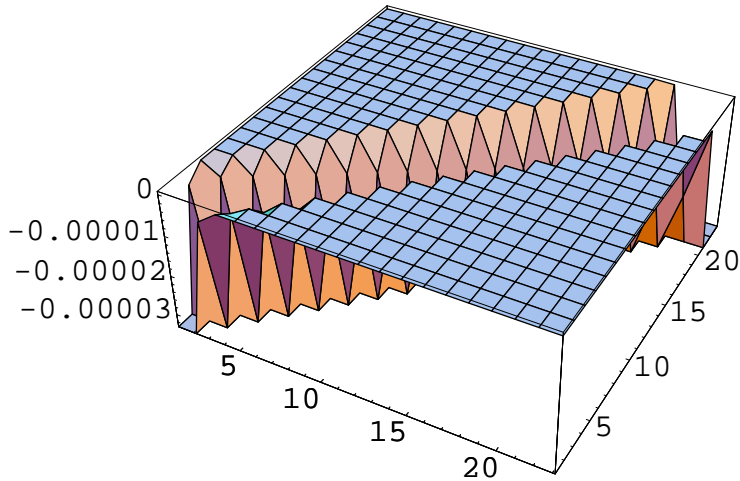}
\label{RS61sp5z}
\end{figure}

\subsection{
App. C.6\ :\  
Warped 5D Massless Vector (Z$_2$-parity Even, Neumann-Neumann b.c., time-like 4 momentum)}
Here we give the P/M propagator behaviour of the warped 5D massless vector
(Z$_2$-parity Even, Neumann-Neumann b.c., time-like). The propagator is given by
\bea
G_p(y,y')=K_p(Y(y,y'),Y'(y,y'))\com\nn
K_p(Y,Y')=-\frac{1}{\om}\exp(k|Y|+k|Y'|)\times\nn
\frac{\{\N_0(\tPla)\J_1(\tPla \mbox{e}^{\om|Y|})-\J_0(\tPla)\N_1(\tPla\mbox{e}^{\om|Y|})\}  
      \{\N_0(\tTev)\J_1(\tPla\mbox{e}^{\om|Y'|})-\J_0(\tTev)\N_1(\tPla\mbox{e}^{\om|Y'|})\}
     }{\N_0(\tPla)\J_0(\tTev)-\J_0(\tPla)\N_0(\tTev)}
\com
\label{VecEnnTM1}
\eea 
where $Y(y,y')$ and $Y'(y,y')$ are defined in (\ref{PMapp16}). 

This is the time-like case of App.C.4. 

(1T) $\phat$ << 1/$l$ < $\om$\com\q Fig.\ref{RS415tm0005}\nl
The situation is quite similar to (1S) of App.C.4. 
%

(2T) 1/$l$ < $\phat$ $\sim$ $\om$\com\q Fig.\ref{RS415tm1}\nl
Wavy behaviour appears. Two types of waves are there. One type has
the small wave-length of order 1/$\phat$=1/$\om$, and the waves of this 
type gather near the 4 corners. The other type has the long wave-length
of order $l$, which comes from the boundary constraint. In particular, 
there exists a very moderate hill around the center. 
The propagator takes nearly 0 value there. 
This is contrasting with the space-like case.
The overall height decreases.

(3T) 1/$l$ < $\om$ < $\phat$\com\q Fig.\ref{RS415tm10}\nl
Two types of waves are there. 
One type has
the small wave-length of order 1/$\phat$, and the waves of this 
type gather near the 4 corners and the 4 rims. 
Their heights differ so much. 
The other type has the long wave-length
of order $\om$ and the very low height. 
These waves gather around the center and form a slightly
wavy plain.  
The propagator takes nearly 0 value there. 
This is contrasting with the space-like case.
The overall height decreases.
Although the scale $l$ looks to appear as the radius
of the plain around the center, the main configuration
is free from the boundary effect. 
This is the "dynamical phase". When $\phat$ becomes further 
larger, the configuration approaches a "flat limit". 
It differs from the flat result, Fig.\ref{yFlatE(NN)tm17}.

\begin{figure}
\caption{
Warped 5D Massless Vector (Z$_2$-parity Even, Neumann-Neumann b.c.), 
$\phat$=0.005 << T$\sim$0.04 << 1/$l$$\sim$0.3 << $\om$=1, time-like. App.C.6(1T). 
}
\includegraphics[height=8cm]{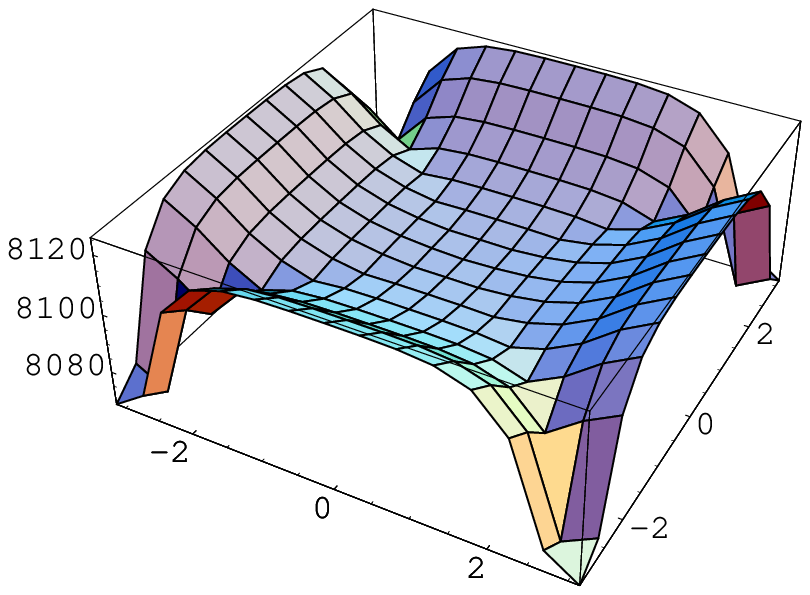}
\label{RS415tm0005}
\end{figure}


\begin{figure}
\caption{
Warped 5D Massless Vector (Z$_2$-parity Even, Neumann-Neumann b.c.), 
T$\sim$0.04 << 1/$l$$\sim$0.3 < $\phat$= $\om$=1, time-like. App.C.6(2T). 
}
\includegraphics[height=8cm]{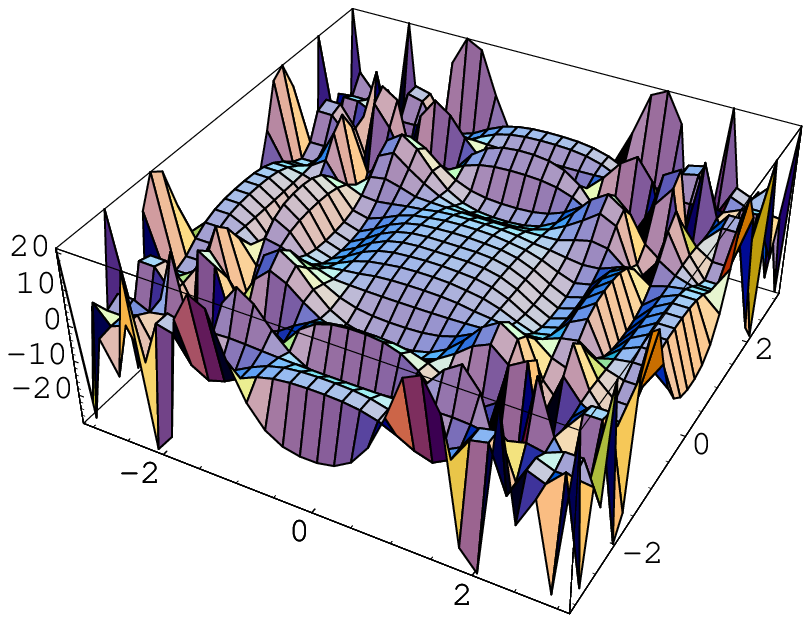}
\label{RS415tm1}
\end{figure}

\begin{figure}
\caption{
Warped 5D Massless Vector (Z$_2$-parity Even, Neumann-Neumann b.c.), 
T$\sim$0.04 << 1/$l$$\sim$0.3 < $\om$=1 <<  $\phat$=10, time-like. App.C.6(3T). 
}
\includegraphics[height=8cm]{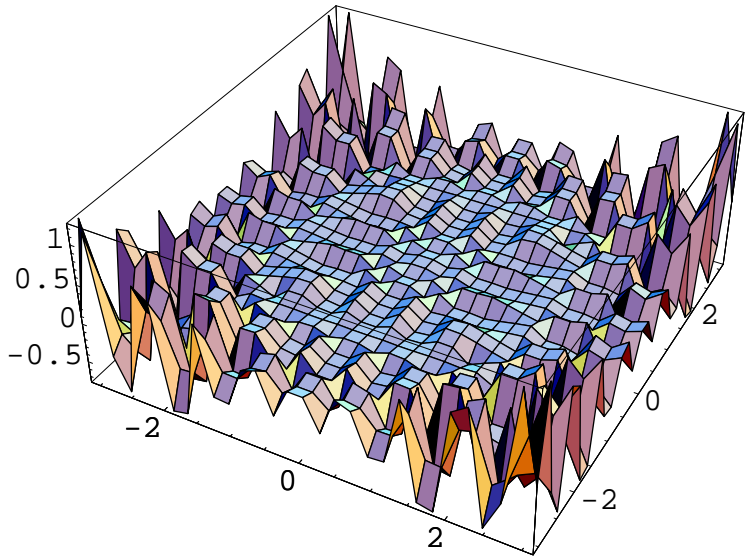}
\label{RS415tm10}
\end{figure}

\newpage
\section{Acknowledgement}
 Parts of the content of this work have been already presented at
YITP workshop on QFT and String (06.9.12, Kyoto, Japan), 
Joint Meeting of Pacific Region Particle Physics Communities
 (06.11.01,Hawaii,Honolulu,USA) and 
RIKEN Seminar(06.11.27, Wako, Japan). 
The authors thank N. Nakanishi, Hiroshi Suzuki and K. Oda for useful comments
on the occasions.

\end{document}